\documentclass[b5paper,final]{tADP2e_my}
\pdfoutput=1
\usepackage[T1]{fontenc}\usepackage{lmodern}\usepackage[latin1]{inputenc}
\usepackage{graphicx,epstopdf,color,bm,booktabs,wrapfig,subfigure,microtype,afterpage,booktabs,fancyhdr,truncate} \graphicspath{{Figures/}}

\graphicspath{{../Figures/}}

\usepackage{hyperref}
\hypersetup{
    colorlinks,
    citecolor=black,
    filecolor=black,
    linkcolor=black,
    urlcolor=black
}

\begin{document}

\pagestyle{fancy}\fancyhead{}\fancyfoot{}
\fancyhead[LO]{\small D.\,S.~Inosov}\fancyhead[RE]{\small D.\,S.~Inosov}%
\fancyhead[C]{\textit{\small Quantum Magnetism in Minerals}}%
\fancyhead[LE]{\small\thepage}\fancyhead[RO]{\small\thepage}
\setlength\belowrulesep{-0.2ex}\setlength\aboverulesep{0.4ex}

\jvol{68} \jnum{01} \jyear{2019} \jmonth{February}
\doi{DOI: \href{http://dx.doi.org/10.1080/00018732.2018.1571986}{10.1080/00018732.2018.1571986}}

\articletype{\vspace*{-1.8em}REVIEW ARTICLE}

\title{Quantum magnetism in minerals}

\author{
\name{D.\,S.~Inosov$^{\ast}$\thanks{$^\ast$Corresponding author. E-mail: \href{mailto:dmytro.inosov@tu-dresden.de}{dmytro.inosov@tu-dresden.de}}}
\affil{\fontsize{9}{11}\selectfont Institut f\"ur Festk\"orper- und Materialphysik, TU Dresden, D-01069 Dresden, Germany}
\received{Received 27 June 2018; final version received 12 January 2019; accepted 14 January 2019}
}

\maketitle

\begin{abstract}
The discovery of magnetism by the ancient Greeks was enabled by the natural occurrence of lodestone\,---\,a magnetized version of the mineral magnetite. Nowadays, natural minerals continue to inspire the search for novel magnetic materials with quantum-critical behaviour or exotic ground states such as spin liquids. The recent surge of interest in magnetic frustration and quantum magnetism was largely encouraged by crystalline structures of natural minerals realizing pyrochlore, kagome, or triangular arrangements of magnetic ions. As a result, names like azurite, jarosite, volborthite, and others, which were barely known beyond the mineralogical community a few decades ago, found their way into cutting-edge research in solid-state physics. In some cases, the structures of natural minerals are too complex to be synthesized artificially in a chemistry lab, especially in single-crystalline form, and there is a growing number of examples demonstrating the potential of natural specimens for experimental investigations in the field of quantum magnetism. On many other occasions, minerals may guide chemists in the synthesis of novel compounds with unusual magnetic properties. The present review attempts to embrace this quickly emerging interdisciplinary field that bridges mineralogy with low-temperature condensed-matter physics and quantum chemistry.
\end{abstract}

\begin{classcode}
\mbox{75.30.-m}\,---\,intrinsic properties of magnetically ordered materials;
\mbox{\hspace{3.65em}75.10.Jm\,---\,models of magnetic ordering, including quantum spin frustration};
\mbox{\hspace{3.65em}91.60.Pn\,---\,magnetic properties of minerals.}
\end{classcode}

\begin{keywords}
quantum magnetism; magnetic frustration; low-dimensional spin models; magnetic minerals\vspace*{-3pt}
\end{keywords}

\fontsize{10}{12}\selectfont

\tableofcontents

\section{Introduction}

\subsection{Historical perspective}

Many developments in solid-state physics and crystallography were historically inspired by observations on naturally occurring minerals. Various phenomena in condensed-matter physics were first discovered in natural samples, which had unsurpassed quality as compared to synthetic materials before the advent of modern chemical and crystal-growth technology \cite{Bohm85, Scheel00}. Among the most eminent examples are the first observations of ferromagnetism in lodestone dating back to the Greek philosopher and engineer \textit{Thales of Miletus} in the 6$^{\rm th}$ century BC \cite{Mattis81}. Alchemists of the early 17$^{\rm th}$ century were awed by Vincenzo Casciorola's discovery of phosphorescence in the ``Bologna stone'', or \textit{lapis solaris} (presumably baryte, BaSO$_4$) \cite{Roda11}. At the beginning of the 19$^{\rm th}$ century, optical activity was discovered by Fran\c{c}ois Arago in one of Earth's most common minerals, quartz \cite{Arago11}. Then, in 1875, Weber noticed a deviation in the heat capacity of diamond from the Dulong-Petit law at room temperature \cite{Weber1875}, which was ultimately explained in Einstein's seminal paper of 1907 that became one of the cornerstones of quantum mechanics \cite{Einstein07}. However, synthetic diamonds remained unavailable until 1955 \cite{BundyHall55} and lacked quality until much later. From more recent examples, the bulk electronic band structure of the prototypical layered material graphite has been studied using quantum oscillations \cite{SouleMcClure64} and modern angle-resolved photoelectron spectroscopy on natural single crystals \cite{ZhouGweon06, GrueneisAttaccalite08}, as they are often superior in quality to synthetic graphite samples.

In modern times, experimental solid-state physics predominantly relies on synthetic crystals, whose quality and chemical purity can be well controlled for an ever increasing number of materials. Therefore, today's condensed-matter physics has largely departed from mineralogy and geophysical sciences in its need for samples. Yet, there have always been exceptions to this general tendency, especially in the field of magnetism. In the early 1950's, the first triple-axis neutron spectrometer was put together by Brockhouse at the Chalk River NRX reactor in Canada \cite{Brockhouse61, BrockhouseWoods64}, which enabled the first direct measurements of phonon and magnon excitations by inelastic neutron scattering (INS). For the spin-wave measurement, Brockhouse needed a large single crystal of a substance in which magnetic scattering dominated over nuclear scattering, with a high enough Curie or N\'eel temperature, and with small neutron absorption and incoherent scattering cross-sections \cite{Brockhouse57}. He rightly judged that magnetite (Fe$_3$O$_4$) is the best possibility in spite of its complex ferrimagnetic structure, which was already known by that time from the works of N\'eel \cite{Neel48, ShullWollan51}. The platelike specimen with a size of 6.3$\times$3.8$\times$0.3~cm$^3$ ($\sim$\,37\,g) was cut from a natural single crystal of magnetite, which contained multiple impurities (Ti, Mn, Al, Si) on the Fe sites but nevertheless facilitated the first direct observation of ferromagnetic (FM) spin waves and the first measurement of their low-energy dispersion. Later, spin-wave measurements on magnetite were repeated at low temperatures below the Verwey transition ($T_{\rm V}\approx120$\,K) but revealed no significant changes in the dispersion \cite{SamuelsenBleeker68}.

\begin{figure}[t!]
\includegraphics[width=\linewidth]{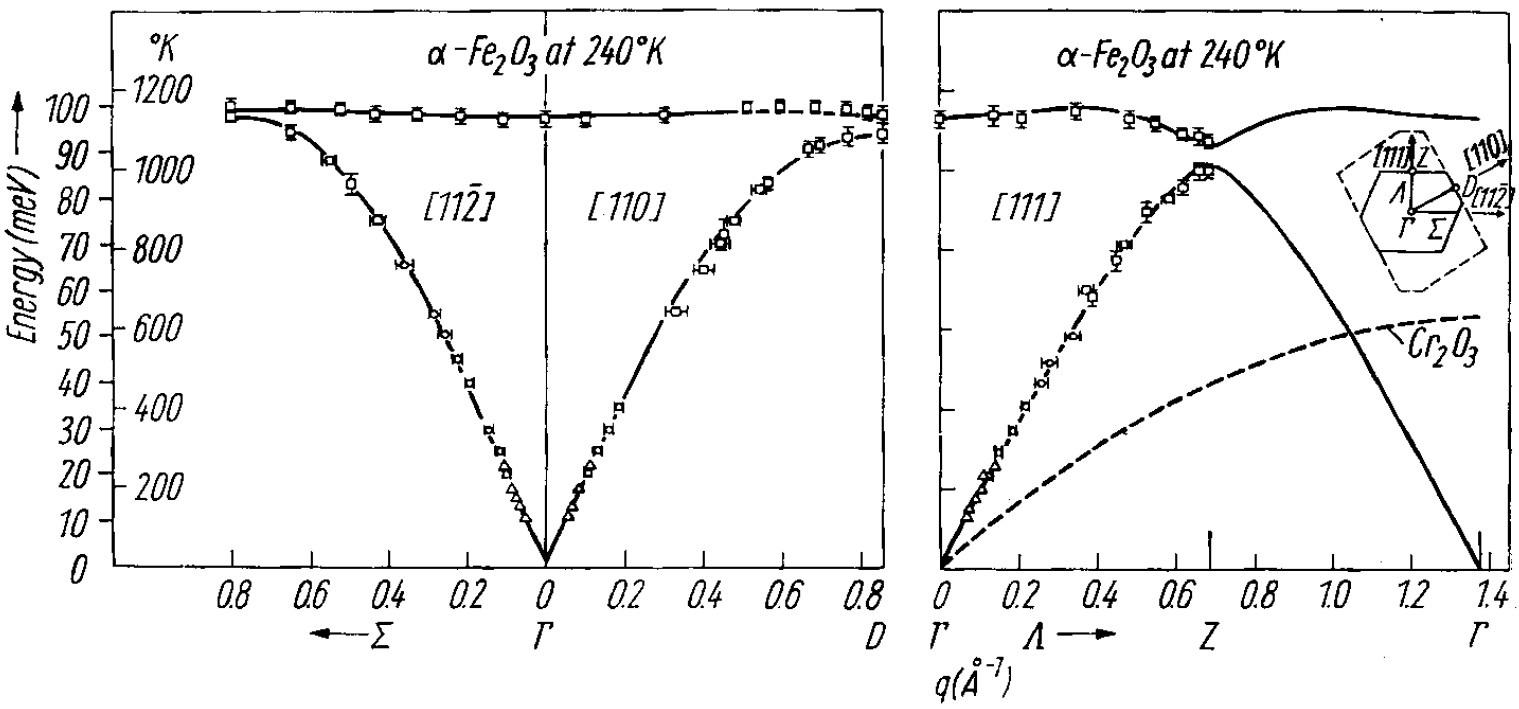}
\caption{First INS measurements of spin-wave dispersions in a 48~g natural single crystal of hematite along three principal directions, which were used to extract exchange parameters of the Heisenberg model (solid curve). The dashed line is a comparison with the dispersion relation of Cr$_2$O$_3$ \cite{SamuelsenHutchings70}. After Samuelsen and Shirane~\cite{SamuelsenShirane70}.}\label{Fig:Hematite}
\end{figure}

Unlike magnetite, another famous iron-bearing mineral hematite ($\alpha$-Fe$_2$O$_3$, with a corundum crystal structure) is an antiferromagnet with a very high N\'eel point, $T_{\rm N}=960$\,K. In addition, weak ferromagnetism is observed above the so-called Morin transition ($T_{\rm M}\approx263$\,K) due to a spin-reorientation transition \cite{Morin50, BesserMorrish67} involving a $90^\circ$ spin flop from a structure with magnetic moments aligned parallel to the $c$ axis below $T_{\rm M}$ to an orthogonal alignment with spins lying in the basal $c$ plane above $T_{\rm M}$. This caused much confusion before the 1950's when its antiferromagnetic (AFM) structure was finally confirmed by neutron diffraction \cite{ShullStrauser51, NathansPickart64}. The dispersion relations for AFM spin waves in hematite ($\alpha$-Fe$_2$O$_3$), shown in Fig.~\ref{Fig:Hematite}, were reported in 1970 by Samuelsen and Shirane \cite{SamuelsenShirane70}, from which they obtained Heisenberg exchange interactions up to the 5$^{\rm th}$ nearest neighbour. They also confirmed linear dispersion relationships for low-energy AFM spin waves, predicted by Hulth\'en \cite{Hulthen36}, and accurately measured the corresponding spin-wave velocities. Hematite is found not only as a pure mineral but also in the form of nanoscale exsolution lamellae in the ilmenite host (FeTiO$_3$). This mineral intergrowth, known as hemoilmenite, possesses a large natural remanent magnetization (NRM) that has been claimed responsible for Earth's local magnetic field anomalies \cite{RobinsonHarrison02, McEnroeHarrison05}. A recent polarized neutron diffraction study on a single-crystalline hemoilmenite specimen from South Rogaland, Norway, has confirmed that lamellar magnetism of hematite can fully account for the observed NRM at ambient temperature \cite{BrokSales14}. In the resulting model, magnetic moments in contact layers between exsolution lamellae that align parallel to the magnetizing field combine with the canted AFM moments above the Morin transition that are approximately orthogonal to the field to produce a net magnetic moment at an intermediate angle with a saturation value of $\sim$\,56$^\circ$.

Another example of a classical antiferromagnet that was investigated by neutron scattering is hauerite (MnS$_2$, pyrite structure), which has a habit of forming large cm-sized octahedral single crystals of high natural quality. It adopts a commensurate collinear AFM structure with a $\mathbf{k}=\left(1\,\frac{1}{2}\,0\right)$ wave vector below $T_{\rm N}=48$\,K \cite{ChattopadhyayBurlet91}. Above this temperature, diffuse critical scattering of magnetic origin was revealed by INS at an incommensurate wave vector $(1~0.44~0)$, with a temperature dependence characteristic of a lock-in transition of a three-dimensionally (3D) ordered system \cite{ChattopadhyayBurlet89, ChattopadhyayBrueckel92}. The same study also found two branches of spin-wave excitations: a nearly flat one at about 2~meV and a strongly dispersing one (2--7~meV) with a spin gap of 2~meV that is unusually large for the Mn$^{2+}$ ion in the spin-$\frac{5}{2}$ state \cite{ChattopadhyayBurlet89}. Very recently, the lock-in transition to the commensurate AFM phase received a natural explanation as a result of a tiny tetragonal distortion of the lattice with the $c/a$ ratio of 1.0006 that could be resolved in a synchrotron x-ray diffraction (XRD) experiment~\cite{KimberChatterji15}. This distortion is sufficient to lift the geometric frustration on the face-centred cubic (fcc) magnetic sublattice of Mn$^{2+}$ ions and stabilize long-range AFM order. Another recent study revealed that hauerite undergoes an unprecedented collapse of the unit cell volume by 22\% under the application of hydrostatic pressure above 11~GPa \cite{KimberSalamat13}. This occurs as a result of a transition to the much denser arsenopyrite structure that favours low-spin state of Mn and leads to the total loss of magnetic moments.

As can be seen from the few examples of classical ferro- and antiferromagnets listed above, until the end of the 20$^{\rm th}$ century neutron scattering had been applied only sporadically to study magnetic dynamics in natural minerals. The applications of neutron scattering to the Earth and mineral sciences in this period have been reviewed in a number of books and articles \cite{Rinaldi02, Dove02, GhoseCoey88}, but their focus was mainly on structural properties, phononic spectra, and magnetic phase transitions. A substantial body of literature also covers magnetic properties of minerals from the perspective of geophysics and geochemistry \cite{StaceyBanerjee74, OReilly84, GhoseCoey88, OReilly76, HuntMoskowitz95, DunlopOezdemir97, Harrison00}, with an emphasis on ferro- and ferrimagnetism at ambient or elevated temperatures and pressures that can occur in the Earth's crust. More recently, some of the iron-bearing compounds that occur as natural minerals also attracted attention of material scientists because of their intriguing magnetic properties that are promising for applications. To name a few examples, the polar magnet Fe$_2$Mo$_3$O$_8$ (kamiokite) has been discussed because of its field-driven multiferroic and pyroelectric properties \cite{WangPascut15, KurumajiIshiwata15, LiGao17}. As a result of its magnetic structure that represents an antiferromagnetic alignment of weakly ferrimagnetic layers \cite{McAlisterStrobel83}, it exhibits giant magnetoelectricity with a differential magnetoelectric coefficient approaching 10$^4$~ps/m \cite{WangPascut15}. Kurumaji \textit{et~al.} \cite{KurumajiTakahashi17} recently revealed an electric-field active magnon mode (electromagnon) in Fe$_2$Mo$_3$O$_8$ using terahertz spectroscopy and suggested possible spin configurations for these excitations. Later, the linear magnetoelectric effect was also observed in $\alpha$-FeOOH (g\"othite), persisting above room temperature because of the high $T_{\rm N}\approx400$~K \cite{TerOganessian17}. Another mineral LiFePO$_4$ (triphylite) is an olivine-type compound that attracted considerable interest as a storage cathode material for rechargeable lithium batteries \cite{WangSun15}. From the point of view of magnetism, it is a collinear $S=2$ antiferromagnet with a N\'eel temperature of 52~K \cite{RousseCarvajal03, LiGarlea06}. It was recently suggested that Fe$^{2+}$ ions in this compound undergo a high-spin to low-spin transition at pressures of the order of 72~GPa, which is unprecedentedly high among Fe-bearing minerals \cite{ValdezEfthimiopoulos18}. About a year ago, high-pressure structural and magnetic properties were also reported for FeCO$_3$ (siderite) \cite{GolosovaKozlenko17}. This material serves as a model example of a 3D magnetic Ising system, in which the application of hydrostatic pressure leads to a rapid increase in the N\'eel temperature with an unusually large rate of 1.8~K/GPa, followed by a subsequent volume collapse and an associated spin-state crossover of the Fe$^{2+}$ ions to the low-spin ($S=0$) state at 40--50~GPa \cite{MattilaPylkkaenen07, ShiLuo08, LavinaDera10, CerantolaMcCammon15}.

However, all the presented examples are ``classical'' magnets, i.e. compounds with well developed magnetic order and high value of the local spin, for which classical (that is, non-quantum) models of magnetism are valid to a good approximation. Despite their abundance in the mineral world, they will not be central for the present review. Here I will focus mainly on ``quantum'' magnets that exhibit emergent phenomena that cannot be captured by the classical models. In a rather vague definition, this includes spin-$\frac{1}{2}$ systems and higher-spin lattices with strongly pronounced magnetic frustration that tends to suppress conventional long-range magnetic order and promote more exotic magnetic states at low temperatures. At the start of the new millennium, we saw a surge of interest in low-temperature quantum magnetism, magnetic frustration, and exotic magnetic phases such as spin liquids \cite{Normand09, Balents10, MendelsWills11, Norman16, MendelsBert16, SavaryBalents17, ZhouKanoda17}. This stimulated an intensive search for new materials realizing complex crystal structures with specific geometries of magnetic lattices. It was noticed that intricately connected \mbox{spin-$\frac{1}{2}$} magnetic sublattices are found in many copper-bearing minerals that are difficult or impossible to reproduce artificially in the form of large single crystals. Among such structures are those that realize various spin-dimer and spin-chain models, two- and three-dimensional geometrically frustrated lattices, as well as quasi-isolated frustrated magnetic clusters that are natural archetypes of quantum molecular magnets. The constantly rising number of experimental and theoretical works focusing on low-temperature magnetism in such natural samples and the even broader variety of minerals that could be of potential interest for future studies manifest the emergence of a new interdisciplinary field that bridges mineralogy with solid-state magnetism. To the best of my knowledge, no published review has so far collected and classified low-temperature studies on magnetic minerals. Therefore, the goal of the present article is to fill this niche by summarizing recent experimental and theoretical work performed on natural samples as well as on their very close synthetic analogs that were motivated by the magnetic minerals. The main emphasis of this review will be on magnetic structure and spin dynamics, with the aim of providing a firm standpoint for future developments in this relatively young field of research.

\subsection{Structure of this review}

This review does not aim to give a complete coverage of the field of quantum magnetism, which would hardly be possible in the limited volume of an article. There are several existing books and reviews that already presented an excellent introduction to this field, such as the lecture notes on \textit{``Quantum Magnetism''} edited by Schollw\"ock \textit{et al.} \cite{SchollwoeckRichter04} and the more recent review volumes on frustrated spin systems edited by Lacroix \textit{et al.} \cite{LacroixMendels11} and by Diep \cite{Diep13}. Neither does it present an exhaustive introduction to magnetism from the usual perspective of geophysics and geochemistry. From a geologist's perspective, rock magnetism would be typically restricted to classical ferro- or ferrimagnetic properties of minerals at ambient or elevated temperatures and pressures. Here an interested reader can also refer to the vast body of published works \cite{StaceyBanerjee74, OReilly84, GhoseCoey88, OReilly76, HuntMoskowitz95, DunlopOezdemir97, Harrison00}, yet from the physics perspective, such classical magnetism is generally considered as a well established and understood field of material science.

Instead, this review outlines an interdisciplinary field that bridges these two seemingly distinct topics by discussing low-temperature properties of quantum spin systems that occur naturally as minerals or were inspired by the naturally occurring mineral structures. Of course, one cannot help noticing the vagueness of this distinction in the choice of presented materials. Quite often, initial low-temperature magnetic characterization performed on natural mineral samples turns out inconclusive because of imperfections in sample quality or insufficient sample size. Whenever synthetic powders or single crystals of the same material are available, they provide complementary data, typically of higher quality. Chemical synthesis may also broaden the range of physical characterization methods that can be applied in order to understand the magnetic properties of minerals, e.g. through isotope substitution or precisely tuned chemical composition. On other occasions, especially for complex minerals with large unit cells, chemical synthesis is highly challenging, so natural specimens remain the only option for physical characterization in spite of all the difficulties associated with impurities and structural defects that are likely in naturally occurring samples. Sometimes, chemically pure compounds are available as powders, whereas single-crystal studies have to be done on naturally grown crystals. This demonstrates that it would be unthinkable to discuss the physical properties of minerals with no reference to their synthetic analogs. Moreover, our theoretical understanding of these properties often benefits from a comparison with related compounds that exhibit similar crystal structures but are not necessarily found in nature or still await their discovery. Such a situation is anticipated for an interdisciplinary subject that is deeply rooted in several closely related and well established fields of knowledge. Therefore, the choice of materials to be covered in this review relies not only on their direct or indirect relationship to minerals, but also on the presence of reliable high-quality experimental results or theoretical predictions and on the exciting new physics demonstrated by these compounds. In marginal cases, I relied on my personal taste, so I feel necessary to apologize to the reader who may not find a reference to their favourite compound on the list.

Even though the presentation is given from an experimentalist's point of view, numerous links to theoretical works are also highlighted whenever the theory was used to interpret, explain, or model the experimental results. Still, one should not expect to find here an exhaustive coverage of the theoretical developments in the field of quantum magnetism. In terms of the physical methods used for experimental characterization of magnetic materials, the emphasis of this review is deliberately shifted to microscopic probes and spectroscopy. Macroscopic magnetic and thermodynamic measurements, such as magnetization and specific heat, undoubtedly represent the most important and basic characterization methods in solid state physics. The majority of physical measurements, especially at the initial stage of studying a newly discovered compound, are done using such macroscopic techniques. Yet, the ultimate goal in the study of magnetic minerals is to establish their microscopic magnetic Hamiltonians and explain them from first principles. From this perspective, the capabilities of macroscopic measurements are limited, especially for complex materials with multiple magnetic interactions. While the analysis of temperature-dependent magnetization can be sufficient to form the initial guess about the system's dimensionality, degree of frustration, and estimate its most dominant magnetic interactions, a much more complete picture emerges when these data are complemented by local-probe and momentum-resolved measurements, including single-crystal diffraction, nuclear magnetic resonance (NMR), muon-spin relaxation ($\mu$SR), and neutron scattering. A combination of all these methods, together with state-of-the-art quantum chemistry calculations, can reveal exciting new details about the system's magnetic properties. Even if the number of cases, when all these techniques have been applied jointly to a given material in their full power, remains scarce, they still represent much more interest than numerous other examples of less-studied systems that were only exposed to a superficial initial characterization. Therefore, these cases were given priority in the review.

The review is structured according to the physics of magnetic subsystems, rather than the chemistry of corresponding compounds. Chapters 2 through 7 discuss different groups of minerals according to the dimensionality and complexity of their magnetic interactions, including zero-dimensional dimer systems (Chapter~2) and molecular magnets (Chapter~7), one-dimensional spin chains (Chapter~3), two-dimensional (2D) layered structures (Chapters~4 and 5), and 3D magnetic lattices (Chapter~6). Among the quasi-2D compounds, kagome-lattice systems are so much more numerous among minerals, that they deserved a separate chapter. Finally, Chapter~8 discusses the 2D magnetic metal FeS, whose naturally occurring analog mackinawite can be seen as the only known natural prototype of a recently discovered iron-based superconductor. Most of the chapters start with an introductory section that summarizes the central results for a specific class of magnetic lattices and gives references for further reading. It is then followed by more specific sections that can be seen as case studies devoted to a certain magnetic mineral or a group of compounds with related properties. Because the main focus of this review is on quantum magnetism, the majority of the presented materials are cuprates (that is, spin-$\frac{1}{2}$ systems). Several Fe-, Mn-, and Co-based magnetic minerals with classical spins, such as jarosites (section \ref{Sec:Jarosites}), delafossites (section \ref{Sec:Delafossite}) and columbites (section \ref{Sec:Columbite}), are also discussed for completeness. These are, however, limited to geometrically frustrated and low-dimensional crystal structures. Strong frustration can partially suppress or distort magnetic order even on lattices with large spins, which we would otherwise consider classical, making them sensitive to quantum-fluctuation effects. The mentioned classification is also not free of ambiguity, as the dimensionality and the essential physics of a magnetic system sensitively depends on the relative magnitude of magnetic interactions and on the temperature and energy range that is taken in consideration. For instance, the majority of quasi-low-dimensional systems exhibit long-range order at sufficiently low temperatures in spite of the smallness of 3D interactions, but display low-dimensional properties above the 3D ordering temperature. The hierarchy of exchange interactions can also be re-evaluated in a successive study, e.g. as a result of better refined details of the crystal structure, which can affect the effective dimensionality of a system, and change its classification. In this review, I give a snapshot of the current state of the art, with the hope to see many new exciting discoveries and developments in this relatively young and rapidly expanding domain of quantum magnetic materials in the coming years.

\section[\vspace{-1.1em}\\Coupled spin dimers]{Coupled spin dimers}\label{Chap:Dimers}

\subsection{Short motivation and physical models}

Dimerized quantum spin systems are among the simplest and best studied physical models that serve as textbook examples for the realization of quantum critical phase transitions \cite{Sachdev11, Carr11}. An isolated magnetic dimer consisting of two antiferromagnetically coupled $S=1/2$ Heisenberg spins with an exchange constant $J>0$ has an $S=0$ singlet ground state and an $S=1$ excited triplet state. Therefore, its magnetic excitation spectrum consists of a triply degenerate singlet-triplet transition at the energy $J$, which can be split by the application of an external magnetic field. With the introduction of weak interdimer interactions, the ground state can be described as a valence bond solid (VBS) \cite{AffleckKennedy87}. The triplet excitations can now propagate through the lattice, thereby acquiring a dispersion. Nevertheless, the spectrum remains gapped and no long-range ordered ground state is stabilized until a quantum critical point (QCP) is reached at a certain critical value of the interdimer coupling, $J^\prime=J_{\rm c}$. The archetypal model materials that realize a quantum-disordered ground state along this scenario (upon neglecting weaker interdimer interactions that are not necessarily unimportant) are the coupled-dimer antiferromagnets KCuCl$_3$ and TlCuCl$_3$ \cite{CavadiniHeigold00, CavadiniHeigold01, OosawaKato02}, characterized by different values of the alternation ratio $\alpha=J^\prime/J=1-|\Delta/J|$ that controls the proximity to the QCP. Here $\Delta$ is the spin gap, and the estimates $\alpha\approx0.3$ and $\alpha\approx0.8$ were obtained for the KCuCl$_3$ and TlCuCl$_3$ systems, respectively \cite{CavadiniHenggeler00}. In an applied magnetic field, the triplet excitation splits into three branches due to the Zeeman effect \cite{CavadiniRueegg02, MatsumotoNormand02}, ultimately leading to a field-tuned QCP associated with the Bose-Einstein condensation of the triplet states at a certain critical field $B_{\rm c}$ where the energies of the lower Zeeman-split triplet component and the ground-state singlet intersect \cite{RueeggCavadini03}. In TlCuCl$_3$, a long-range magnetic order is already induced at 6~T, so that this QCP is easily accessible in an experiment.

\begin{figure}[t!]
\begin{center}
\includegraphics[width=0.83\linewidth]{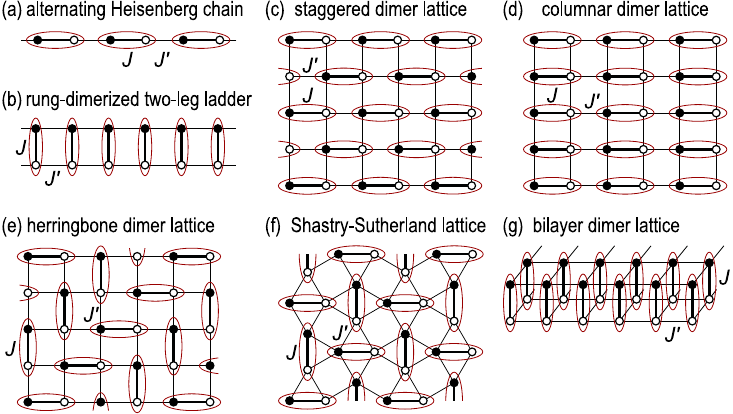}
\end{center}
\caption{Some of the most commonly studied dimer models \cite{ChitovRamakko08, FritzDoretto11, ShastrySutherland81, KogaKawakami00}.}\label{Fig:DimerModels}
\end{figure}

A variety of models with different critical alternation ratios $\alpha_{\rm c}$ arise depending on the specific lattice of interdimer interactions, several of them shown in Fig.~\ref{Fig:DimerModels}. For one-dimensional (1D) spin-chain compounds, one arrives at the alternating Heisenberg chain model \cite{JohnstonKremer00} embodied in the above-mentioned copper chlorides, with a QCP at \mbox{$\alpha_{\rm c}=1$}, or various dimerized spin-ladder models \cite{Dagotto99, BatchelorGuan07, ChitovRamakko08}. A two-leg spin ladder emerges, for instance, in the famous ``telephone number compound'' Sr$_{14}$Cu$_{24}$O$_{41}$ \cite{EcclestonUehara98}. Two-dimensional (2D) models include the exactly solvable Shastry-Sutherland lattice model with orthogonal dimers \cite{ShastrySutherland81}, manifested in SrCu$_2$(BO$_3$)$_2$ \cite{KageyamaOnizuka98, KageyamaYoshimura99}, which has two quantum-critical transitions at $\alpha_{\rm c1}\approx0.675$ and $\alpha_{\rm c2}$ that has varying estimates from 0.765 \cite{CorbozMila13} to 0.86 \cite{KogaKawakami00}. Numerous theoretical studies exist also for various other dimerized antiferromagnets arranged in a staggered ($\alpha_{\rm c}\approx0.397$), columnar ($\alpha_{\rm c}\approx0.524$), herringbone ($\alpha_{\rm c}\approx0.4$), or bilayer ($\alpha_{\rm c}\approx0.397$) dimer lattices (see Ref.~\cite{FritzDoretto11} and references therein).

From the materials perspective, compounds that can serve as model systems for the realization of these assorted dimer lattices in the proximity to criticality are of the greatest interest. Therefore, the search continues among synthetic compounds as well as natural minerals. Particularly exciting are those systems that exhibit high sensitivity of the exchange constants to external parameters, such as hydrostatic pressure, offering the possibility to vary the alternation ratio experimentally in order to approach the QCP or stabilize new quantum ground states. As will become clear from the following examples, this situation is common in natural copper minerals, where the superexchange coupling sensitively depends on the Cu--O--Cu bond angles that can be tuned by pressure. In addition, some minerals realize less common and more complicated dimer structures that provide a new playground for theorists and serve as real-world challenges for testing the accuracy and reliability of different computational methods beyond simple purist models.

\subsection{Malachite: a 2D network of antiferromagnetically coupled dimers}

\begin{figure}[t!]
\includegraphics[width=\linewidth]{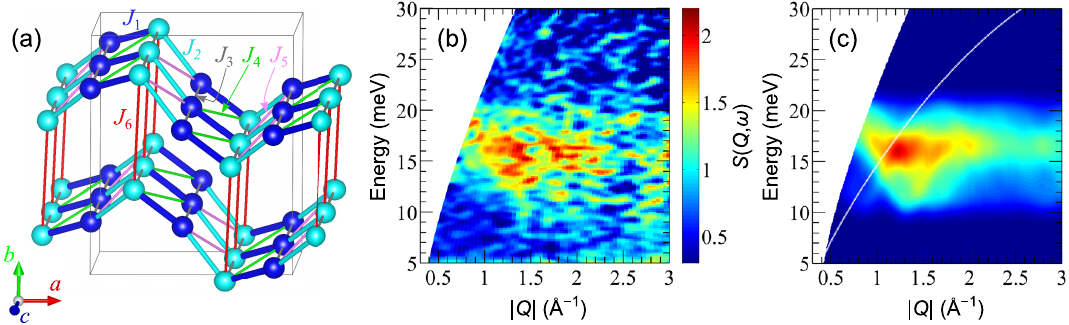}
\caption{(a)~Structure of the magnetic sublattice and the network of exchange pathways in malachite. (b)~Background-subtracted neutron scattering intensity, $S(Q,\omega)$, measured on a fully deuterated powder sample of malachite using the MARI time-of-flight spectrometer at ISIS, UK, operated with an incident neutron energy $E_{\rm i}=80$\,meV. (c)~Powder-averaged model calculations that reproduce the experimental data within the harmonic triplon approximation \cite{FritzDoretto11}. After Can\'evet~\textit{et~al.}~\cite{CanevetFak15}.}\label{Fig:Malachite}
\end{figure}

The carbonate mineral malachite, with the formula Cu$_2$CO$_3$(OH)$_2$, is famous as a gemstone and has been mined as a copper ore since antiquity. It has a monoclinic crystal structure (space group $P2_1/a$) in which pairs of edge-sharing planar CuO$_4$ plaquettes form Cu$_2$O$_6$ dimers that are arranged into buckled dimer chains [see Fig.~\ref{Fig:Malachite}\,(a)]. The intradimer exchange interaction between the spin-$\frac{1}{2}$ Cu$^{2+}$ ions is AFM with $J_1 \equiv J \approx 190$\,K, while the interdimer exchange within the chains is about twice weaker, $J_2 \equiv J^\prime \approx 90$\,K, as estimated from susceptibility data that could be well fitted by the alternating Heisenberg chain model \cite{JanodLeonyuk00, LeberneggTsirlin13a, CanevetFak15}. While reporting correct values for the leading exchange interactions, the first work by Janod \textit{et al.} misplaced the spin dimers in the lattice, assuming that the strongest $J_1$ coupling acts between the structural dimers, as one might expect from the larger Cu--O--Cu bridging angle of 122.1$^\circ$ in accordance to the Goodenough-Kanamori-Anderson (GKA) rules \cite{Goodenough55, Kanamori59, Anderson63}. Lebernegg \textit{et al.} later demonstrated that malachite is an exception among Cu$^{2+}$ systems, in which structural and magnetic dimers coincide \cite{LeberneggTsirlin13a}. Thermodynamic measurements reveal no signatures of magnetic order or spin freezing in this system down to temperatures as low as 0.4~K \cite{CanevetFak15}. The value of the spin gap, $\Delta\approx130$~K (or 11~meV), was estimated from susceptibility data on three natural specimens of malachite \cite{JanodLeonyuk00, LeberneggTsirlin13a} and was recently confirmed by direct neutron-spectroscopy measurements on a deuterated powder sample \cite{CanevetFak15} that are shown in Fig.~\ref{Fig:Malachite}\,(b), while single-crystal INS measurements are still lacking.

Malachite is a rare example among Cu minerals, for which accurate hydrogen positions in the crystal structure, whose bonding angles have a strong influence on the exchange couplings, are available both from experiments and from structural optimization within density functional theory (DFT). This enabled reliable theoretical calculations of the exchange constants, showing good quantitative agreement with the experimental values \cite{LeberneggTsirlin13a, CanevetFak15}. Further, DFT+$U$ was also used to calculate isotropic interchain interactions $J_3$\,...\,$J_6$, resulting in a value of $J_6 \equiv J_\perp$ along the $b$ axis of similar magnitude to $J_2$, while all remaining exchange constants were found to be considerably weaker. Based on their neutron-scattering results, the authors of Ref.~\cite{CanevetFak15} come to the conclusion that this coupling scheme places malachite ``between strongly coupled alternating chains, square lattice antiferromagnets, and infinite-legged ladders''. They find that a 2D depleted staggered dimer model (in contrast to the 1D spin-chain model originally suggested by Janod \textit{et al.} \cite{JanodLeonyuk00}) is well suited for the description of malachite [see Fig.~\ref{Fig:Malachite}\,(c)] because of the similar values of interdimer couplings parallel ($J^\prime$) and perpendicular ($J_\perp$) to the chains: $J^\prime/J=0.34$ and $J_\perp/J=0.26$. These values are smaller than the critical interdimer coupling strength $J_{\rm c}^\prime$ at the QCP separating a dimer state and AFM order, $J_{\rm c}^\prime/J=1/2$, which explains the large size of the spin gap and the absence of a long-range magnetic order in malachite \cite{CanevetFak15}.

In addition, full-relativistic calculations show that the lack of inversion symmetry along the exchange pathways gives rise to non-negligible antisymmetric Dzyaloshinskii-Moriya interactions (DMI), which are important for understanding the finite values of the low-field uniform magnetization that are observed in experiments despite the large value of the spin gap \cite{LeberneggTsirlin13a, MiyaharaFouet07}. There are two nonequivalent DMI vectors, $\mathbf{D}_1$ and $\mathbf{D}_2$, for the two Cu--Cu bonds within the chain, with the relatively large $|\mathbf{D}|/J$ and $|\mathbf{D}^\prime|/J^\prime$ ratios of 0.11 and 0.26, respectively \cite{LeberneggTsirlin13a}. The expected influence of DMI on the excitation spectrum is the splitting of the triplets of the order of $10^{-3}\,J_1\approx0.016$\,meV, which is too small to be detectable on a polycrystalline sample but could possibly be resolved in future experiments on single crystals.

Finally, Lebernegg \textit{et al.} took advantage of the experimental high-pressure structural re\-fine\-ment available from XRD data, which they complemented with calculated hydrogen positions optimized within the generalized gradient approximation (GGA), to estimate the expected changes in the individual exchange parameters and the spin gap energy under the application of hydrostatic pressure \cite{LeberneggTsirlin13a}. They found that the intradimer exchange interaction $J_1$ tends to decrease, while the interdimer coupling $J_2$ remains nearly constant within 15\%, so that both become approximately equal at pressures of the order of 5~GPa. This implies a nearly twofold increase in the $J^\prime/J$ ratio, which may lead to the stabilization of a long-range magnetic order. At the same time, the dominant interchain coupling $J_\perp$ is considerably suppressed by pressure according to the calculations, indicating a crossover from the 2D dimer network to quasi-1D uniformly coupled spin chains. As a result, the calculated value of the spin gap is also suppressed under pressure with an average rate of $\sim$\,1~meV/GPa. The predicted sensitivity of the intradimer coupling to pressure should serve as a motivation for more detailed experimental studies of malachite, in particular to establish its magnetic phase diagram under pressure and to search for low-temperature magnetically ordered phases.

\subsection{Callaghanite: nearly compensated solitary dimers}

The empirical GKA rules \cite{Goodenough55, Kanamori59, Anderson63} predict how superexchange depends on the ion-ligand-ion bridging angle. According to these rules, a 180$^\circ$ superexchange of two magnetic ions with partially filled $d$ shells is antiferromagnetic, whereas a 90$^\circ$ superexchange is ferromagnetic. At intermediate angles the AFM and FM interactions tend to compensate each other, resulting in very weak coupling even for closely spaced ions. For Cu--O--Cu bonds, this transition occurs in the range of 95$^\circ$--98$^\circ$ \cite{BradenWilkendorf96} and corresponds to a situation where first-principles calculations become less reliable in predicting not only the absolute value but even the sign of the magnetic interactions \cite{LeberneggTsirlin14}. Moreover, the suppression of the nearest-neighbour exchange through this mechanism enhances the relative importance of further-neighbour exchange paths that could be otherwise neglected. Therefore, systems that realize structures with compensated exchange are expected to show complicated magnetic behaviour as a result of frustration among multiple weak interactions, demonstrate high sensitivity of the ground state to external parameters such as pressure, and represent ideal systems for testing the accuracy and reliability of different computational methods.

\begin{figure}[t!]
\begin{center}
\includegraphics[width=0.78\linewidth]{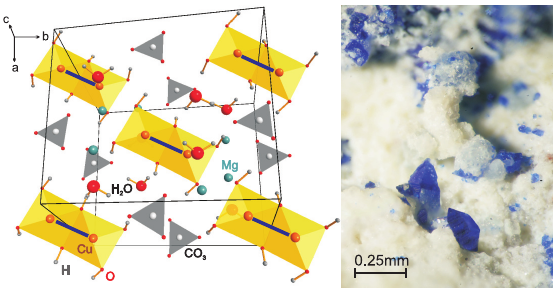}\vspace{-2pt}
\end{center}
\caption{The dimer structure of callaghanite (left) and a photo of natural callaghanite crystals (right). After Lebernegg \textit{et~al.}~\cite{LeberneggTsirlin14}.\vspace{-2pt}}\label{Fig:Callaghanite}
\end{figure}

According to Lebernegg \textit{et al.}, the dimer compound Cu$_2$Mg$_2$(CO$_3$)(OH)$_6$\,$\cdot$\,2H$_2$O, which is found in nature in the form of a violet-blue mineral known as callaghanite (Fig.~\ref{Fig:Callaghanite}), is a perfect realization of the de\-scribed scenario \cite{LeberneggTsirlin14}. Unlike malachite, this min\-er\-al is very rare and has been much less studied by physical methods. Its structure, refined from XRD data apart from the hydrogen positions \cite{Brunton73}, is characterized by the monoclinic space group $C2/c$ and hosts Cu$_2$(OH)$_6$ structural dimers separated by weakly bonded carbonate groups and water molecules. The interdimer interactions are ferromagnetic and very weak, not exceeding 1.5\,K (0.12\,meV). The lattice symmetry also forbids DMI within the dimers. However, magnetization and specific-heat measurements reveal that even the strongest AFM intradimer coupling $J$ is very weak in this compound, of only 7\,K (0.6\,meV), and is therefore difficult to evaluate from first principles \cite{LeberneggTsirlin14}. This results in a relatively low value of magnetic field at which the magnetization saturates, of the order of 14~T.

Such an unusual behaviour is a consequence of the Cu--O--Cu bridging angle of 96.14$^\circ$ within the dimers, which falls within the range where the compensation of AFM and FM exchange interactions is expected. At ambient pressure, the system shows quantum paramagnetic behaviour and no signs of magnetic ordering down to the lowest measured temperatures. Still, the sensitivity of the Cu--O--Cu bridging angle to pressure may result in a situation when the AFM interaction is reduced even further by very moderate pressures, leading to a magnetic phase transition and a pressure-driven QCP. While experimental investigations of callaghanite under pressure are not yet available, it definitely represents a very interesting model compound for studying the subtle balance of AFM and FM interactions close to compensation. A similar network of dimers is also present in the mineral liroconite, Cu$_2$Al(AsO$_4$)(OH)$_4$$\,\cdot\,$4H$_2$O \cite{BurnsEby91}, which would deserve a similarly detailed investigation.

\subsection{Urusovite: corrugated honeycomb planes prone to dimerization}

\begin{figure}[t!]
\begin{center}
\includegraphics[width=0.55\linewidth]{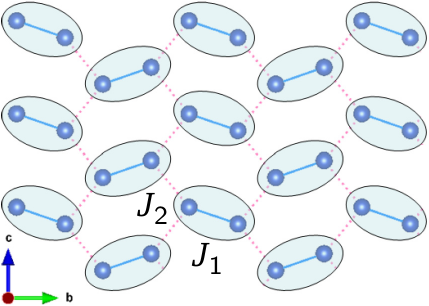}
\end{center}
\caption{Schematic depiction of the VBS ground state in urusovite after Vasiliev \textit{et~al.}~\cite{VasilievVolkova15}.}\label{Fig:Urusovite}
\end{figure}

A layered structure consisting of corrugated honeycomb layers is realized in the rare copper-aluminium-arsenate mineral urusovite, CuAl(AsO$_4$)O, discovered about 20 years ago on Kamchatka \cite{VergasovaFilatov00, KrivovichevMolchanov00}. Its crystal structure is monoclinic, with the space group $P2_1/c$. Synthetic powders of this compound were recently investigated by magnetization, specific heat, and electron spin resonance (ESR) \cite{VasilievVolkova15}, showing a spin-gap behaviour with a gap of about 30~meV. Analysis of the exchange constants derived from DFT calculations revealed that this gap is due to a strong dimerization on one of the bonds in the distorted honeycomb plane. The dominant edge-shared superexchange interaction $J_1\approx30$~meV on one of the bonds is mediated by two Cu--O--Cu paths with a bond angle of 102.6$^\circ$. The corner-shared superexchange paths on two other bonds have a notably larger bond angle of 111.3$^\circ$ but despite that have a vanishingly small exchange constant $J_2\approx0.05$~meV, which is even weaker than the interlayer super-superexchange $J_3\approx0.7$~meV. It has been therefore concluded that urusovite realizes a model of very weakly interacting spin dimers with a VBS ground state, schematically depicted in Fig.~\ref{Fig:Urusovite}, where the ellipses represent valence-bond dimers in a singlet state. At elevated temperatures, thermal expansion leads to a gradual modification of the bond angles and, consequently, to a substantial increase in the in-plane exchange parameters. These thermal effects were claimed responsible for the observed deviations of the high-temperature experimental data from the predictions of a simple weakly interacting dimer model~\cite{VasilievVolkova15}.

Naturally, the lowest-energy excitation in CuAl(AsO$_4$)O, observed by ESR, corresponds to the singlet-triplet transition with an energy given by the spin gap, $\Delta\approx30$~meV. The average effective $g$ factor was found to be $g=2.05$ and remained constant down to $\sim$\,90~K, shifting afterwards to higher values. The latter artefact has been ascribed to paramagnetic impurities, which were also evidenced by the static magnetic susceptibility measurements~\cite{VasilievVolkova15}. Estimates of the $g$-factor anisotropy from the parallel
and perpendicular components of the hyperfine structure resulted in the value for $(g_\parallel-g_\perp)/g_\perp\approx0.17$.

\subsection{Clinoclase: nonequivalent dimers with two energy scales}

\begin{figure}[t!]
\begin{center}
\includegraphics[width=0.51\linewidth]{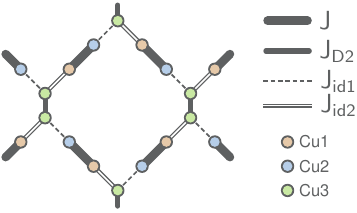}
\end{center}
\caption{The planar network of exchange interactions in clinoclase. Intra- and interdimer interactions are shown with thick and thin lines, respectively. After Lebernegg \textit{et~al.}~\cite{LeberneggTsirlin13}.}\label{Fig:Clinoclase}
\end{figure}

Another example of a system with weakly coupled magnetic dimers is clinoclase, Cu$_3$(AsO$_4$)(OH)$_3$, whose magnetic properties also lack in-depth experimental investigations to date.
X-ray structural refinement \cite{GhoseFehlmann65, EbyHawthorne90} revealed an intricate structure with the monoclinic $P2_1/c$ space group, in which Cu$^{2+}$ ions occupy three inequivalent Wyckoff positions: Cu(1) and Cu(3) have strongly distorted octahedral coordinations approaching square pyramidal, while Cu(2) is coordinated by 5 anions only \cite{EbyHawthorne90}. These spin-$\frac{1}{2}$ ions are arranged into two types of structural dimers, Cu(1)\,--\,Cu(2) and Cu(3)\,--\,Cu(3), which are both built of two edge-sharing CuO$_4$ plaquettes but with different geometrical parameters and, consequently, different Cu--O--Cu bridging angles. The Cu(1)\,--\,Cu(2) dimers share corners and form zigzag chains along the [001] direction, connected by AsO$_4$ tetrahedra into layers parallel to the $bc$ plane, with Cu(3)\,--\,Cu(3) dimers sandwiched in between these layers. According to the DFT+$U$ calculations within the local spin-density approximation (LSDA+$U$) by Lebernegg \textit{et al.} \cite{LeberneggTsirlin13}, this complex structure hosts four AFM exchanges that exceed 100\,K, of which the two leading interactions operate between the Cu(1)\,--\,Cu(2) dimers ($J\approx700$\,K) and within the Cu(3)\,--\,Cu(3) dimers ($J_{\rm D2}\approx300$\,K). This situation implies that the spin dimer with the strongest interaction $J$ does not coincide with any of the structural dimers but acts between the corner-sharing CuO$_4$ plaquettes that have a much larger Cu--O--Cu bridging angle of 150$^\circ$. The interaction within the Cu(1)\,--\,Cu(2) dimer ($J_{\rm D1}$) is nearly compensated because of the much smaller Cu--O--Cu bridging angle $<100^\circ$, in agreement with the GKA rules. At the same time, contrary to the GKA rules, the second-leading interaction $J_{\rm D2}$ is nearly twice larger than the interaction between the Cu(3)\,--\,Cu(3) dimers despite the smaller bridging angle. As a result, the authors of Ref.\,\cite{LeberneggTsirlin13} arrive at the planar network of magnetic interactions in clinoclase that is presented in Fig.~\ref{Fig:Clinoclase}. It contains two types of AFM dimers defined by the leading exchange terms $J$ and $J_{\rm D2}$, connected by weaker nonfrustrated AFM exchange paths $J_{\rm id1}$ and $J_{\rm id2}$, both of~about~160\,K.

Susceptibility measurements performed on natural clinoclase crystals \cite{LeberneggTsirlin13} show perfect agreement with the output of theoretical calculations summarized above. An attempt to fit the data with a single dimer contribution clearly fails, but they can be perfectly described with the ``2+1'' dimer model in which strong and weak dimers are included in the 2:1 ratio as in Fig.~\ref{Fig:Clinoclase}. The experimentally determined intradimer exchange constants of $J_1=703.5$\,K and $J_2=289.3$\,K perfectly match with the $J$ and $J_{\rm D2}$ values resulting from the DFT+$U$ calculation. The inclusion of interdimer interactions in the framework of a Quantum Monte Carlo (QMC) simulation leads to no observable changes in the susceptibility.

The presence of weakly interacting nonequivalent spin dimers in clinoclase implies that its spin-excitation spectrum should consist of two gapped branches of triplet excitations similar to those observed in the prototypical coupled-dimer antiferromagnets KCuCl$_3$ or TlCuCl$_3$ \cite{CavadiniHeigold00, CavadiniHeigold01, OosawaKato02}, with characteristic energy scales $J_{1,2}$ that correspond to singlet-triplet transitions of both dimers \cite{CavadiniHenggeler00}. The dispersion of triplet excitations, and consequently their spin gaps, would sensitively depend on interdimer interactions. Each of these triply degenerate excitations would then split into three branches in an external magnetic field \cite{CavadiniRueegg02, MatsumotoNormand02}, possibly leading to a field-tuned QCP via the Bose-Einstein condensation of the triplet states \cite{RueeggCavadini03}. Further, sufficiently strong interdimer interactions may lead to a field-dependent hybridization of propagating triplet excitations from different dimers that would represent an interesting topic for future spectroscopic studies.

\section{Quantum spin chains}

\subsection{The assortment of spin chains and ladders}\label{Sec:Chains}

\begin{figure}[b!]
\begin{center}\vspace{-2pt}
\includegraphics[width=0.7\linewidth]{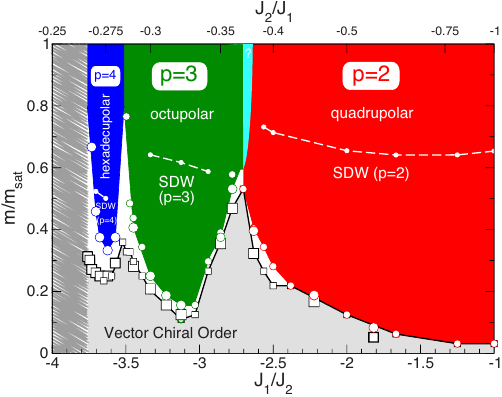}\vspace{-2pt}
\end{center}
\caption{Numerical phase diagram for the frustrated FM spin chain with AFM second-nearest-neighbour interactions. The vertical axis represents~the field-induced magnetization normalized to its saturation value. The coloured regions denote spin-multipolar Luttinger liquids of multimagnon bound states composed of $p=2,\,3,\,4$ spin flips. Below the dashed crossover lines, spin-density-wave (SDW) correlations start to dominate. After Sudan \textit{et~al.}~\cite{SudanLuescher09}.}\label{Fig:SpinChainDiagram}
\end{figure}

Quantum spin chain and ladder models \cite{MikeskaKolezhuk04, Lecheminant13} serve for the description of magnetism in a variety of quasi-1D systems that commonly occur in minerals. Of our primary interest are lattice configurations with magnetic frustration realized via the competition of nearest- and further-neighbour exchange interactions. The simplest model of this kind is a uniform spin-$\frac{1}{2}$ chain with a FM nearest-neighbour (NN) coupling $J_1<0$ and AFM next-nearest-neighbour (NNN) coupling $J_2>0$, which is relevant for the minerals tolbachite (CuCl$_2$), trippkeite (CuAs$_2$O$_4$), linarite $\bigl($PbCuSO$_4$(OH)$_2\bigr)$, and szenicsite $\bigl($Cu$_3$(MoO$_4$)(OH)$_4\bigr)$ as well as for a number of archetypal low-dimensional cuprates CuBr$_2$ \cite{ZhaoHung12, LeeLiu12, ToledanoAyala17}, Li$_2$CuO$_2$ \cite{BoehmCoad98, GraafMoreira02, XiangLee07, LorenzKuzian09}, LiCu$_2$O$_2$ \cite{MasudaZheludev04, MasudaZheludev05, ParkChoi07, SekiYamasaki08}, NaCu$_2$O$_2$ \cite{DrechslerRichter06}, LiCuVO$_4$ \cite{EnderleMukherjee05, BuettgenKrugvonNidda07, BuettgenKraetschmer10, MourigalEnderle11, NawaTakigawa17}, LiCuSbO$_4$ \cite{DuttonKumar12, DeyKumar18}, and Li$_2$ZrCuO$_4$ \cite{DrechslerVolkova07, VavilovaMoskvin09}, composed of edge-sharing copper-oxygen chains. The ground state of such a chain depends on the level of frustration that is uniquely determined by the ratio $\alpha=J_2/|J_1|$. In the isotropic exchange case, for \mbox{$0<\alpha<1/4$} a FM ground state occurs, whereas for $\alpha>1/4$ various exotic quantum states including a chiral nematic spin liquid \cite{Chubukov91} and a VBS with a hidden topological order \cite{FurukawaSato12, AgrapidisDrechsler19} have been predicted. In a magnetic field, the lowest-lying excitations represent multimagnon bound states that are subject to Bose-Einstein condensation and may lead to fancy types of quasi-long-range order (nematic, antiferrotriatic, etc.) and novel Tomonaga-Luttinger liquids with multipolar spin correlations \cite{KeckeMomoi07, VekuaHonecker07, HikiharaKecke08, SudanLuescher09, BalentsStarykh16}. The resulting phase diagram in Fig.~\ref{Fig:SpinChainDiagram} that was famously presented by Sudan, L\"uscher, and L\"auchli \cite{SudanLuescher09} highlights the sequence of spin-multipolar Luttinger liquid phases (quadrupolar, octupolar, hexadecapolar, etc.) in dependence on the frustration ratio $\alpha$ and the magnetic field. The inclusion of interchain interactions into the model, which cannot be fully excluded in any real material, stabilizes an incommensurate spin-spiral order that is experimentally found in many of the mentioned compounds. The pitch angle of the spiral therefore provides a sensitive measure of the weak interchain interactions, ever more so in the proximity to the QCP at $\alpha=1/4$ \cite{NishimotoDrechsler12}.

Magnetic 1D chains with higher values of the local spin can also be found in some Mn- and Fe-based minerals. For example, the borate mineral gaudefroyite, Ca$_4$(MnO)$_3$(BO$_3$)$_3$CO$_3$, features ferromagnetic linear chains of edge-shared MnO$_6$ octahedra along [001] that are arranged on a kagome lattice in the $ab$ plane \cite{YakubovichSimonov75, HoffmannArmbruster97}. These chains can be described by the classical Heisenberg model with a small single-ion anisotropy \cite{LiGreaves04}, whereas the AFM coupling between the chains leads to geometric frustration, resulting in a macroscopic ground-state degeneracy and a suppression of long-range magnetic order. It was recently realized that gaudefroyite also has remarkable low-field magnetocaloric properties and could be suitable for refrigeration applications such as liquefaction of hydrogen \cite{LiLi18}. These studies have been performed on natural mineral samples, because the synthesis of gaudefroyite even in polycrystalline form represents a challenge due to the presence of the carbonate group that decomposes at high temperatures.

\begin{figure*}[b!]
\begin{center}
\includegraphics[width=0.92\textwidth]{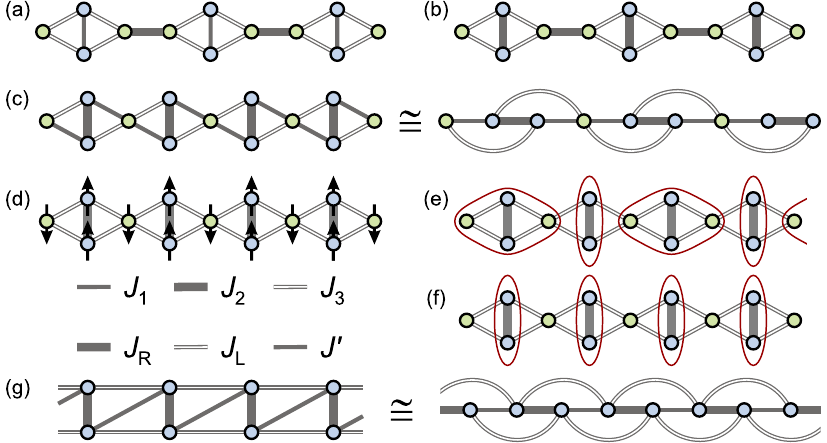}
\end{center}
\caption{Examples of decorated spin-chain models: (a)~dimer-plaquette chain; (b)~orthogonal-dimer chain; (c)~distorted diamond chain and its equivalent representation as a trimerized linear chain with NNN exchange \cite{OkamotoTonegawa03}. (d--f)~The three ground states of the symmetric diamond chain after Takano, Kubo and Sakamoto \cite{TakanoKubo96}: ferrimagnetic (d), tetramer-dimer (e), and dimer-monomer (f). (g)~Zigzag spin ladder and its equivalent representation as an alternating chain with NNN exchange.}\label{Fig:SpinChains}
\end{figure*}

Quantum effects, however, have much more drastic effects in low-dimensional antiferromagnets, even for systems with a large absolute value of the spin. In particular, an AFM spin chain with any integer spin, in contrast to the case of half-integer spin, has a singlet ground state that is separated by an energy gap, $\Delta$, from the excited states. This has been first conjectured by Haldane \cite{Haldane83, Haldane83a} for a Heisenberg chain with nearest-neighbour AFM interactions. Both single-ion anisotropy and interchain interactions tend to reduce the Haldane gap and can suppress the singlet state if they are large enough. The Haldane gap can also be closed by a sufficiently strong magnetic field, as soon as the Zeeman splitting of the excited triplet state becomes equal to~$\Delta$. Some of the most important generalizations of the Haldane model have been reviewed, for instance, by Renard, Regnault, and Verdaguer~\cite{RenardRegnault02}.

The next level in the hierarchy of simple 1D spin systems is represented by decorated spin chains, where more complicated structural units are inserted in the 1D arrangement. The dimer-plaquette chain \cite{RichterIvanov98}, its special case the orthogonal-dimer spin chain \cite{KogaOkunishi00, VerkholyakStrecka16}, and the dimer-monomer (diamond) chain \cite{TakanoKubo96, OkamotoTonegawa03} can be ascribed to this class of models [see Fig.~\ref{Fig:SpinChains}\,(a\,--\,c)]. The symmetric diamond chain obtained by setting $J_1=J_3$ is a limiting case of the more general distorted diamond chain model illustrated in Fig.~\ref{Fig:SpinChains}\,(c) \cite{OkamotoTonegawa03}, which is relevant for the copper-carbonate mineral azurite. For a symmetric spin-$\frac{1}{2}$ chain, three distinct ground states have been identified by Takano, Kubo and Sakamoto \cite{TakanoKubo96} as a function of the parameter $\lambda=J_2/J_1$, namely the ferrimagnetic state for negative or positive but small $\lambda < 0.909$, the tetramer-dimer (TD) state for $0.909<\lambda<2$, and the dimer-monomer (DM) state for $\lambda>2$. These states are shown schematically in Fig.~\ref{Fig:SpinChains}\,(d\,--\,f). The dimer-monomer state, representing a kind of spin fluid, is further stabilized by the increased asymmetry between the $J_1$ and $J_3$ interactions \cite{OkamotoTonegawa03}, as seen in the phase diagram in Fig.\,\ref{Fig:AzuriteStruct}\,(b). It is believed to be realized in azurite and in the structurally related compound Bi$_4$Cu$_3$V$_2$O$_{14}$ \cite{SakuraiYoshimura02, ZhouChoi10} above their corresponding 3D ordering temperatures.

Two or more coupled spin chains form a spin ladder, which takes an intermediate position between 1D and 2D spin systems. The principal difference between a solitary spin-$\frac{1}{2}$ chain and a spin ladder is that a spin-$\frac{1}{2}$ ladder with AFM couplings is a spin liquid with a singlet ground state, exhibiting a Haldane-type energy gap due to the dimerization of half-integer spins \cite{VekuaHonecker06}. It therefore displays the characteristic properties of an integer-spin chain. The same result can be generalized to any half-odd-integer spin ladder with an even number of legs \cite{DagottoRice96}. The simplest spin ladder is governed by only ``rung'' ($J_{\rm R}$) and ``leg'' ($J_{\rm L}$) exchange interactions. This model can be further generalized by considering additional exchange paths as in the case of a zigzag spin ladder, which is supplemented by a diagonal exchange $J^\prime$ between the opposite spins on the neighbouring rungs \cite{MikeskaKolezhuk04}. It can be alternatively viewed as an alternating Heisenberg chain with NNN interactions, as shown in Fig.~\ref{Fig:SpinChains}\,(g). Along the $J_{\rm R}=J^\prime$ line, this model reduces to the well-studied uniform Heisenberg chain with NN and NNN interactions, which undergoes a quantum phase transition to the gapped twofold-degenerate dimer-crystal ground state at $J_{\rm L}/J_{\rm R}=\alpha_{\rm c}=0.2411$. In this regime, one finds the Majumdar-Ghosh point at $J_{\rm L}/J_{\rm R}=1/2$, where the two degenerate ground states can be factorized into a product of singlets formed on NN dimers \cite{MajumdarGhosh69}. By including alternation ($J_{\rm R}\neq J^\prime$), the degeneracy is removed, but one of the Majumdar-Ghosh states remains the exact ground state of the system along the so-called Shastry-Sutherland lines, $J_{\rm L}=\frac{1}{2}\min{}(J_{\rm R},\,J^\prime)$ \cite{ShastrySutherland81prl, MikeskaKolezhuk04}. It was recently realized that the proximity to the Majumdar-Ghosh point, which requires a delicate balance among the exchange parameters, is an essential ingredient of the effective model for the copper-molybdate mineral szenicsite \cite{LeberneggJanson17}.\vspace{-2pt}

\subsection{Tolbachite: a cycloidal helimagnet with field-driven ferroelectricity}

\begin{figure}[t]
\begin{center}
\includegraphics[width=0.62\linewidth]{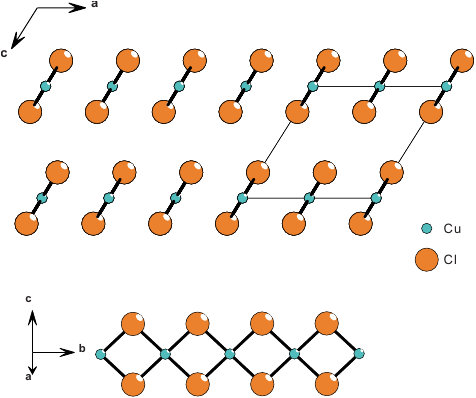}
\end{center}
\caption{The crystal structure of anhydrous cupric chloride CuCl$_2$ (tolbachite), featuring one-dimensional frustrated spin chains. After Banks \textit{et~al.} \cite{BanksKremer09}.}
\label{Fig:Tolbachite}
\end{figure}

The mineral tolbachite, or anhydrous cupric chloride CuCl$_2$, forms brown to golden-brown monoclinic crystals characterized by the space group $C2/m$ \cite{VergasovaFilatov83, BurnsHawthorne93}. The mineral is unstable in air as it readily absorbs water, becoming hydrated to eriochalcite, CuCl$_2$\,$\cdot$\,2H$_2$O. Its crystal structure contains corrugated sheets of Jahn-Teller distorted Cu$^{2+}$Cl$_6$ octahedra, bonded by weak Van der Waals forces (see Fig.~\ref{Fig:Tolbachite}). This effectively results in a chemically simple quasi-1D AFM spin-$\frac{1}{2}$ quantum chain system. According to neutron powder and single-crystal diffraction, anhydrous CuCl$_2$ orders magnetically below a N\'eel temperature of $T_{\rm N}=23.9$~K, forming an incommensurate cycloidal spin-spiral structure \cite{StoutChisholm62, BanksKremer09}. This spiral propagates along the chains ($b$ axis), while the magnetic moments are confined in the $bc$ crystallographic plane, so that the angle between the neighbouring moments along the $b$ direction constitutes approximately 81$^\circ$.

According to DFT calculations \cite{BanksKremer09, SchmittJanson09}, the spin spiral results from competing ferromagnetic NN and antiferromagnetic NNN interactions along the spin chains, causing multiferroic behaviour. The frustration ratio $\alpha=J_2/|J_1|$ varies between 1.3 and 1.8, depending on the value of $U$ assumed in the GGA+$U$ calculations \cite{BanksKremer09}. In other words, the NNN antiferromagnetic spin-exchange coupling dominates. Schmitt \textit{et~al.} \cite{SchmittJanson09} additionally considered the role of crystal water in CuCl$_2$\,$\cdot$\,2H$_2$O as compared to anhydrous CuCl$_2$ and showed that it leads to a dramatic change in magnetic behaviour. Eriochalcite was known as a 3D antiferromagnet \cite{HandelGijsman52, Marshall58}, which can be explained by the relative orientation of the isolated CuCl$_2$O$_2$ plaquettes. The loss of water leads to a switch of the magnetically active orbital, thereby causing a dramatic change in magnetic properties between both compounds.

Soon afterwards, the multiferroic properties of CuCl$_2$ were experimentally investigated by Seki \textit{et al.} \cite{SekiKurumaji10}. They found ferroelectric polarization emerging along the $c$ axis that sensitively depends on the external magnetic field. For a magnetic field applied along the $b$ axis, a spin-flop transition occurs around 4~T that flips the magnetic moments forming the spiral into the $ac$ plane, resulting in a simultaneous suppression of ferroelectricity. For magnetic fields orthogonal to the $b$ axis, the spin-spiral plane depends on the field direction, and the associated polarization shows a continuous dependence on the field angle. This complex magnetoelectric behaviour could be successfully explained using the inverse Dzyaloshinskii-Moriya (IDM) model \cite{KatsuraNagaosa05, Mostovoy06, SergienkoDagotto06, TokuraSeki14}, demonstrating that it is applicable even to quantum-spin systems in spite of strong quantum fluctuations. A more recent theoretical work by Tol\'edano \textit{et~al.} \cite{ToledanoAyala17} analyzed the magnetoelectric effects observed in CuCl$_2$ and CuBr$_2$ and worked out their phase diagram in zero and applied magnetic fields. Contrary to the conclusions of Seki \textit{et al.} \cite{SekiKurumaji10}, they suggested that the emergence of ferroelectric polarization is only partly due to DMI but otherwise results from two distinct AFM spin-density wave order parameters which lead, across a first-order transition, to an incommensurate polar phase with a typical improper ferroelectric behaviour.

\subsection{Trippkeite: a linear-chain ferromagnet close to criticality}

\begin{figure}[b]
\begin{center}
\includegraphics[width=0.52\linewidth]{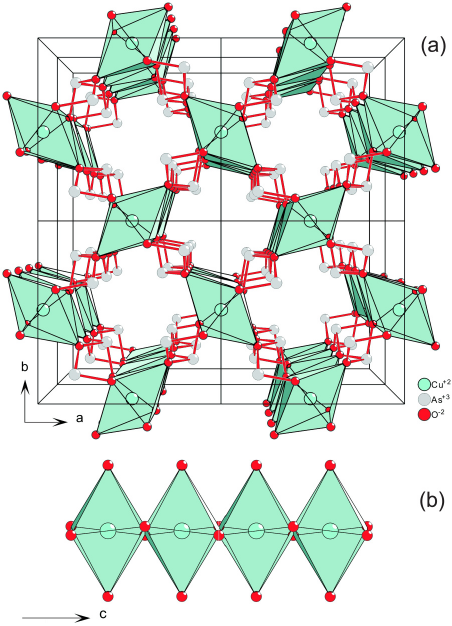}
\end{center}
\caption{(a)~The crystal structure of CuAs$_2$O$_4$ (trippkeite), viewed along the $c$ axis. (b)~Fragment of a spin chain consisting of edge-sharing CuO$_6$ octahedra. After Caslin \textit{et~al.} \cite{CaslinKremer14}.}
\label{Fig:Trippkeite}
\end{figure}

The greenish-blue mineral trippkeite, CuAs$_2$O$_4$, crystallizing in the tetragonal space group $P4_2/mbc$ \cite{Pertlik75} as shown in Fig.~\ref{Fig:Trippkeite}, represents another realization of FM spin-$\frac{1}{2}$ chains with frustrated interactions. However, in contrast to tolbachite, where the antiferromagnetic NNN interactions dominate ($\alpha\approx1.5$), trippkeite is characterized by a small value of the frustration parameter $\alpha\approx0.24$ and orders ferromagnetically at $T_{\rm C}\approx7.4$~K under ambient pressure \cite{CaslinKremer14, CaslinKremer16}. It therefore represents a rare example of a FM spin chain close to the QCP at $\alpha_{\rm c}=1/4$. As already mentioned above, the ground state of the system in this regime sensitively depends on the weak interchain interactions. In other nearly critical spin-chain systems with similar values of $\alpha$, such as Ca$_2$Y$_2$Cu$_5$O$_{10}$ ($\alpha\approx0.19$) \cite{KuzianNishimoto12} and Li$_2$CuO$_2$ ($\alpha\approx0.33$) \cite{XiangLee07, DrechslerMalek10}, the interchain interactions enforce an antiparallel alignment of the spins, resulting in a long-range AFM ordering. In this respect, trippkeite represents an exceptional (and so far possibly unique) case, where the theoretically expected FM ground state for $\alpha<1/4$ is actually realized \cite{CaslinKremer14}.

From measurements of the magnetization, magnetic susceptibility, and electron paramagnetic resonance, a positive Curie-Weiss temperature of $\Theta_{\rm CW}\approx40$~K (consistent with the predominantly FM exchange interactions) and a saturated ordered moment of $\sim$\,1\,$\mu_{\rm B}$ have been estimated. Initial measurements under hydrostatic pressure have indicated an increase in the Curie temperature with a rate of 1.35~K/Gpa, with no evidence for pressure-induced phase transitions up to 1.2~GPa \cite{CaslinKremer14}. However, in a follow-up work, Caslin \textit{et~al.} \cite{CaslinKremer16} extended their measurements up to 11.5~GPa and detected a structural phase transition at 9.2~GPa into the lower-symmetry tetragonal space group $P\overline{4}2_1c$, associated with an increased twisting in the CuO$_2$ ribbon chains and accompanied by a large drop in $T_{\rm C}$. These changes result from the suppression of the axial Jahn-Teller elongations of the CuO$_6$ octahedra. According to DFT calculations, the high-pressure phase is actually characterized by a stronger ferromagnetic NN interaction and a reduced frustration ratio of $\alpha\approx0.17$. Therefore, the drop in $T_{\rm C}$ cannot be explained within the $J_1$-$J_2$ model only and is most probably associated with subtle changes in the interchain or further-neighbour interactions.

\subsection{Linarite: frustrated spin chains with helimagnetic order}

\begin{figure}[b!]
\begin{center}
\includegraphics[width=0.75\linewidth]{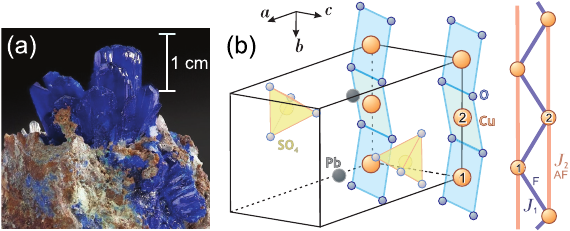}
\end{center}
\caption{(a)~The photographic image of large linarite crystals, after Sch\"apers \textit{et~al.}~\cite{SchaepersWolter13}. (b)~The crystal structure of linarite (hydrogen atoms are omitted for clarity) and the diagram of two intrachain interactions: ferromagnetic $J_1$ and anti\-ferro\-magnetic $J_2$. After Povarov \textit{et~al.}~\cite{PovarovFeng16}.}\label{Fig:LinariteStruct}
\end{figure}

The translucent blue mineral linarite $\bigl($PbCuSO$_4$(OH)$_2$, space group $P2_1/m\bigr)$ is arguably the most studied natural quantum magnetic system, and for a good reason. The high degree of frustration in this system is evidenced by the large ratio of the Curie-Weiss and N\'eel temperatures, $\Theta_{\rm CW}/T_{\rm N}\approx10$ \cite{WolterLipps12}. The first studies of linarite's magnetic and thermodynamic behaviour at low temperatures date back to the early 2000's \cite{KamieniarzBielinski02, BaranJedrzejczak06}. Its mono\-clinic crystal structure contains magnetically frustrated buckled CuO$_2$ ribbons controlled by the leading FM nearest-neighbour interaction $J_1$ and subleading AFM next-nearest-neighbour exchange $J_2$ (see Fig.~\ref{Fig:LinariteStruct}), while the interchain interaction $J_{\rm ic}$ is at least one order of magnitude weaker than $J_1$ \cite{WolterLipps12, SchaepersRosner14, RuleWillenberg17}. In linarite, the ratio $\alpha=J_2/|J_1|\approx0.27$\,--\,0.37 (depending on the source and measurement method \cite{RuleWillenberg17, CemalEnderle18}) is close to the QCP at $\alpha_{\rm c}=1/4$, where in an isolated spin chain the FM interaction takes over and leads to a fully polarized ground state. Consequently, the saturation field is below 10~T, and the entire magnetic phase diagram of linarite is easily accessible with many experimental probes~\cite{WillenbergSchaepers12, SchaepersWolter13, WillenbergSchaepers16, PovarovFeng16}.

\begin{figure}[t]
\begin{center}
\includegraphics[width=0.57\linewidth]{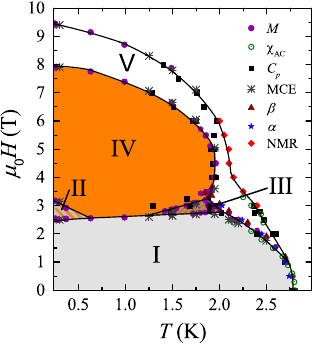}\vspace{-2pt}
\end{center}
\caption{The magnetic phase diagram of linarite after Willenberg \textit{et~al.}~\cite{WillenbergSchaepers12, SchaepersWolter13, WillenbergSchaepers16}.}\label{Fig:LinaritePhaseDiagram}
\end{figure}

The frustration among $J_1$ and $J_2$ leads to a long-range elliptical-spiral ordering below the N\'eel temperature $T_{\rm N}=2.8$\,K, which causes electric polarization of the lattice and leads to a ferroelectric behaviour \cite{YasuiSato11, YasuiYanagisawa11} manifested in signatures of $T_{\rm N}$ in the dielectric constant, electric polarizability, and the pyroelectric and magnetoelectric current measurements~\cite{PovarovFeng16}. Several other multiferroic phases can be stabilized in linarite by the application of magnetic field, resulting in the rich phase diagram depicted in Fig.~\ref{Fig:LinaritePhaseDiagram}. The distinct thermo\-dynamic phases are represented here by the elliptical spiral~(I), a circular helix coexisting with collinear ordering~(III), a collinear N\'eel AFM phase~(IV), and a spin-density wave~(V)~\cite{WillenbergSchaepers16}. The small pocket of phase II is supposedly a metastable mixture of phases I and IV.

Most recently, Y.~Feng \textit{et al.} \cite{FengPovarov18} also in\-ves\-ti\-gated the phase diagram of linarite in tilted magnetic fields applied at an arbitrary angle to the chain axis. According to their orientational phase diagrams, the spiral phase I is nearly isotropic with respect to the field direction. The high-field SDW phase V is also robust against deviations of the field directions away from the $b$ axis. However, the N\'eel phase IV is destabilized at moderate deflection angles, giving way to a new conical phase that exists in magnetic fields applied perpendicular to the chains.

\begin{figure}[t]
\begin{center}
\includegraphics[width=0.665\linewidth]{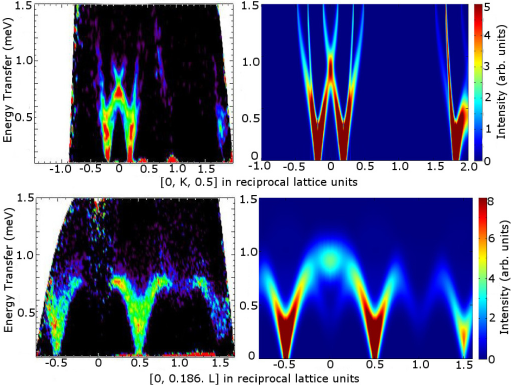}
\end{center}
\caption{The measured (\mbox{$T~=~0.5$~K}, left) and calculated (LSWT, right) magnon spectra of linarite. After Rule \textit{et~al.}~\cite{RuleWillenberg17}.}\label{Fig:Linarite}
\end{figure}

Recent INS measurements of spin dynamics in linarite \cite{RuleWillenberg17} have revealed spin-wave excitations below 1.5~meV, dispersing both parallel and perpendicular to the chain direction as shown here in Fig.~\ref{Fig:Linarite}. The low-energy spin-wave branches emanate from the incommensurate $(0~0.186~0.5)$ wave vector that corresponds to the elliptical helical order of phase~I established from earlier neutron diffraction data \cite{WillenbergSchaepers12}. The presence of dispersion in the direction orthogonal to the chains emphasizes the importance of interchain coupling $J_{\rm ic}$. However, its exact experimental value turned out to be somewhat model dependent. The comparison of INS data with calculations in the framework of linear spin-wave theory (LSWT) presented in Fig.~\ref{Fig:Linarite} results in $J_{\rm ic}\approx0.34$~meV, which constitutes only 3.5\% of $|J_1|$. On the other hand, a similar comparison with the dynamical density-matrix renormalization group (DDMRG) calculations performed with a $32\times2$ cluster yields a higher value of $J_{\rm ic}\approx0.6$~meV and somewhat lower value of $|J_1|\approx6.7$~meV, thereby increasing the $J_{\rm ic}/|J_1|$ ratio to 9\%. The values reported earlier from bulk magnetic susceptibility measurements and LSDA+$U$ calculations \cite{WolterLipps12} fall within the same range of uncertainty. The authors of Ref.~\cite{RuleWillenberg17} emphasize that the agreement between LSWT and DDMRG approaches should generally improve near the critical point at $\alpha=1/4$ and below due to the suppression of spin fluctuations that are ignored by LSWT. As linarite falls close to this limit, LSWT can be used for a semi-quantitative analysis of the INS data.

By extending the INS measurements above the saturation field, Cemal~\textit{et al.} \cite{CemalEnderle18} re-evaluated the magnetic interactions in linarite, including less significant terms such as direct, diagonal, and third-nearest-neighbour interchain exchange couplings. The authors fitted globally the spin-wave dispersions measured along high-symmetry directions at elevated magnetic fields of 10, 11, and 14.5~T applied along the $a$ axis. The resulting ratio of $J_2/|J_1|\approx0.27$ places linarite even closer to the QCP between the FM state and spin-multipolar phases than anticipated in earlier works. This implies that despite the smallness of all interchain interactions and anisotropy terms, they can be highly relevant in suppressing the higher-order spin-multipolar quantum phases. The new data also demonstrated the sensitivity of the complex magnetic phase diagram and the ordering wave vector itself to the field direction, as entirely different phase sequences were found along the three orthogonal crystal directions \cite{CemalEnderle18}.

\subsection{Azurite: realization of the generalized diamond-chain model}

\begin{figure}[b!]
\begin{center}
\includegraphics[width=0.44\linewidth]{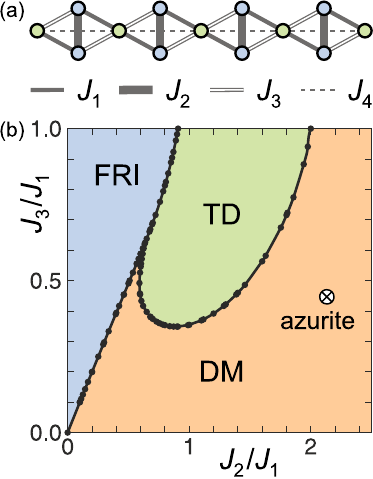}\vspace{-2pt}
\end{center}
\caption{(a)~Schematic structure of the generalized diamond chain in azurite with the most important exchange interactions. (b)~The location of azurite in the numerical phase diagram of a distorted diamond chain \cite{OkamotoTonegawa03} according to the exchange parameters from Ref.~\cite{JeschkeOpahle11} (neglecting~$J_4$).}\label{Fig:AzuriteStruct}
\end{figure}

Frustrated spin chains of a different type are realized in another quantum spin system, Cu$_3$(CO$_3$)$_2$(OH)$_2$, known as azurite. This deep-blue copper carbonate mineral, closely associated with green malachite, has a monoclinic crystal structure, and its precise atomic positions are well known from both x-ray and neutron diffraction experiments \cite{GattowZemann58, ZiganSchuster72, BelokonevaGubina01, GibsonRule10, RuleReehuis11}. Originally, its structure was described in the centrosymmetric space group $P2_1/c$, but more recent neutron-diffraction measurements \cite{RuleReehuis11} revealed an additional distortion that lowers the lattice symmetry to its enantiomorphic subgroup $P2_1$. The copper atoms are arranged into quasi-1D diamond chains consisting of corner-sharing rhombic plaquettes, schematically depicted in Fig.~\ref{Fig:AzuriteStruct}\,(a). The chain is dimerized across the rungs that are formed by the dominant AFM exchange interaction $J_2$ between the nearest Cu(2) sites. Its absolute value was reported to be between 2.1 and 3.8~meV depending on the specific model or measurement method \cite{JeschkeOpahle11, HoneckerHu11}. The dimer-monomer interactions are given by the two nonequivalent Cu(1)\,--\,Cu(2) coupling constants $J_1\approx0.47 J_2$ and $J_3\approx 0.21 J_2$. Further, an additional direct monomer-monomer exchange \mbox{$J_4\equiv J_{\rm m}\approx0.14 J_2$} is often considered as well~\cite{RuleWolter08, JeschkeOpahle11}. The interchain coupling $J_\perp$ is usually neglected, except in a recent neutron-scattering study where a weak anisotropic (Ising-type) interchain interaction has been claimed responsible for the multiple spin gaps observed in the magnon spectrum \cite{RuleTennant11}.

\begin{figure}[t]
\begin{center}
\includegraphics[width=0.85\linewidth]{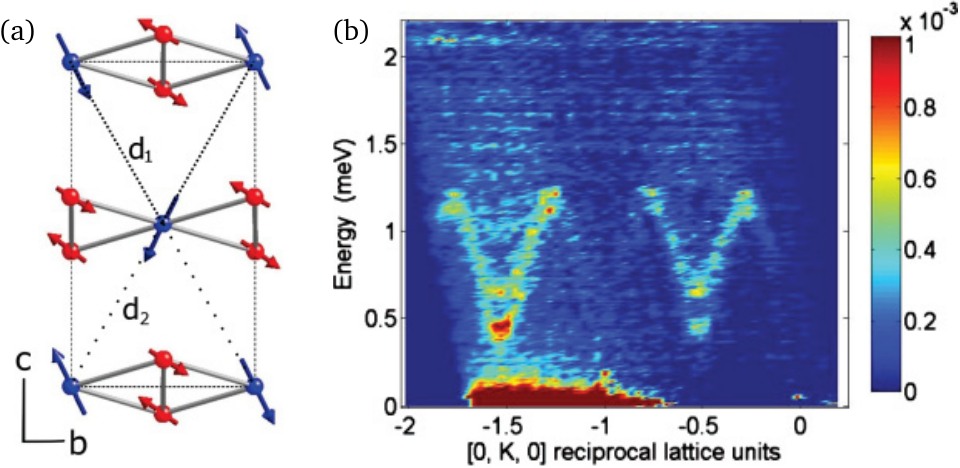}
\end{center}
\caption{(a)~The magnetic structure of azurite after Rule \textit{et~al.} \cite{RuleReehuis11}. The magnetic moments on the Cu(1) sites (0.68\,$\mu_{\rm B}$) and Cu(2) sites (0.26\,$\mu_{\rm B}$) are shown in blue and red, respectively. (b)~INS data on azurite, measured below $T_{\rm N}$ at a temperature of 60~mK, revealing two spin gaps due to the 3D interchain interactions. After Rule \textit{et~al.}~\cite{RuleTennant11}.}\label{Fig:Azurite}
\end{figure}

Neutron diffraction measurements allowed Rule \textit{et~al.} \cite{RuleReehuis11} to solve the magnetic structure of azurite that sets in below $T_{\rm N}\approx1.9$\,K. It is comprised of two inequivalent ordered magnetic moments on Cu(1) and Cu(2) sites of magnitudes $\sim$\,0.68\,$\mu_{\rm B}$ and $\sim$\,0.26\,$\mu_{\rm B}$, respectively. They form a commensurate noncollinear spin structure with the propagation vector $\mathbf{q}=\left(\frac{1}{2}\,\frac{1}{2}\,\frac{1}{2}\right)$ \cite{GibsonRule10, RuleReehuis11}, which is shown in Fig.~\ref{Fig:Azurite}\,(a). The noncollinearity of the magnetic structure cannot be explained by isotropic interactions alone, suggesting additional DMI terms in the spin Hamiltonian that act along the $J_1$ and $J_3$ superexchange paths \cite{RuleReehuis11}.

The symmetric diamond chain obtained by setting $J_1=J_3$ and $J_4=0$ has attracted much attention from the theory side \cite{TakanoKubo96} as a limiting case of the more general distorted diamond chain ($J_1\neq J_3$, $J_4=0$) \cite{OkamotoTonegawa03}. According to the numerical phase diagram reproduced from Ref.~\cite{OkamotoTonegawa03} in Fig.~\ref{Fig:AzuriteStruct}\,(b), the exchange parameters of azurite \cite{JeschkeOpahle11} would place it within the region corresponding to the DM (spin fluid) state above its 3D ordering temperature. Its characteristic feature is the appearance of the 1/3-magnetization-plateau phase \cite{KikuchiFujii05}, related to the polarization of the monomers by the external magnetic field. The applicability of this model to azurite gained support, in particular, from the $^{63,65}$Cu NMR data \cite{AimoKraemer09} that indicate full polarization of the monomer spins within the plateau phase, whereas the local spin polarization of the dimers is limited to a few percent, indicating that they essentially remain in the singlet state. In azurite, the 1/3 plateau is observed in fields between $B_{\rm c1}=11$\,T and $B_{\rm c2}=30$\,T and is characterized by a spin gap opening at the AFM point $\mathbf{q}=\left(1\,\frac{1}{2}\,0\right)$. The excitations are represented by ferromagnons with an approximately cosinusodial dispersion, originating from the interacting monomers, and a weakly dispersive Zeeman-split singlet-triplet excitation at higher energies, originating from the dimers \cite{RuleWolter08}. These experimental observations at high magnetic fields are well captured by the results of exact diagonalization and DDMRG calculations \cite{HoneckerHu11, JeschkeOpahle11}, lending further support to the dimer-monomer model.

Honecker~\textit{et~al.}~also presented simulations of the dynamic structure factor in zero magnetic field using DDMRG on a chain with $N=60$ sites~\cite{HoneckerHu11}. The spectrum consists of a gapless two-spinon continuum at low energies ($E\leq2$\,meV), well separated from a broad band of higher-energy dimer excitations, in qualitative agreement with the earlier neutron-spectroscopy data \cite{RuleWolter08}. More recently, finer details of the magnetic excitation spectrum were revealed below the 3D ordering temperature, where azurite enters an AFM-ordered state. The combination of an anisotropic staggered field with a weak interchain coupling leads to the opening of two spin gaps in the spectrum, $\Delta_1\approx0.4$\,meV and $\Delta_2\approx0.6$\,meV. These features are not captured by the diamond-chain model and highlight the significant role of anisotropy terms and interchain interactions in the spin Hamiltonian of azurite \cite{RuleTennant11}. Upon application of magnetic fields up to 7~T (which is below the critical field for the 1/3 plateau phase), the lower gap remains at its commensurate position, whereas the upper gap splits in the longitudinal direction. This effect allowed the authors of Ref.~\cite{RuleTennant11} to associate these two contributions with the transverse and longitudinal continua, originating from spin correlations perpendicular, $S^\perp(\mathbf{Q},\omega)$, and parallel, $S^\parallel(\mathbf{Q},\omega)$, to the magnetic field direction.

Application of a magnetic field also leads to very notable and rather intricate changes in the energies of the spin excitations. According to the INS data \cite{RuleTennant11}, reproduced here in Fig.\,\ref{Fig:Azurite}, the zone-boundary mode at 1.2~meV splits into two modes at low fields with a third branch appearing above 3~T, which do not follow a Zeeman-type splitting. At the zone centre, the lower gap $\Delta_1$ shows a clear decrease in energy, while the upper gap $\Delta_2$ remains approximately constant with field. This unusual behaviour emphasizes that the character of these continuum-boundary excitations is distinct from that of conventional spin waves. The higher-energy mode corresponding to the low-lying $\left|\uparrow\uparrow\right\rangle$ triplet branch of the dimers shows a linear-in-field decrease as a function of the magnetic field for $B<B_{\rm c1}$ due to the Zeeman effect. Above $B_{\rm c1}$, within the 1/3 plateau phase, the linear slope ($g$ factor) changes, and the mode reaches zero energy at $B_{\rm c2}\approx30$\,T. This behaviour has been observed consistently both by neutron spectroscopy \cite{RuleTennant11} and by ESR \cite{OhtaOkubo03, OkuboKamikawa04, Kamenskyi12}.

However, more recent ESR experiments \cite{Kamenskyi12} identified a number of previously unknown resonances, some of which are likely responsible for the broad unresolved peaks seen earlier by neutron scattering at energies of about 4\,--\,6~meV \cite{RuleWolter08}. The complexity of the ESR spectrum indicates deviations from the simple dimer-monomer model but can be remarkably well explained by an energy-level diagram involving a ground-state doublet, an excited doublet with an energy of $\Delta_{\rm ED}\approx0.76$~meV (183~GHz), and an excited quadruplet state at $\Delta_{\rm ES}\approx4.3$~meV (1043~GHz), which undergo a Zeeman splitting in magnetic field. This led Kamenskyi \cite{Kamenskyi12} to propose a more elaborate model, consisting of six interacting spins forming a segment of the diamond chain, which likely constitutes the minimal backbone that is required to explain all the experimental observations.

\subsection{Euchroite: frustrated sawtooth chains with a spin gap}

\begin{figure}[b]
\begin{center}
\includegraphics[width=0.52\linewidth]{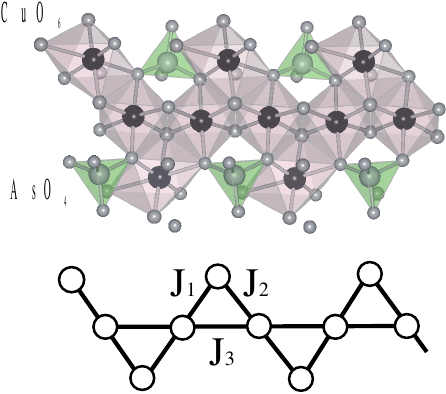}
\end{center}
\caption{The crystal structure of euchroite (top) and the schematic view of the sawtooth chains with three relevant exchange interactions (bottom). After Kikuchi \textit{et~al.}~\cite{KikuchiFujii11}.}\label{Fig:Euchroite}
\end{figure}

The quasi-1D Heisenberg AFM spin chains found in the copper arsenate hydroxide mineral euchroite, Cu$_2$(AsO$_4$)(OH)\,$\cdot$\,3H$_2$O (orthorhombic space group $P2_12_12_1$) \cite{Finney66}, consist of corner-sharing triangles, as shown in Fig.~\ref{Fig:Euchroite}. Such structures are known as sawtooth chains (or~$\Delta$~chains) and have been previously considered in relationship to the compound YCuO$_{2.5}$, derived from a delafossite-type structure by doping additional oxygen atoms that mediate AFM super-exchange interactions in the Cu plane \cite{SenShastry96, LeBacqPasturel05, OlariuBono06}. In euchroite, which is essentially a hydrated analog of olivenite \cite{Heritsch38}, magnetic Cu$^{2+}$ ions are surrounded by six oxygen atoms that mediate superexchange interactions. A slight difference in the bond lengths (3.01, 3.06, and 3.17~\AA) and, consequently, in the Cu--O--Cu bond angles along the sides of every unit triangle guarantees that the three exchange paths $J_1$, $J_2$, and $J_3$ are non\-equivalent.

Natural samples of euchroite have been studied by magnetic susceptibility, specific heat, high-field magnetization, and $^{1}$H-NMR measurements~\cite{KikuchiFujii11}. The susceptibility data show a maximum around 85~K, characteristic of a low-dimensional antiferromagnet, and a spin-gap behaviour at low temperatures with a small upturn due to impurities. The high-temperature range can be fitted to a Curie-Weiss law with $\Theta_{\rm CW}\approx-50$~K. The authors of Ref.~\cite{KikuchiFujii11} employed a simple spin-dimer model to extract the approximate scale of the dominant exchange constant, $J\approx11.6$~meV, by fitting the susceptibility data. A reasonably good quality of the fit suggests that euchroite may form a singlet ground state due to dimerization. The high-field magnetization of euchroite revealed a plateau followed by a steep increase above 40~T, which can be explained by the presence of a spin gap with a magnitude of $\sim$\,5~meV. A more accurate estimate of the spin gap could be obtained from the temperature dependence of the spin-lattice relaxation rate $T_1^{-1}$, fitted to an Arrhenius-type equation with an activation energy of $\sim$\,8~meV \cite{KikuchiFujii11}. Further details about this dimerized $\Delta$~chain system could be obtained in future high-field magnetization and spectroscopic measurements.

\subsection[Fedotovite: Haldane chains of edge-shared tetrahedral spin clusters]{\mbox{\hspace{-3pt}Fedotovite: Haldane chains of edge-shared tetrahedral spin clusters}}

\begin{figure}[t]
\begin{center}
\includegraphics[width=0.61\linewidth]{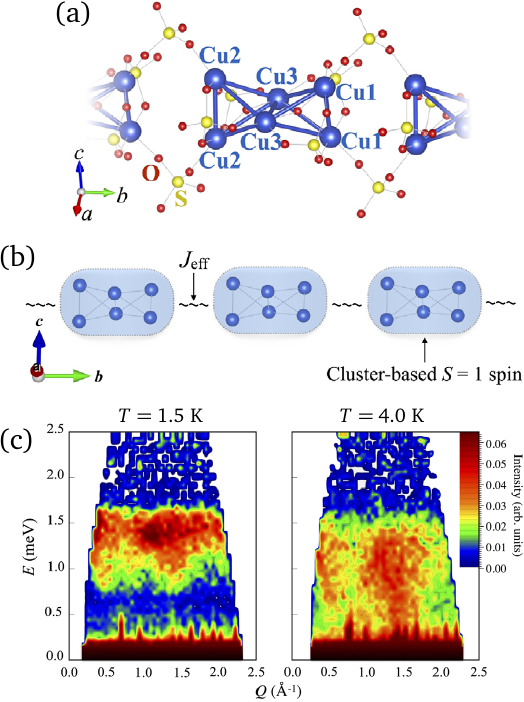}\vspace{-2pt}
\end{center}
\caption{(a)~The crystal structure of fedotovite in the vicinity of a spin cluster. (b)~An effective model of the cluster-based $S=1$ Haldane chain. (c)~The INS data, measured on a synthetic powder sample of fedotovite at $T=1.5$~K (left) and 4.0~K (right). After Fujihala \textit{et~al.}~\cite{FujihalaSugimoto18}.}\label{Fig:Fedotovite}
\end{figure}

In a recent study, Fujihala \textit{et~al.} \cite{FujihalaSugimoto18} considered the magnetic properties of the sulfate mineral fedotovite, K$_2$Cu$_3$O(SO$_4$)$_3$. This compound forms emerald-green crystals with a monoclinic crystal structure (space group $C2/c$), in which magnetic Cu$^{2+}$ ions are grouped by six into edge-sharing tetrahedral clusters \cite{StarovaFilatov91}. These clusters are then connected to each other by SO$_4^{2-}$ ions to form chains along the $b$ axis, as shown in Fig.~\ref{Fig:Fedotovite}\,(a,b), whereas interchain coupling along the $a$ and $c$ axes is negligible. The effective model of such a hexamer cluster, proposed by Fujihala \textit{et~al.} \cite{FujihalaSugimoto18}, includes three superexchange interactions: the weakly ferromagnetic rung interaction $J_1\approx-3$~meV, which acts between the structurally equivalent Cu sites, and two nearly equal diagonal AFM interactions $J_2\approx J_3\approx 10.8$~meV. These parameters result in a triplet ground state of the cluster, which has been confirmed by magnetic susceptibility and high-field magnetization measurements. The latter exhibits a 1/3 plateau above $\sim$\,30~T, which is small compared to the $J_2\approx J_3$ energy scale. Consequently, an effective $S=1$ degree of freedom can be associated with every cluster in its ground state. It is therefore expected that such clusters, connected into a 1D chain by an effective intercluster interaction $J_{\rm eff}$, would realize a Haldane state by analogy with integer-spin chains or AFM spin-$\frac{1}{2}$ ladders \cite{MasudaZheludev06, VekuaHonecker06}. In the same work \cite{FujihalaSugimoto18}, the authors extend this case to a more general situation of edge-shared tetrahedral clusters with an arbitrary number of rungs and come to the conclusion that the ground state would alternate between singlet and triplet as the number of tetrahedra is increased. Therefore, any cluster with an even number of tetrahedra (i.e. odd number of rungs) would realize a Haldane state. In this respect, fedotovite can be considered as a generalization of the $S=\frac{1}{2}$ AFM spin-ladder model, where a cluster trivially consists of a single rung.

To estimate the value of the Haldane gap, Fujihala \textit{et~al.} also measured the low-energy spin-excitation spectra of a synthetic K$_2$Cu$_3$O(SO$_4$)$_3$ powder sample by time-of-flight neutron spectroscopy \cite{FujihalaSugimoto18}. The colour maps illustrating the distribution of powder-averaged INS intensity in the energy-momentum space are reproduced here in Fig.~\ref{Fig:Fedotovite}\,(c). At low temperatures, the spectrum is fully gapped below $\sim$\,0.6~meV\,$\approx$\,7~K, which is consistent with the value of the excitation gap of $\sim$\,6~K that was obtained from magnetization measurements. A relatively small increase in temperature to 4~K is already sufficient to close the energy gap in the INS spectrum. Later, Furrer~\textit{et al.} \cite{FurrerPodlesnyak18} extended the INS data to higher energies and observed additional transitions to the lowest-lying excited hexamer states in both fedotovite and its isostructural analog Na$_2$Cu$_3$O(SO$_4$)$_3$. This enabled a more accurate estimate of the exchange parameters within the cluster that differentiates between all three nonequivalent FM coupled rungs and four nonequivalent AFM coupling paths. In contrast to the originally proposed model, the FM exchange on the inner rung turned out to be at least twice larger in comparison to the outer rungs. Similarly, a significant difference was also revealed among the diagonal AFM interactions, yet the original conclusion about an $S=1$ triplet ground state of the cluster remained unaffected.

\subsection{Dioptase: a hexagonal network of ferromagnetically coupled AFM chains}

\begin{figure*}[b]
\includegraphics[width=\textwidth]{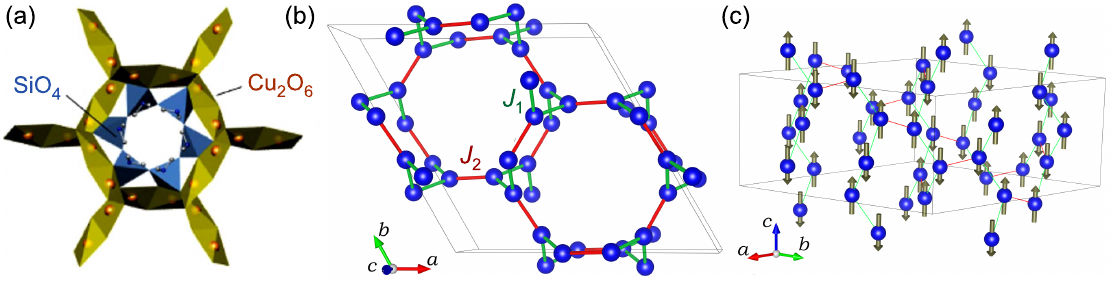}
\caption{The crystal structure, magnetic interactions, and AFM order in green dioptase. The leading AFM coupling $J_1$ along the chain direction and the weaker ferromagnetic coupling $J_2$ between the chains are shown with green and red lines, respectively. After Janson \textit{et~al.}~\cite{JansonTsirlin10} and Podlesnyak \textit{et~al.}~\cite{PodlesnyakAnovitz16}.}\label{Fig:DioptaseStructure}
\end{figure*}

Another copper mineral that attracted recent attention is the green dioptase, Cu$_6$Si$_6$O$_{18}\cdot6$H$_2$O. Its rhombohedral crystal structure, described by the $R\overline{3}$ space group \cite{HeideBollDornberger55, RibbeGibbs77}, consists of spiral chains running along the $c$ axis, which are arranged into closely packed 12-membered Cu rings in the $a{\kern-.6pt}b$ plane, with smaller nonmagnetic beryl-type Si$_6$O$_{18}$ rings located inside [see Fig.~\ref{Fig:DioptaseStructure}\,(a)]. It can be therefore classified as a member of the cyclosilicate family of minerals. Despite the seemingly complicated lattice, copper ions occupy a single Wyckoff site and are therefore structurally equivalent. Magnetic interactions are represented by the stronger AFM coupling $J_1\equiv J_c>0$ along the chains and weaker FM coupling $J_2\equiv J_{a{\kern-.3pt}b}<0$ between the chains, as shown in Fig.~\ref{Fig:DioptaseStructure}\,(b), resulting in a very elegant magnetic Hamiltonian with only one tuning parameter, $\alpha=J_2/J_1$. At the same time, there has been a lot of controversy about the values and even the sign of the interchain interaction until recently.

\begin{figure*}[t!]
\includegraphics[width=\textwidth]{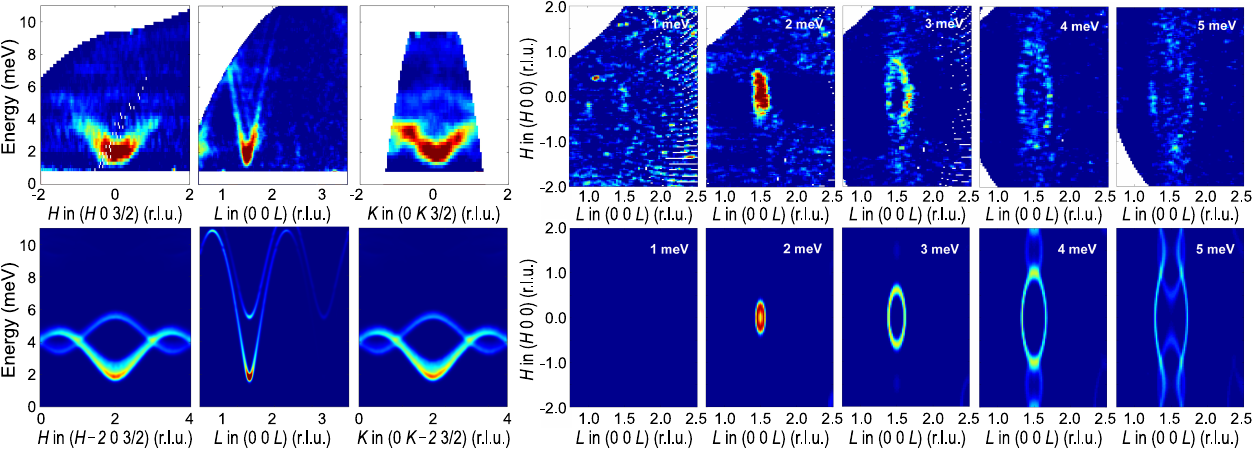}\vspace{6pt}
\caption{Selected energy-momentum (left) and constant-energy (right) cuts through the INS data measured on a natural sample of green dioptase in the AFM state at $T=1.7$\,K. The experimental data shown at the top are compared with the spin-wave model convoluted with spectrometer resolution, shown for the same cuts below each panel. After Podlesnyak \textit{et~al.}~\cite{PodlesnyakAnovitz16}.}\label{Fig:DioptaseINS}
\end{figure*}

It has been known since the 1960s that dioptase orders antiferromagnetically at low temperatures, though the reported values of the N\'{e}el temperature, $T_{\rm N}$, varied widely  \cite{SpenceMuller58, EisenbergForstat64, NewnhamSantoro67, KiselevaOgorodova93, WintenbergerAndre93}. The magnetic structure has been reported in a number of works and is illustrated here in Fig.~\ref{Fig:DioptaseStructure}\,(c). It is characterized by the ordered moment of only $0.55\,\mu_{\rm B}$ \cite{WintenbergerAndre93, BelokonevaGubina02}, which is unusually low for a 3D antiferromagnet and can be traced back to the low connectivity of the lattice. For the anhydrous version of the same mineral, known as black dioptase, Wintenberger \textit{et al.} proposed that the intrachain exchange interactions are AFM, \mbox{$J_1>0$}, whereas the interchain ones are FM, $J_2<0$, and much weaker \cite{WintenbergerAndre93}. Later, it was demonstrated using theoretical calculations that $J_2$ in black dioptase is weaker than $J_1$ by two orders of magnitude~\cite{LawHoch10}. On the other hand, it was suggested that in the fully hydrated green dioptase both $J_1$ and $J_2$ are AFM and comparable \cite{GrosLemmens02}, which would place it in the proximity of a QCP at $\alpha_{\rm c}=1.86$ that separates an AFM ordered state from a quantum spin liquid. However, recent $\mu$SR experiments revealed no evidence for quantum magnetic fluctuations \mbox{either above or below $T_{\rm N}$ \cite{BerlieTerry17}}.

Gross and co-workers \cite{GrosLemmens02} parameterized the dioptase Hamiltonian with $\delta=(\alpha-1)/(\alpha+1)$, which is equivalent to $J_1=J(1-\delta)$ and $J_2=J(1+\delta)$ with $J>0$, and studied the phase diagram of the resulting model in the range $-1<\delta<1$ by QMC simulations. They found that for sufficiently strong $J_2$, namely for $\delta>0.3$, the AFM order is destroyed due to the dimerization on the $J_2$ bonds, which leads to a gapped singlet-dimer state. As $J_2$ is reduced, the maximum of $T_{\rm N}$ is reached near $\delta\approx0$, but upon approaching $\delta=-1$ the AFM chains become decoupled, which again leads to a suppression of long-range order according to the Mermin--Wagner theorem. Unfortunately, the region of negative $\delta$, which would correspond to FM interchain coupling, was not explored in this work.

More recently, however, full-potential DFT calculations and neutron-spectroscopy measurements consistently demonstrated that the interchain coupling in dioptase is actually ferromagnetic. The theoretical calculations resulted in the exchange parameters $J_1=6.72$ and $J_2=-3.19$~meV \cite{JansonTsirlin10}, whereas the experimental values are $J_1=10.6$ and $J_2=-1.2$~meV \cite{PodlesnyakAnovitz16}. The latter values, which correspond to $\alpha\approx-0.11$, are consistent with the observed reduction of the N\'eel temperature $T_{\rm N}\approx15$~K with respect to $J_1$ and the reduced ordered moment of $\sim$\,0.55\,$\mu_{\rm B}$ \cite{BelokonevaGubina02} by quantum spin fluctuations expected in the quasi-1D system of weakly coupled chains. The ordering in dioptase could be additionally impeded by the low coordination number of the lattice \cite{JansonTsirlin10}. The consistency between the INS data and the new model is illustrated by the comparison of the time-of-flight data with the spin-spin correlation function calculated from linear spin-wave theory \cite{PodlesnyakAnovitz16}, as shown in Fig.~\ref{Fig:DioptaseINS}. These new results clearly exclude the scenario of an AFM interchain coupling. The uncertainty surrounding the value of interchain interactions is understandable if one considers the intradimer Cu--O--Cu bridging angle of 97.4$^\circ$, which falls within the range where the FM and AFM contributions tend to compensate each other according to the GKA rules \cite{LeberneggTsirlin14}. It is also worth noting that the crystal symmetry allows for DMI between the Cu spins, which have not been typically included in the models but were claimed responsible for a broad spectral feature observed in ESR \cite{OhtaOkubo09}.

\begin{figure}[b!]
\begin{center}
\includegraphics[width=0.65\linewidth]{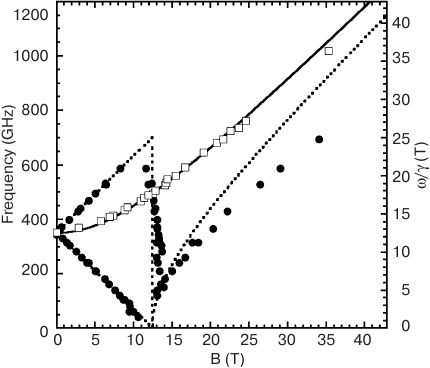}
\end{center}
\caption{Low-temperature (\mbox{$T=4.2$\,K}) ESR data for dioptase, measured in pulsed magnetic fields applied parallel~($\bullet$) and perpendicular ($\square$) to the chains. The solid and dotted lines describe two-sublattice AFM resonances under uniaxial anisotropy for fields along the hard and easy axis, respectively. After Ohta \textit{et~al.}~\cite{OhtaOkubo09}.}\label{Fig:DioptaseESR}
\end{figure}

The value of the spin gap in the zero-field AFM state, estimated from the INS data, amounts to $\sim$\,1.5~meV \cite{PodlesnyakAnovitz16}, which agrees perfectly with the zero-field AFM gap of $\Delta_{\rm AF}=350$~GHz (1.45~meV) measured by ESR \cite{OhtaOkubo09}, as shown in Fig.~\ref{Fig:DioptaseESR}. The AFM resonance in ESR has been tracked in high magnetic fields up to 35~T for fields applied both perpendicular and along the chains at $T=4.2$~K. For $\mathbf{B}\perp\mathbf{c}$ (hard axis), the energy of the resonance monotonically increases without any notable anomalies, whereas for $\mathbf{B}\parallel\mathbf{c}$ (easy axis) it initially splits into two branches as expected for the two-sublattice AFM resonance in the presence of a uniaxial anisotropy. At a critical field of $B_{\rm sf}=12.5$~T, the spin gap closes as the lower branch hits zero energy, and a spin-flop transition occurs \cite{OhtaOkubo09}. Taking into account that the spin-flop field is within the reach of most experimental probes, the high-field phase would certainly warrant a more detailed investigation by other methods, including neutron scattering, as it could help to pinpoint magnetic interactions in dioptase with higher accuracy.

\subsection{Szenicsite and antlerite: frustrated spin chains near the Majumdar-Ghosh point}

\begin{figure}[b!]
\begin{center}
\includegraphics[width=0.7\linewidth]{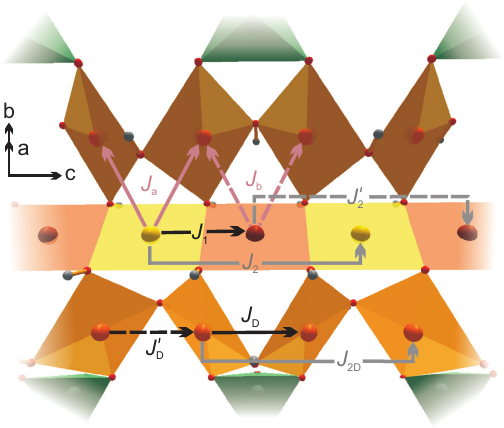}
\end{center}
\caption{Exchange parameters within the triple copper-oxide chain of szenicsite. After Lebernegg \textit{et~al.}~\cite{LeberneggJanson17}.}\label{Fig:Szenicsite}
\end{figure}

Interesting physical aspects of frustrated spin chains were recently revealed in the rare copper-molybdate mineral szenicsite by Lebernegg \textit{et al.} \cite{LeberneggJanson17}. This system, with the chemical formula Cu$_3$(MoO$_4$)(OH)$_4$, crystallizes in an orthorhombic structure described by the space group $Pnnm$, in which CuO$_4$ plaquettes form triple chains that are parallel to the $c$ axis \cite{Burns98, StolzArmbruster98}. In this respect, the structure resembles that of the sulfate mineral antlerite, Cu$_3$(SO$_4$)(OH)$_4$, in which the edge-sharing Cu(1) central chain shows an idle-spin behaviour, and only the corner-sharing Cu(2) outer chains order magnetically below $T_{\rm N}\approx5.3$\,K into a collinear structure characterized by an antiparallel orientation of spins in the opposite FM aligned chains \cite{VilminotRichardPlouet03, VilminotAndre07, BissengaliyevaBekturganov13}. An isostructural synthetic compound Cu$_3$(SeO$_4$)(OH)$_4$, on the other hand, stabilizes a more complex cycloidal magnetic structure among the FM chains that locks into the commensurate $\left(\frac{1}{7}\,0\,0\right)$ propagation vector at low temperatures \cite{VilminotAndre07}. A more recent work has cast doubt on the idle-spin behaviour of the central chain, and an alternative model with multiple AFM interactions between the chains was put forward \cite{KooKremer12}. Despite the structural similarities among these compounds, the magnetic behaviour of szenicsite turns out to be very different, as its ground state shows no long-range magnetic order. Its central chain represents a realization of the frustrated AFM $J_1$-$J_2$ chain model with the NN coupling $J_1\approx5.86$~meV and a sizable NNN coupling $J_2$ that is approximately twice weaker, $J_2/J_1\approx0.5$ \cite{LeberneggJanson17}. It was argued that the interchain frustration essentially decouples this central chain from the two side chains that can be in turn described by AFM coupled dimers \cite{LeberneggJanson17}, rather than by the uniform chain model proposed earlier \cite{FujisawaKikuchi11, FujiiKikuchi15}.

As discussed in section \ref{Sec:Chains} above, the uniform AFM Heisenberg chain with NN and NNN interactions features a gapped dimer-crystal ground state for $J_2/J_1>\alpha_{\rm c}=0.2411$, and the exchange parameters estimated for the central chain in szenicsite certainly place it within this regime. Moreover, its $J_2/J_1$ ratio is very close to the Majumdar-Ghosh point ($\alpha_{\rm MG}=0.5$), where two degenerate ground states can be exactly represented by a superposition of spin singlets \cite{MajumdarGhosh69}. A maximum observed in the specific heat divided by temperature, $C_p/T$, at 1.2\,K is an additional indication of the szenicsite's proximity to the Majumdar-Ghosh point, pointing to an $\alpha$ ratio between 0.4 and 0.5 \cite{LeberneggJanson17}. In total, the microscopic magnetic model derived in Ref.~\cite{LeberneggJanson17} from LSDA+$U$ calculations combined with susceptibility, high-field magnetization, and thermodynamic measurements includes 8 exchange parameters as shown in Fig.~\ref{Fig:Szenicsite}. Nevertheless, combining the dimer contribution ($J_{\rm D}$) from the outer chains with the uniform-Heisenberg-chain contribution ($J_1$, $J_2$) from the inner chain is already sufficient to fit the susceptibility data \cite{LeberneggJanson17}. Lebernegg \textit{et al.} also analyzed the specifics of the exchange network in szenicsite and its differences as compared to antlerite, which are observed in the absence of long-range order, non-idle-spin behaviour, and a large positive Curie-Weiss temperature of 68\,K indicating the predominance of AFM interactions. These differences are explained mainly by the $J_{\rm D}$ and $J^\prime_{\rm D}$ couplings in the side chains, which in the case of antlerite are both ferromagnetic \cite{VilminotRichardPlouet03, KooKremer12, FujiiIshikawa13}. At the same time, the interactions within the central chain in both szenicsite and antlerite are similar, and the debated idle-spin behaviour (absence of an ordered moment) obtains a natural explanation as a consequence of the gapped spin-singlet state.\vspace{-2pt}

\subsection{Columbite niobates: from Heisenberg to Ising chains}\label{Sec:Columbite}

The black mineral columbite, also known as niobite, is mined as a niobium ore. In 1801, the chemist Charles Hatchett analyzed a columbite specimen from the mineral collection of the British Museum, which then led to the discovery of the element niobium \cite{GriffithMorris03}. Columbite has the general chemical composition of $A$Nb$_2$O$_6$ ($A$\,=\,Fe,\,Mn) and an orthorhombic crystal structure described by the space group \textit{Pbcn} \cite{NielsenLebech76}. Isostructural niobates can be also synthesized with other transition metals on the $A$ site (e.g. Co or Ni) \cite{Pullar09}, and the corresponding solid solutions, such as Fe$_{1-x}$Mn$_x$Nb$_2$O$_6$ \cite{TealdiMozzati04, TarantinoGhigna05, TarantinoZema05}, Fe$_{1-x}$Co$_x$Nb$_2$O$_6$  \cite{SarvezukKinast11} and Fe$_{1-x}$Ni$_x$Nb$_2$O$_6$ \cite{SarvezukGusmao12}, are known to be stable in the whole range of concentrations between the stoichiometric end members. The magnetic transition-metal cations form weakly coupled quasi-1D chains that run along the $c$ axis, which is a common feature of all columbite compounds. However, the degree of magnetic anisotropy is strongly dependent on the choice of the magnetic ion. On the Mn-rich side, the spin chains are of Heisenberg type and antiferromagnetic \cite{HnedaNeto18}, whereas on the Fe- and Co-rich side they are of Ising type and ferromagnetic, interacting via much weaker AFM interchain couplings \cite{KobayashiMitsuda00, SarvezukKinast11}.

\begin{figure}[b!]
\begin{center}
\includegraphics[width=0.77\linewidth]{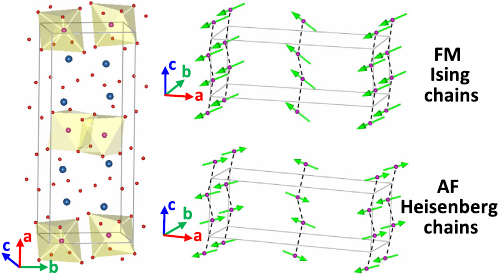}\vspace{-2pt}
\end{center}
\caption{The crystal structure of columbite and the difference in magnetic structures of FeNb$_2$O$_6$ (top) and MnNb$_2$O$_6$ (bottom). After Hneda \textit{et~al.}~\cite{HnedaNeto18}.}\label{Fig:Columbites}
\end{figure}

In particular, CoNb$_2$O$_6$ attracted a lot of recent interest as a model material for the realization of a frustrated antiferromagnet consisting of ferromagnetic Ising chains \cite{ColdeaTennant10, LeeKaul10, CabreraThompson14, KinrossFu14, LiangKoohpayeh15}. In a transverse magnetic field of $H_{\rm c}\approx5.24$~T, this compound exhibits a quantum critical point associated with the suppression of FM intrachain correlations, which marks the transition to a quantum paramagnetic phase at higher fields. Below this critical point, the spin dynamics is governed by the propagating domain-wall quasiparticles known as ``kinks''. In a freestanding Ising chain, these fractionalized excitations are delocalized and form a gapped continuum in the neutron-scattering spectrum due to scattering by kink pairs. However, weak interactions with the neighbouring chains introduce an effective attractive potential between the kinks that leads to their confinement. As a result, deep within the magnetically ordered phase, the continuum splits into a sequence of sharp dispersing modes, as it was elegantly demonstrated in an INS experiment by Coldea \textit{et al.} \cite{ColdeaTennant10}. In the quantum paramagnetic phase above the critical point, the excitation spectrum changes to a single sharp mode that corresponds to scattering by spin-flip quasiparticles \cite{CabreraThompson14}.

In spite of the weakness of interchain interactions, they determine the magnetic ordering pattern, which differs among the columbite compounds. Although the order is globally antiferromagnetic and noncollinear in all cases, in Heisenberg-type MnNb$_2$O$_6$ a two-sublattice ``$\mathbf{q}=0$'' magnetic structure is realized, whereas in FeNb$_2$O$_6$ the FM Ising chains are arrange alternatingly in the $ab$ plane with a propagation vector $\mathbf{k}=(0\,\frac{1}{2}\,0)$ \cite{Weitzel71, NielsenLebech76, HnedaNeto18}. These magnetic structures are illustrated in Fig.~\ref{Fig:Columbites}. In CoNb$_2$O$_6$, the presence of two phases with propagation vectors $\mathbf{k}_1=(0\,\frac{1}{2}\,0)$ and $\mathbf{k}_2=(\frac{1}{2}\,\frac{1}{2}\,0)$ has been reported \cite{SarvezukKinast11jap}. The noncollinearity of the spin structure is dictated by the direction of the local Ising axis in the distorted $A$O$_6$ octahedra. In all of the studied solid solutions with random distribution of the $A$ cations, the non-stoichiometric composition rapidly suppressed long-range magnetic order \cite{TealdiMozzati04, SarvezukKinast11, SarvezukGusmao12, HnedaNeto18}. This evidences a substantial drop of the average exchange strength both within and between the chains and a tendency for enhanced frustration among the interchain interactions as a result of disorder.\vspace{-2pt}

\section{Kagome systems}\vspace{-1pt}\label{Chap:Kagome}

\subsection{The perplexing magnetism in kagome layers}\label{Sec:KagomeIntro}

\begin{figure}[b!]
\begin{center}
\includegraphics[width=0.85\linewidth]{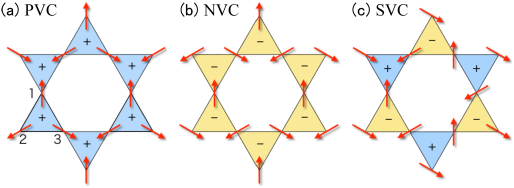}
\end{center}
\caption{Possible coplanar types of magnetic order on a kagome lattice, characterized by the positive, negative, and staggered vector chirality. The direction of the vector
chirality on each triangle is shown by ``+'' (up) or ``--'' (down). After Okuma \textit{et~al.}~\cite{OkumaYajima17}.}\label{Fig:KagomeStruct}
\end{figure}

Periodic 2D arrangements (tilings) of regular polygons are exemplified by 11 uniform Archimedean lattices, which serve as prototypes of 2D interacting spin systems realized in various crystal structures \cite{RichterSchulenburg04, FarnellGoetze14, Yu15, SchmidtThalmeier17, FarnellGoetze18}. Among them, the kagome lattice plays an exceptional role, because it is characterized by the combination of strong frustration (as strong as in the triangular lattice) and low coordination number, $z=4$. It is composed of corner-sharing triangles and can be obtained by a 1/4 site depletion of the triangular lattice. Along with the more exotic star lattice \cite{RichterSchulenburg04prb, ZhengTong07, ChoyKim09, YangParamekanti10}, in which the triangles are additionally separated by a dimer, the Heisenberg model on a kagome antiferromagnetic (KAFM) lattice has no semiclassical ordered ground state. It is therefore an ideal candidate for realizing a quantum-disordered state or spin liquid. In addition, it is also commonly found in minerals and a variety of crystal structures that have attracted much recent interest and has already become the subject of several dedicated review articles \cite{MendelsWills11, MendelsBert16, Norman16}.

The corner-sharing geometry of KAFM layers supports an extensive classical ground-state manifold that leads to a finite residual entropy at \mbox{$T=0$}. A coplanar or slightly canted spin configuration is then usually selected by the longer-range Heisenberg couplings \cite{HarrisKallin92, JansonRichter08, SuttnerPlatt14, IqbalJeschke15}, anti\-symmetric DMI \cite{ElhajalCanals02, BallouCanals03, GroholNocera03, CepasFong08, OkumaYajima17}, or via the order-by-disorder mechanism \cite{Chalker11}. This classical limit is relevant, for instance, for the $A{\kern1pt}B_3$(SO$_4$)$_2$(OH)$_6\kern.5pt$ jarosites, where the $B^{3+}$ cation can be Fe$^{3+}$ ($S=5/2$) \cite{GroholNocera03, GroholMatan05} or Cr$^{3+}$ ($S=3/2$) \cite{TownsendLongworth86, NishiyamaMorimoto01, OkutaHara11}. The behaviour of quantum-spin Heisenberg KAFM systems, however, still represents an open area of research, and even the exact ground state of the $S=1/2$ Heisenberg KAFM model remains debated \cite{MisguichLhuillier13, LiaoXie17}. Possible classes of solutions include valence-bond crystals (gapped) and several types of quantum spin liquids (gapped or gapless) \cite{MendelsBert16}. These possible~ground states are very close in energy and pose a challenge for their conclusive~experimental discrimination in real materials because of the limited number of observables that are accessible in the absence of any long-range magnetic order. That's why the celebrated copper mineral herbertsmi\-thite, $\gamma$-ZnCu$_3$(OH)$_6$Cl$_2$, with a number of its structural relatives hosting nearly perfect \mbox{$S=1/2$} kagome layers \cite{Norman16, PuphalZoch18} came into the focus of current research in quantum magnetism. Herbertsmithite was the first kagome-layered compound to feature a perfect equilateral geometry and no order down to $T=0$, while its metastable polymorph kapellasite with ferromagnetic NN interactions was later shown to host a gapless spin-liquid ground state due to the frustrated further-neighbour interactions \cite{FakKermarrec12, BernuLhuillier13}.

Such quantum-disordered ground states are generally fragile. Even in structurally perfect kagome layers, the presence of weak magnetic anisotropy and DMI would promote 120$^\circ$ spin structures characterized by the ``positive'' (PVC), ``negative'' (NVC), or ``staggered'' (SVC) vector chirality in the classical limit \cite{YildirimHarris06, CepasFong08, OkumaYajima17}. These structures are illustrated in Fig.~\ref{Fig:KagomeStruct}. Further, the frustration can be partially relieved by structural distortions \cite{WangVishwanath07, KanekoMisawa10, MasudaOkubo12} or disorder \cite{DommangeMambrini03, LaeuchliDommange07}, precluding a spin-liquid ground state. For instance, a low-temperature ordered state is found in volborthite, Cu$_3$V$_2$O$_7$(OH)$_2$\,$\cdot$\,2H$_2$O \cite{BertBono05, YoshidaTakigawa09, YoshidaTakigawa12}, and vesignieite, BaCu$_3$V$_2$O$_8$(OH)$_2$ \cite{OkamotoYoshida09}, which has been explained by the non-equivalent exchange on the triangles forming a distorted kagome lattice as a result of orbital ordering \cite{OkamotoIshikawa12, BoldrinKnight16}. In this chapter, we discuss the specific behaviour of various minerals realizing different magnetic models on the kagome lattice. While most of the discussed compounds are found in nature and became known first as minerals, the majority of the experimental results discussed here were obtained on synthetic single crystals. This is explained by the extreme sensitivity of the observed magnetic phenomena to impurities or crystalline defects that are impossible to control in natural mineral samples, which can be in addition extremely rare or small in size.\vspace{-1pt}

\subsection{Jarosites: classical spins on the kagome lattice}\label{Sec:Jarosites}

\begin{figure}[t!]
\begin{center}
\includegraphics[width=0.5\linewidth]{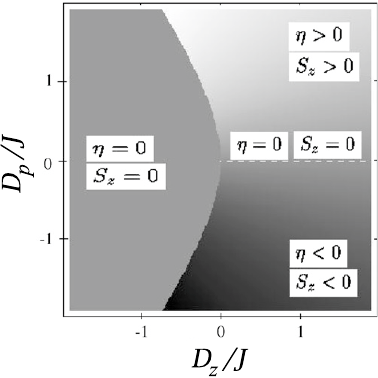}\vspace{-2pt}
\end{center}
\caption{The mean-field phase diagram of the KAFM model with a NN Heisenberg exchange $J$ and a DMI consisting of the in-plane ($D_p$) and out-of-plane ($D_z$) components. The grey scale represents the angle $\eta$ between the magnetic moments and the kagome plane. After Elhajal \textit{et~al.}~\cite{ElhajalCanals02} and Ballou \textit{et~al.}~\cite{BallouCanals03}.\vspace{-3pt}}\label{Fig:Jarosites}
\end{figure}

\begin{figure}[t]
\includegraphics[width=\textwidth]{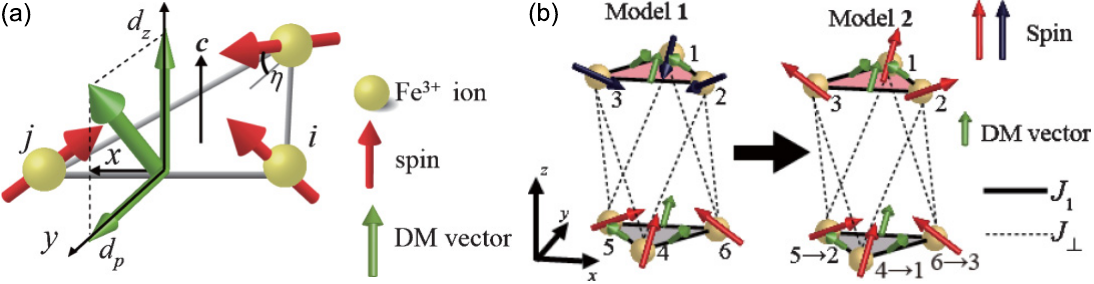}
\caption{(a)~The DMI vectors and the canted spin structure of a K-Fe-jarosite layer. (b)~The change in the magnetic structure associated with the spin-flip transition of the weak FM component of the magnetic moments. After Fujita \textit{et~al.}~\cite{FujitaYamaguchi12}.}\label{Fig:JarositeStructure}
\end{figure}

The jarosite family of minerals with the general formula $A$Fe$_3$(SO$_4$)$_2$(OH)$_6$ (with $A$ being an alkali metal) are model kagome antiferromagnets with the classical spin $S=5/2$ \cite{MendelsWills11}. Their Cr and V analogues with $S=3/2$ and $S=1$, respectively, are also well known \cite{TownsendLongworth86, NishiyamaMorimoto01, Wills01, KatoaHoria03, OkutaHara11, OkuboNakata17}. In the naturally occurring iron jarosites, long-range N\'eel magnetic order is induced by the DMI that is allowed by the low symmetry of the kagome lattice, according to the classical Monte-Carlo simulations \cite{ElhajalCanals02, BallouCanals03, GroholNocera03}. These interactions, in turn, consist of two components: $D_z$ that is orthogonal to the kagome plane and $D_p$ in the kagome plane, perpen\-dicular to the NN bonds. Their ratios to the AFM exchange coupling $J$ between the NN spins, $D_z/J$ and $D_p/J$, parameterize the magnetic phase diagram of jarosites (Fig.\,\ref{Fig:Jarosites}). For $D_p=0$, the coplanar NVC and PVC magnetic structures are stabilized for \mbox{$D_z<0$} and \mbox{$D_z>0$}, respectively. For finite values of $D_p$, the NVC structure remains coplanar, whereas the PVC structure develops an out-of-plane magnetization component and becomes weakly ferromagnetic within each layer \cite{BallouCanals03}, with the angle $\eta$ between the moments and the kagome plane given by $\tan(2\eta)=2D_p/(\sqrt3J+D_z)$, as illustrated in Fig.~\ref{Fig:JarositeStructure}\,(a). The canted components of the magnetic moments are then stacked antiparallel to give a zero net moment in the bulk magnetic structure \cite{NishiyamaMorimoto01, BuurmaHandayani12}. According to recent calculations, this noncoplanar magnetic order supports topological magnon bands that should give rise to a finite magnon thermal Hall conductivity that can be tuned continuously by magnetic field \cite{LaurellFiete18}.

On the experimental front, ``triangular-spin AFM ordering'' was first observed in a synthetic sample of K-Fe-jarosite back in the 1980's by powder neutron diffraction, M\"ossbauer, and magnetic susceptibility measurements \cite{TownsendLongworth86}. However, the exact magnetic structure, and in particular the way in which magnetic layers are stacked in the direction perpendicular to the kagome planes remained a question of debate and was only settled much later \cite{InamiMaegawa98, InamiNishiyama00}. It is now established that the 120$^\circ$ ordering pattern of the PVC type is realized below $T_{\rm N}\approx65$\,K, while the estimated Curie-Weiss temperature is $\Theta_{\rm CW}\approx-800$\,K \cite{InamiNishiyama00, GroholMatan05, YildirimHarris06}. Both temperatures are practically insensitive to the choice of the $A$ cation \cite{GroholNocera03}. Along the $c$ axis, the magnetic unit cell is doubled, suggesting an AFM coupling between the planes. These results are consistent with NMR measurements, which additionally provided an estimate of the spin-wave energy gap of 15~K (1.3~meV) \cite{NishiyamaMaegawa03}. As expected for the PVC structure, a finite field-dependent out-of-plane magnetization has been measured, resulting in the weakly noncoplanar umbellate spin structure \cite{GroholMatan05}. At high magnetic fields above 16.4~T, the weak ferromagnetic spin component flips, as shown in Fig.~\ref{Fig:JarositeStructure}\,(b), causing an abrupt step in the $\mathbf{B}\parallel\mathbf{c}$ magnetization \cite{MatanBartlett11, FujitaYamaguchi12}.

\begin{figure}[t]
\begin{center}
\includegraphics[width=0.65\textwidth]{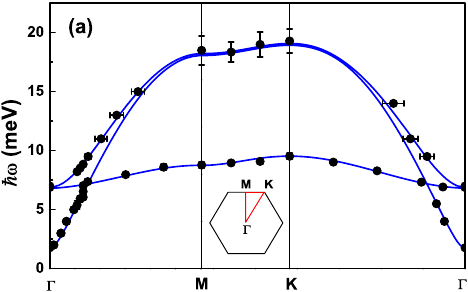}
\end{center}
\caption{Spin-wave dispersion in potassium jarosite along the high-symmetry directions in the 2D Brillouin zone at \mbox{$T=10$\,K}. The data points from single-crystal INS measurements are fitted to the 4-parameter DMI model (solid lines). After Matan \textit{et~al.}~\cite{MatanGrohol06}.}\label{Fig:JarositeDispersion}
\end{figure}

The quantitative understanding of magnetic interactions in jarosite was enabled by the INS measurements performed independently by two groups in 2006. First, Matan \textit{et al.} reported the spin-wave spectrum measured on a deuterated powder sample and a smaller (100~mg) nondeuterated synthetic single crystal of KFe$_3$(SO$_4$)$_2$(OH)$_6$ \cite{MatanGrohol06}. In the powder data, they found a peak in the magnon density of states at $\hslash\omega_0\approx8$~meV and a second broader peak at about $2\omega_0$. Details of the spin-wave dispersion were then verified from the single-crystal measurement and could be perfectly described using a generic Hamiltonian with NN and NNN exchange terms, $J_1=3.18$~meV and $J_2=0.11$~meV, and the DMI parameters $D_z=-0.196$~meV and $|D_p|=0.197$~meV as presented in Fig.~\ref{Fig:JarositeDispersion}. An alternative model, which attributes all of the anisotropy to the single-ion crystal field (CF model) \cite{NishiyamaMaegawa03}, resulted in an equally good fit of the magnon dispersion along all principal directions. The strong effect of the spin canting angle due to the umbellate structure on the spin gap and the splitting of spin-wave branches at high-symmetry points enabled an accurate fit of its value to \mbox{$\eta=1.9^\circ$} within the DMI model. The weak NNN in\-teraction $J_2$ was shown to be positive and responsible for the observed dispersion of the ``flat'' magnon modes. At the zone centre, two spin gaps with energies \mbox{$\Delta_1=1.8$~meV} and \mbox{$\Delta_2=6.7$~meV} were observed, in reasonable agreement with the previous NMR result \cite{NishiyamaMaegawa03} and with the subsequent ESR measurements~\cite{FujitaYamaguchi12}. Motivated by these results, a much more detailed theoretical analysis that compared the spin-wave spectrum measured by Matan~\textit{et~al.} with numerical calculations for a large number of model Hamiltonians with increasing complexity has been carried out by Yildirim and Harris \cite{YildirimHarris06}. They arrived at the slightly modified values of Hamiltonian parameters ($J_1=3.225$, $J_2=0.11$, $D_z=-0.195$, and $|D_p|=0.218$~meV) and argued in favour of the DMI model as opposed to the CF model, as one would not expect the single-ion anisotropy $D$ for the Fe$^{3+}$ ion to be as high as 10\% of $J_1$.

Shortly afterwards, another set of INS measurements on the K-Fe-jarosite has been published by Coomer \textit{et al.} \cite{CoomerHarrison06}. In this work, both a natural single crystal and a synthetic deuterated powder sample have been used. The dispersion of the ``flat'' magnon mode in the natural mineral sample from Eureka (Utah, USA) has been identified using the high-flux IN8 triple-axis spectrometer at the Institut Laue--Langevin (Grenoble, France), which is remarkable in view of the very small sample mass of only 14.9~mg. Unfortunately, the small crystal size and the high background from hydrogen in the sample impeded a more complete mapping of other spin-wave excitations. Nevertheless, comparison with the powder data resulted in the interaction parameters that are very close to those listed above, and the results could be also described equally well by the DMI and CF models.

More recent works focused on the origin of spin anisotropy in K-Fe-jarosite, in order to distinguish the effects of single-ion anisotropy from those of DMI. The single-ion anisotropy energy could be first accurately measured using x-ray absorption spectroscopy \cite{VriesJohal09}, resulting in a value of $D=0.5$~meV that is close to those obtained by Matan \textit{et al.} (0.428~meV) and Coomer \textit{et al.} (0.47~meV) within the CF model of Ref.~\cite{NishiyamaMaegawa03}. This important result implies an unusually large easy-plane anisotropy of the Fe$^{3+}$ spins, up to 15\% of $J_1$. Further evidence arrived from a polarized neutron-scattering study \cite{MatanHelton09}, which revealed an out-of-plane gap that is significantly larger than the in-plane gap, and the persistence of the out-of-plane gap above $T_{\rm N}$. Therefore, the quasielastic scattering above $T_{\rm N}$ consists of in-plane-only spin fluctuations that are characteristic of an \textit{X\hspace{0.5pt}Y}-type spin anisotropy. According to these results, K-Fe-jarosite should belong to the 2D \textit{X\hspace{0.5pt}Y} universality class of magnetic systems. The debate continues, as the most recent high-field ESR results could not be reconciled with the CF model because it failed to fit the data at high frequencies \cite{FujitaYamaguchi12}. The conclusion was that the DMI should still be considered as the dominant perturbation term, and to explain the spin-reorientation transition, an additional interplanar coupling term $J_\perp\approx8.3\cdot10^{-3}$~meV is required. At the same time, some difficulties of the DMI model have also been emphasized.

Finally, magnetization and neutron-diffraction measurements have indicated the possible existence of a multicritical point in the magnetic~field\,--\,temperature phase diagram of iron jarosites on the spin-reorientation transition line \cite{MatanBartlett11}. In the recent work by Freitas and Albuquerque \cite{FreitasAlbuquerque15}, a possible interpretation of these results has been proposed. They investigated the $B\kern1pt$--${\kern1pt}T$ phase diagram for the \textit{X\hspace{0.5pt}Y} model with DMI using the mean-field theory and suggested the existence of an additional high-field magnetic phase with a nonhelical (collinear) spin structure, forming a narrow wedge between the low- and high-field canted PVC phases. This new phase, which still awaits a direct experimental confirmation, would persist only at low temperatures and may therefore lead to the \mbox{appearance of a tricritical point}.

In contrast to the alkali-metal jarosites discussed above, synthetic samples of hydronium jarosite (H$_3$O)Fe$_3$(SO$_4$)$_2$(OH)$_6$ \cite{WillsHarrison96} show a markedly different behaviour. Instead of the long-ranged N\'eel state, this compound displays an unconventional spin glass transition at $T_{\rm g}\approx17$~K \cite{WillsHarrison98, WillsDupuis00, BissonWills08}. Neutron diffraction indicated the presence of only short-range magnetic correlations \cite{WillsHarrison98}, and critical slowing down of spin fluctuations was evidenced by the frequency dependence of the ac susceptibility peak \cite{WillsDupuis00}. On the one hand, the quadratic temperature dependence of the specific heat is indicative of Goldstone modes in a 2D antiferromagnet, in contrast to a linear dependence of conventional spin glasses \cite{WillsDupuis00}. On the other hand, the spin-spin correlation length determined from neutron scattering is only $\sim$\,19~\AA~\cite{WillsHarrison98}. These observations obtained an elegant explanation in the ``kagome spin glass'' model \cite{BissonWills08}. This unconventional form of spin glass with non-Abelian properties is facilitated by the planar spin anisotropy and has an emergent symmetry among the partially disordered spin configurations that preserves the low-energy Goldstone modes \cite{RitcheyChandra93}.\vspace{-2pt}

\subsection{Herbertsmithite and related $S=1/2$ kagome-layer antiferromagnets}

Herbertsmithite, $\gamma$-ZnCu$_3$(OH)$_6$Cl$_2$, is a copper-hydroxychloride mineral \cite{BraithwaiteMereiter04} that became famous in the physics community as a realization of the structurally perfect spin-$\frac{1}{2}$ kagome antiferromagnet \cite{ShoresNytko05} and a likely candidate of a quantum spin liquid \cite{Norman16}. It derives from the mineral paratacamite, a rhombohedral polymorph of Cu$_2$(OH)$_3$Cl discussed in section \ref{Sec:Atacamite}, in which one quarter of the Cu atoms has been substituted by Zn to stabilize the rhombohedral herbertsmithite structure (space group $R\overline{3}m$) consisting of nearly decoupled undistorted kagome planes. Although in the perfect herbertsmithite lattice zinc should go solely onto the copper intersite positions, real samples available for magnetic studies vary both in the number of zinc ions sitting on the copper kagome sites and in the zinc deficiency on the intersites~\cite{Norman16}. Synthetic single crystals that became recently available from the hydrothermal growth method \cite{ChuMueller11, HanHelton11} may thus show deviations from the perfectly stoichiometric compositions, which are difficult to reveal using standard x-ray diffraction.

The most remarkable property of herbertsmithite is the absence of any long-range magnetic order down to the lowest measured temperatures, despite the large negative Curie-Weiss temperature of $\Theta_{\rm CW}\approx-314$\,K \cite{ShoresNytko05} and the exchange energy scale of $J\approx170$\,--\,200~K \cite{RigolSingh07, MisguichSindzingre07, MendelsBert07, HeltonMatan07, ImaiNytko08, ZorkoNellutla08, HanChu12, JeschkeSalvat13}. This conclusion has been consistently confirmed down to millikelvin temperatures by $\mu$SR \cite{MendelsBert07} and neutron scattering \cite{HeltonMatan07, NilsendeVries13}, whereas measurements at elevated pressures found a pressure-driven phase transition from the quantum-disordered spin-liquid phase to a long-range ordered AFM phase with $T_{\rm N}\approx6$~K at 2.5~GPa \cite{KozlenkoKusmartseva12}. The absence of magnetic order at ambient pressures stimulated a considerable experimental effort and raised intense theoretical discussions regarding the true ground state of herbertsmithite\,---\,whether it is gapped or gapless \cite{ShaginyanAmusia18}, and whether the singlet valence bonds form a solid (valence bond solid, VBS) \cite{NikolicSenthil03, HermeleRan08} or fluctuate (resonating valence bonds, RVB) \cite{Anderson73, MambriniaMila00}. For a long time, inadvertent magnetic disorder has hindered the detection of a spin gap at low temperatures until it was finally observed in recent NMR measurements, evidenced by a sharp drop in the nuclear magnetic relaxation rate, $1/T_1$, and in the $^{17}$O NMR frequency shift \cite{FuImai15, ShermanImai16}. The resulting value of the spin gap, $\Delta\approx10$~K, favours theories promoting a $\mathbb{Z}_2$ (topological) quantum spin-liquid ground state with a spin-gap of order $0.1J$ \cite{JiangWeng08, YanHuse11, DepenbrockMcCulloch12, JiangWang12, MessioBernu12, NishimotoShibata13, JuBalents13, DoddsBhattacharjee13, IqbalPoilblanc15}, supplanting alternative scenarios like valence-bond solids or gapless (algebraic) spin liquids \cite{MarstonZeng91, SinghHuse92, ElstnerYoung94, LecheminantBernu97, WaldtmannEverts98, Mila98, SyromyatnikovMaleyev02, NikolicSenthil03, BudnikAuerbach04, WangVishwanath06, RanHermele07, SinghHuse07, NakanoSakai11, IqbalBecca13, IqbalPoilblanc14}.

\begin{figure}[t]
\begin{center}
\includegraphics[width=0.58\textwidth]{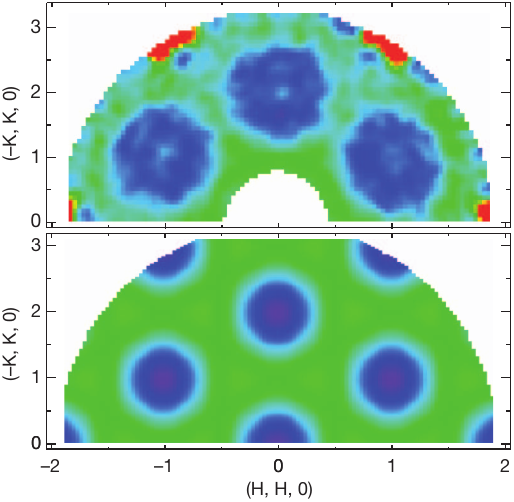}
\end{center}
\caption{The continuum of fractionalized $S=1/2$ excitations in herbertsmithite. The top panel shows INS data in the $(HK0)$ scattering plane, measured at $T=1.5$~K at an energy transfer of 5~meV on a synthetic deuterated single crystal of $\gamma$-ZnCu$_3$(OD)$_6$Cl$_2$. The bottom panel shows the calculated magnetic structure factor, $S_{\rm mag}(Q)$, for a simple model of uncorrelated nearest-neighbour dimers, similar to a short-range RVB state. After Han \textit{et~al.}~\cite{HanHelton12}.}\label{Fig:Herbertsmithite}
\end{figure}

The momentum-space structure of spin correlations in herbertsmithite was investigated by Han \textit{et al.} using neutron scattering \cite{HanHelton12}. The experiment on a synthetic deuterated single-crystal sample of $\gamma$-ZnCu$_3$(OD)$_6$Cl$_2$ revealed an exceedingly diffuse signal spanning a large fraction of the Brillouin zone at low energies (Fig.~\ref{Fig:Herbertsmithite}). It has been associated with a continuum of fractionalized $S=1/2$ excitations on the kagome lattice, similar to spinons, in strong contrast to sharp spin-wave excitations in nonfrustrated magnets. Figure~\ref{Fig:Herbertsmithite} compares these INS data with the results of a model calculation for a collection of uncorrelated nearest-neighbour singlets on the kagome lattice. This model qualitatively captures the magnetic spectral-weight distribution but overestimates the width of peaks in reciprocal space, indicating that spin-spin correlations beyond the nearest neighbours also play an important role in herbertsmithite \cite{HanHelton12}. In follow-up theoretical works, the influence of additional Hamiltonian parameters, such as Ising and DMI terms \cite{DoddsBhattacharjee13, MessioBieri17}, as well as second-neighbour AFM Heisenberg interactions \cite{DoddsBhattacharjee13}, on the magnetic structure factor, $S_{\rm mag}(Q)$, was also considered.

\begin{figure}[t!]
\begin{center}
\includegraphics[width=0.67\textwidth]{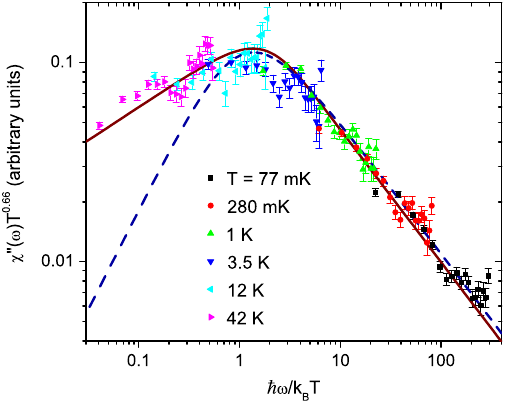}
\end{center}
\caption{The universal $\omega/T$ scaling of the dynamic spin susceptibility in herbertsmithite. The measured values of $\chi''(\omega)T^\alpha$ with $\alpha\approx2/3$, plotted against $\hslash\omega/k_{\rm B}T$ on a log-log scale, collapse onto a single curve. After Helton \textit{et~al.}~\cite{HeltonMatan10}.}\label{Fig:HerbertsmithiteScaling}
\end{figure}

Another important ingredient influencing the dynamical spin structure factor is chemical disorder, which can lead to a randomness-induced quantum spin liquid behaviour (random-singlet state) above a certain critical value \cite{ShimokawaWatanabe15}. In a more recent work, Han \textit{et al.} \cite{HanNorman16} argued that the low-energy magnetic fluctuations also contain a considerable contribution from antiferromagnetically correlated magnetic Cu impurities on the nominally nonmagnetic Zn intersites. According to single-crystal $^2$D NMR \cite{ImaiFu11} and resonant x-ray diffraction measurements \cite{FreedmanHan10}, the concentration of such weakly interacting Cu$^{2+}$ defect spins reaches 15\%. On the other hand, the possibility of Zn$^{2+}$ impurities occupying the kagome sites can be essentially eliminated in the best of the studied samples. The authors of Ref.~\cite{HanNorman16} indicate a good agreement between predictions of their impurity model and the low-energy neutron data that reveal a small spin gap of $\sim$\,0.7~meV after subtraction of the impurity signal. The same model is also able to explain the divergent response in the dynamic spin susceptibility below 1~meV \cite{HeltonMatan10} and is consistent with the NMR and specific-heat data \cite{HeltonMatan07}.

The bulk ac susceptibility and the dynamic susceptibility measured by neutron spectroscopy show a universal scaling relationship in herbertsmithite that is typical for systems with disorder or in proximity to a QCP \cite{HeltonMatan10}. In Fig.~\ref{Fig:HerbertsmithiteScaling}, the dynamic spin susceptibility extracted from INS data, $\chi''(\omega,T)$, is plotted for various temperatures. The quantity $\chi''(\omega,T)T^\alpha$ with $\alpha\approx2/3$ collapses on a single curve as a function of $\hslash\omega/k_{\rm B}T$. It can be well described by the same functional form $F(\omega/T)=(T/\omega)^\alpha\tanh(\omega/\beta T)$ as the one used to fit the dynamic susceptibility in La$_{1.96}$Sr$_{0.04}$CuO$_4$ \cite{KeimerBirgeneau91} and in some heavy-fermion metals such as Ce(Rh$_{0.8}$Pd$_{0.2}$)Sb \cite{ParkMcEwen02} and UCu$_{5-x}$Pd$_x$ \cite{AronsonOsborn95} (solid line). On the other hand, an alternative scaling that was successful in the case of CeCu$_{5.9}$Au$_{0.1}$ \cite{SchroederAeppli00, SchroederAeppli98} (dashed line) fails to describe the data. The observed scaling behaviour in herbertsmithite could be attributed to a critical spin-liquid ground state \cite{Sachdev08} but could also be related to disorder. It was pointed out \cite{HeltonMatan10} that it shares common features with disordered heavy-fermion metals and random magnetic systems featuring a Griffiths phase \cite{Griffiths69, CastroNetoCastilla98} or a random-singlet phase \cite{BhattLee82}. The randomness among valence-bond singlets formed by neighbouring Cu spins in the spin-liquid regime was also claimed responsible for the recently observed broadening of an infrared phonon \cite{SushkovJenkins17}, as expected for a VBS state \cite{HermeleRan08}.

The situation with understanding the low-temperature magnetism in herbertsmithite is further complicated by the recent results of Zorko \textit{et al.} \cite{ZorkoHerak17} who observed two distinct types of defects with different magnetic couplings to the kagome spins in their ESR data. This magnetic response contradicts the threefold symmetry of the ideal kagome lattice, suggesting a minor deviation from the perfect symmetry due to a structural distortion at low temperatures. To what extent this structural distortion is governed by the presence of impurities, and what is its role in stabilizing the spin-liquid ground state, remains to be seen.

Apart from the most studied herbertsmithite, structurally perfect $S=1/2$ KAFM lattices are also realized in its isostructural Mg, Ni, and Co analogues tondiite, $\gamma$-MgCu$_3$(OH)$_6$Cl$_2$ \cite{ColmanSinclair11, MalcherekBindi14}, gillardite, $\gamma$-NiCu$_3$(OH)$_6$Cl$_2$ \cite{ClissoldLeverett07}, and leverettite, $\gamma$-CoCu$_3$(OH)$_6$Cl$_2$ \cite{KampfSciberras13}. The lattice parameters of tondiite and leverettite are very close to those of herbertsmithite, whereas gillardite has a smaller $c$ lattice constant due to the smaller radius of the Ni$^{2+}$ ion as compared to the Zn$^{2+}$ ion. Even before the discovery of natural tondiite was reported in 2014 \cite{MalcherekBindi14}, the magnetic properties of this compound were investigated on a synthetic powder sample. Initial characterization revealed a Curie-Weiss temperature of $\Theta_{\rm CW}\approx-284$\,K and a weak ferromagnetic ordering below $T_{\rm C}\approx4$\,--\,5~K, which was attributed to impurities~\cite{ColmanSinclair11}. The absence of any magnetic ordering or spin freezing down to 20~mK ($\sim$\,$10^{-4}J$) was demonstrated with $\mu$SR \cite{KermarrecMendels11}. In the same experiment, the low-temperature relaxation rate exhibited a power-law scaling, $T_1 \propto \omega^\alpha$ with $\alpha\approx0.63$, pointing to an exotic relaxation channel for intersite defects. This power law can be reconciled with the above-mentioned critical $\omega/T$ scaling with $\alpha=2/3$ in the neutron data on herbertsmithite, shown in Fig.~\ref{Fig:HerbertsmithiteScaling}~\cite{HeltonMatan10}.

The magnetic properties of $\gamma$-NiCu$_3$(OH)$_6$Cl$_2$ (gillardite) and $\gamma$-CoCu$_3$(OH)$_6$Cl$_2$ (leverettite) have been probed by magnetization and susceptibility measurements \cite{LiZhang13}. In contrast to herbertsmithite, the interlayer sites in these compounds are occupied by magnetic Ni$^{2+}$ ($S=1$, $\mu_{\rm eff}=3.5\mu_{\rm B}$) and Co$^{2+}$ ($S=3/2$, $\mu_{\rm eff}=4.9\mu_{\rm B}$) ions, respectively. This opens up new magnetic interaction paths between kagome layers that still await a systematic exploration. A much stronger interlayer coupling is evidenced by the reduced Curie-Weiss temperatures ($\Theta_{\rm CW}=-100$~K and $-40$~K, respectively), as compared to herbertsmithite, and is expected to alter the low-temperature magnetic behaviour substantially in both compounds~\cite{LiZhang13}.

\subsection{Kapellasite vs. haydeeite: the role of further-neighbour exchange}

\begin{figure}[t]
\begin{center}
\includegraphics[width=0.65\textwidth]{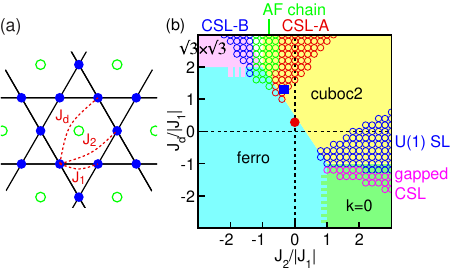}
\end{center}
\caption{(a)~Main exchange interactions on the Cu$^{2+}$ kagome lattice of kapellasite and hay\-dee\-ite. (b)~Estimated location of kapellasite (square) and haydeeite (circle) on a classical phase diagram of the kagome lattice with FM $J_1$ interactions. After Boldrin \textit{et~al.}~\cite{BoldrinFak15}.}\label{Fig:KapellasiteHaydeeite}
\end{figure}

The mineral kapellasite is a metastable polymorph of herbertsmithite, in which Zn ions occupy positions in the kagome layers rather than at the intersites \cite{ColmanRitter08, ColmanSinclair10, Norman16}. It is described by the space group $P\overline{3}m1$ and is commonly referred to\footnote{Alternative naming for polymorphs based on the chronological order of their discovery can also be found in the literature \cite{NilsenSimonet17}, but the one adopted here is by far the most common \cite{ColmanSinclair10, ColmanSinclair11, KermarrecZorko14, BoldrinFak15, PuphalBolte17}.} as $\alpha$-ZnCu$_3$(OH)$_6$Cl$_2$. Its isostructural Mg- and Mn-rich analogues haydeeite, $\alpha$-MgCu$_3$(OH)$_6$Cl$_2$ \cite{SchluterMalcherek07}, and misakiite, $\alpha$-MnCu$_3$(OH)$_6$Cl$_2$ \cite{NishioHamane17}, are also known. In the first comparative study of kapellasite and haydeeite, Janson \textit{et al.} \cite{JansonRichter08} employed electronic-structure and exact-diagonalization calculations to argue that further-neighbour exchange interactions across the kagome hexagons, shown in Fig.~\ref{Fig:KapellasiteHaydeeite}\,(a), may play a crucial role in determining the magnetic ground states of both minerals. The corresponding exchange constants, $J_{\rm d}\approx0.9$~meV for kapellasite vs. 0.8~meV for haydeeite, are nearly the same. However, the nearest-neighbour AFM interaction $J_1$ along the sides of the kagome triangles is 3 times stronger in kapellasite than in haydeeite. At the same time, the second-nearest-neighbour coupling $J_2$ is negligible. This implies that the ground state is solely determined by the ratio $\alpha=J_{\rm d}/J_1$ that is very different for the two compounds: $\alpha\approx0.36$ for kapellasite and $\alpha\approx1$ for haydeeite. The diagonal coupling lifts the degeneracy of the pure KAFM ground state ($\alpha=0$) in favour of a complex noncoplanar ``cuboc'' order with 12 magnetic sublattices similar to the one proposed for the $J_1$-$J_2$ kagome Heisenberg model \cite{DomengeSindzingre05}. For strong $J_{\rm d}$, magnetic correlations along the chains built by diagonal bonds dominate, suggesting that low-energy spin dynamics in haydeeite may be represented by fractionalized spinon excitations as in 1D spin-chain models.

Experimental results available to date suggest that kapellasite develops no long-range order but rather short-range magnetic correlations with the noncollinear ``cuboc2'' structure \cite{FakKermarrec12}, whereas haydeeite orders ferromagnetically at $T_{\rm C}=4.2$~K \cite{ColmanSinclair10}. The importance of the diagonal AFM interaction $J_{\rm d}\approx1.3$~meV was confirmed by the analysis of susceptibility and specific-heat data \cite{FakKermarrec12, BernuLhuillier13}. However, in contrast to the theoretical prediction of Janson \textit{et al.} \cite{JansonRichter08}, the NN exchange constant was found to be ferromagnetic, $J_1\approx-1$~meV for kapellasite \cite{FakKermarrec12, BernuLhuillier13, KermarrecZorko14} and $J_1\approx-3.3$~meV for haydeeite \cite{BoldrinFak15}. Figure~\ref{Fig:KapellasiteHaydeeite}\,(b) shows the phase diagram of classical ground states for a Heisenberg model on the kagome lattice with $J_1<0$ and two further-neighbour couplings $J_2$ and $J_{\rm d}$ \cite{FakKermarrec12, BieriMessio15, BoldrinFak15, IqbalJeschke15, BieriLhuillier16}. According to the exchange constants estimated from neutron-spectroscopy data, kapellasite and haydeeite are located in the ``cuboc2'' and ferromagnetic phases, respectively, very close to the transition line that separates them \cite{BoldrinFak15}. This result is consistent with direct experimental measurements of their magnetic ground states.

\begin{figure}[t!]
\begin{center}
\includegraphics[width=\textwidth]{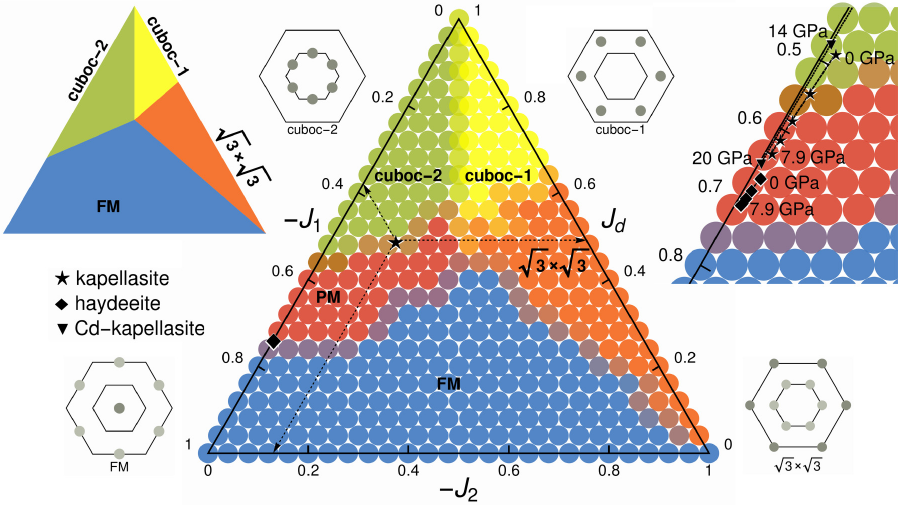}
\end{center}
\caption{Quantum phase diagram of the $J_1$-$J_2$-$J_{\rm d}$ Heisenberg model on a kagome lattice. The classical phase diagram is shown at the left for comparison. Black symbols mark the experimentally obtained exchange couplings for kapellasite and haydeeite \cite{FakKermarrec12, BernuLhuillier13, BoldrinFak15}. The dependence of coupling constants in different materials on pressure is shown in the enlarged region on the right. After Iqbal \textit{et~al.}~\cite{IqbalJeschke15}.}\label{Fig:QuantumKapellasite}
\end{figure}

A more recent theoretical work by Iqbal \textit{et al.} \cite{IqbalJeschke15} extended these results to investigate possible regions in the phase diagram traversed by $\alpha$-(Zn,Mg,Cd)Cu$_3$(OH)$_6$Cl$_2$ compounds as a function of pressure, based on \textit{ab initio} DFT calculations. In the quantum phase diagram shown in Fig.~\ref{Fig:QuantumKapellasite}, they found an extended region of a paramagnetic domain for intermediate values of $J_{\rm d}$ that separates the FM and ``cuboc2'' phases. It has been predicted that kapellasite, located at the onset of the ``cuboc2'' phase, would enter this region under elevated hydrostatic pressure. An even stronger pressure dependence covering a vast extent along the $J_{\rm d}$ axis was predicted for the then unknown Cd-kapellasite, $\alpha$-CdCu$_3$(OH)$_6$Cl$_2$~\cite{IqbalJeschke15}. Meanwhile, a very similar compound with this structure, CdCu$_3$(OH)$_6$(NO$_3$)$_2$\,$\cdot$\,H$_2$O \cite{Oswald69}, also termed Cd-kapellasite, has been grown in single-crystal form by the hydrothermal transport method~\cite{OkumaYajima17}. At ambient pressure, it develops an NVC spin order [as shown in Fig.~\ref{Fig:KagomeStruct}\,(b)] below $T_{\rm N}\approx4$\,K, accompanied by a small spontaneous magnetization. Its origin is related to the DMI, which is considered negligible (with $D/J_1\approx3$\% \cite{KermarrecZorko14}) for kapellasite and haydeeite. Another important difference lies in the NN exchange constant that was estimated to be antiferromagnetic, $J_1\approx3.9$\,meV, whereas further-neighbour interactions $J_2$ and $J_{\rm d}$ could not yet be reliably determined from the existing data.

Finally, it is also worthwhile to mention the effect of disorder on the magnetic properties of kapellasite. In a study that employed a combination of $^{35}$Cl NMR, ESR, and $\mu$SR, Kermarrec \textit{et al.}~\cite{KermarrecZorko14} revealed a severe random depletion of the magnetic kagome lattice by 27\% due to Cu/Zn site intermixing even in high-quality synthetic powder samples. Their surprising finding was that the high-temperature magnetic response of the system remains homogeneous in spite of the structural disorder. The variety of local magnetic environments leads to an appearance of local low-energy excitation modes that broaden the distribution of spin-lattice relaxation times at low temperatures. Due to the metastability of the $\alpha$-ZnCu$_3$(OH)$_6$Cl$_2$ polymorph and the proximity of ionic radii of Zn$^{2+}$ and Cu$^{2+}$ ions, it remains a real chemical challenge to decrease the amount of structural disorder in the kagome layers and thereby improve the quality of synthetic samples.

\begin{figure}[t!]
\begin{center}
\includegraphics[width=0.6\linewidth]{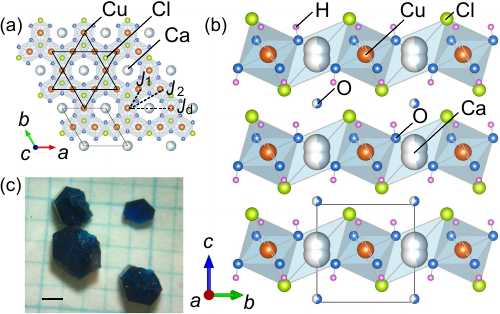}
\end{center}
\caption{Crystal structure and large synthetic single crystals of centennialite (shown on mm paper). After Yoshida \textit{et~al.}~\cite{YoshidaNoguchi17}.}\label{Fig:Centennialite}
\end{figure}

One possible way to circumvent this problem is to replace Zn$^{2+}$ with another nonmagnetic cation. Such a possibility is offered by the newly discovered mineral centennialite of the atacamite group, CaCu$_3$Cl$_2$(OH)$_6$\,$\cdot$\,0.7H$_2$O \cite{CrichtonMueller17}, which is also structurally related to kapellasite and haydeeite but is much less studied. In synthetic samples synthesized by a solid-state reaction method \cite{SunHuang16}, the intersite mixing between Ca and Cu atoms was limited to <\,5\%, which is a significant improvement in comparison to Zn. According to susceptibility data, centennialite is characterized by a Curie-Weiss temperature of $\Theta_{\rm CW}\approx-56$\,K and orders antiferromagnetically below $T_{\rm N}\approx7$\,K, implying a moderate frustration parameter of $|\Theta_{\rm CW}|/T_{\rm N}\approx8$ \cite{SunHuang16}. The successful growth of mm-sized single crystals (Fig.~\ref{Fig:Centennialite}) by a hydrothermal reaction in a two-zone tube furnace \cite{YoshidaNoguchi17} enabled an estimation of the exchange constants $J_1\approx4.5$, $J_2\approx-0.6$, and $J_{\rm d}\approx1.0$~meV from the high-temperature series expansion fitting of the magnetic susceptibility data. \textit{Ab initio} DFT calculations predict that $J_1$ should change sign at a bond angle of 108.5$^\circ$ in the kapellasite family \cite{IqbalJeschke15}, and therefore the AFM character of the NN exchange constant in centennialite is consistent with the large Cu--O--Cu bond angle of 113.9$^\circ$ that is influenced by the ionic radius of the nonmagnetic cation. The estimated parameters place this compound near the multi-critical point of the $\sqrt{3}\times\sqrt{3}$, ``$q=0$'', and ``cuboc1'' phases in the magnetic phase diagram for an isotropic classical Heisenberg model with $J_1>0$ \cite{BernuLhuillier13, BieriLhuillier16}. This implies that novel magnetic ground states, distinct from those found in other atacamite minerals, can be realized in this new kagome-lattice compound with practically no disorder.

An alternative possibility to avoid disorder in the $S=1/2$ kagome lattice is offered by the telluride mineral quetzalcoatlite, Zn$_6$Cu$_3$(TeO$_6$)$_2$(OH)$_6$\,$\cdot$\,Ag$_x$Pb$_y$Cl$_{x+2y}$ \cite{BurnsPluth00}. As noticed by Norman \cite{Norman18}, its crystal structure contains perfect kagome layers with an AA stacking, like in kapellasite (as opposed to ABC stacking in herbertsmithite). The tetrahedral coordination of the Zn ions within the ZnO$_2$(OH)$_2$ units, located on interlayer positions, should prevent disorder resulting from the Cu ions going to the intersites. One should therefore expect considerable interest to this mineral from both chemists and solid-state physicists.

\subsection[Volborthite and vesignieite: Nonequivalent bonds resulting from lattice distortions]{Volborthite, vesignieite, and alike: Nonequivalent bonds resulting from lattice distortions}

In some of the $S=1/2$ kagome minerals, frustration is relieved by structural distortions within the kagome plane, which leads to non-equivalent exchange on the triangles. Historically, the mineral volborthite, Cu$_3$V$_2$O$_7$(OH)$_2$\,$\cdot$\,2H$_2$O with monoclinic space group $C2/m$, was among the first candidate compounds for the realization of a $S=1/2$ KAFM model system \cite{HiroiHanawa01, OkuboOhta01, FukayaFudamoto03}. Its monoclinic distortion was initially considered small enough to keep the essential physics. However, it soon became clear that volborthite orders antiferromagnetically at temperatures near 1~K \cite{BertBono05, YoshidaTakigawa09}. In magnetic fields above $B_{\rm s1}\approx4.5$~T, a second ordered phase (phase~II) with a coplanar cycloidal structure forms, persisting to $B_{\rm s2}\approx25.6$~T, where it gives way to a high-field collinear ferrimagnetic phase (phase~III) with a much higher ordering temperature of $\sim$\,26~K \cite{YoshidaOkamoto09, YoshidaTakigawa12}. Later, measurements in pulsed magnetic fields up to 68~T \cite{OkamotoTokunaga11} revealed an additional transition at $B_{\rm s3}\approx47$~T. A cycloidal-spiral ground state similar to the phase~II of volborthite was theoretically predicted for the spatially anisotropic magnetic model on a distorted kagome lattice \cite{SchnyderStarykh08}.

\begin{figure}[t]
\begin{center}\vspace{-1pt}
\includegraphics[width=0.58\linewidth]{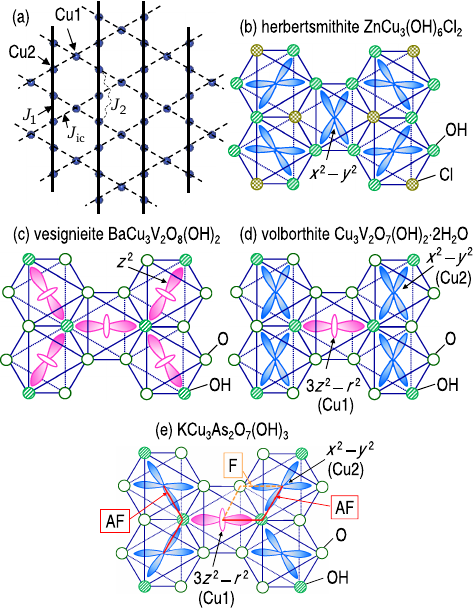}\vspace{-1pt}
\end{center}
\caption{(a)~Exchange interactions on a kagome lattice with an anisotropic distortion. (b--e)~Orbital arrangements in various copper minerals. After Okamoto \textit{et~al.}~\cite{OkamotoIshikawa12}.\vspace{-1pt}}\label{Fig:KagomeOrbitals}
\end{figure}

The first theoretical analysis of the microscopic magnetic model for volborthite by Janson \textit{et~al.} \cite{JansonRichter10} revealed that its physics may be largely different from that of the initially assumed kagome model. An orbital ordering on the Cu$^{2+}$ sublattice leads to a structural distortion and breaks the equivalency between the Cu(1) and Cu(2) sites with the occupied $3d_{z^2-r^2}$ and $3d_{x^2-y^2}$ orbitals, respectively, as illustrated in Fig.~\ref{Fig:KagomeOrbitals}\,(d). The corresponding magnetic model presumably requires three relevant couplings: the nearest-neighbour FM exchange $J_1\approx-6.9$~meV along the base of the isosceles triangles, the strong AFM coupling $J_{\rm ic}\approx8.6$~meV along the two other sides of the triangles, and the NNN exchange interaction $J_2\approx3$~meV, as shown in Fig.~\ref{Fig:KagomeOrbitals}\,(a). In the proposed model, both $J_1$ and $J_{\rm ic}$ with opposite signs supported each other and did not give rise to frustration, hence the only source of frustration came from $J_2$. The resulting model was therefore thought to be much closer to frustrated coupled chains than to KAFM \mbox{lattices}.

\begin{figure}[b]
\begin{center}\vspace{-2pt}
\includegraphics[width=0.51\linewidth]{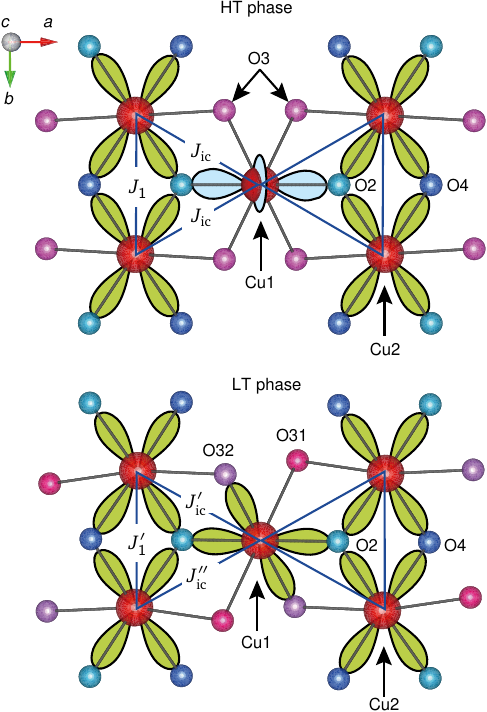}\vspace{-1pt}
\end{center}
\caption{(a)~Orbital switching between the high-$T$ (top) and low-$T$ (bottom) phases of volborthite, breaking the equivalency of the $J_{\rm ic}$ bonds. After Yoshida \textit{et~al.}~\cite{YoshidaYamaura12}.\vspace{-2pt}}\label{Fig:Volborthite}
\end{figure}

Powder neutron-scattering measurements that followed shortly afterwards~\cite{NilsenCoomer11} indicated the presence of short-range magnetic order emerging in zero magnetic field below 5~K at two characteristic wave vectors, $Q_1=0.65$\,\AA$^{-1}$ and $Q_2=1.15$\,\AA$^{-1}$, two steep spin-wave branches dispersing from these wave vectors, and a flat mode at an energy of 5.0~meV. However, the experimental $Q_1$ and $Q_2$ values could not be reconciled with any of the magnetic structures expected for the theoretically predicted $J_1$-$J_2$-$J_{\rm ic}$ model \cite{SchnyderStarykh08, JansonRichter10}.

The situation cleared somewhat when thermodynamic measurements on volborthite single crystals became available. The specific-heat data of Yoshida \textit{et al.} \cite{YoshidaYamaura12} revealed a first-order orbitally driven structural phase transition at $T_{\rm s}\approx300$~K and two successive low-temperature magnetic phase transitions at 1.19~K and 0.81~K instead of a broad kink seen previously in a powder sample~\cite{YamashitaMoriura10}. The phase transition at $T_{\rm s}$ has been associated with an orbital switching \cite{YoshidaYamaura12, SugawaraSugimoto18}, as~the unpaired electron on the Cu(1) site ``switches'' from the $d_{z^2-r^2}$ orbital in the high-temperature phase to the $d_{x^2-y^2}$ orbital in the low-temperature phase. Therefore, at low temperatures, the spin-carrying orbital on both Cu$^{2+}$ sites is $d_{x^2-y^2}$, as in herbertsmithite, but the orientations of the orbitals with respect to each other are different between the two compounds. The consequence of this new orbital configuration for the magnetic model is the broken equivalency between the two $J_{\rm ic}$ bonds, as illustrated in Fig.~\ref{Fig:Volborthite}. In addition, superlattice reflections were observed below $T_{\rm s}$ by x-ray diffraction, indicating a doubling of the unit cell along the $c$ axis \cite{YoshidaYamaura12}. Furthermore, a second high-temperature polymorph with a flipped $d$-orbital orientation on the Cu(1) site and a different $C2/m$ lattice symmetry has been reported \cite{IshikawaYamaura12, SugawaraSugimoto18}. The space group of the low-temperature phase was determined to be $I2/a$, implying that the magnetic model of volborthite had to be modified.

\begin{figure}[t]
\begin{center}
\includegraphics[width=0.54\linewidth]{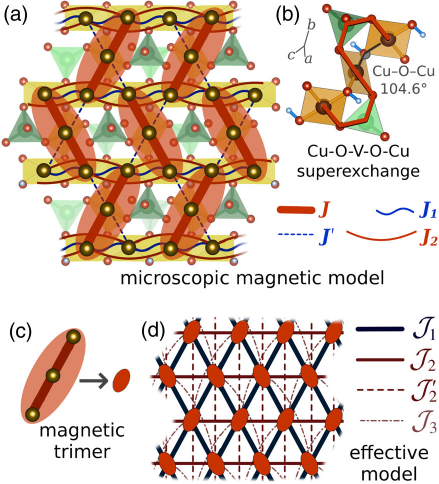}
\end{center}
\caption{(a)~The new microscopic magnetic model of volborthite, featuring magnetic trimers (shaded ovals), formed by the dominant AFM interaction $J$. (b)~The Cu--O--V--O--Cu superexchange paths within the trimer. (c,d)~An effective model of interacting trimers with ferromagnetic $J_1$ and antiferromagnetic $J_2$, $J_2^\prime$, and $J_3$. After Janson \textit{et~al.}~\cite{JansonFurukawa16}.}
\label{Fig:Volborthite_NewModel}
\end{figure}

Soon after the new data on volborthite single crystals became available, Janson \textit{et~al.} \cite{JansonFurukawa16} updated their magnetic model of volborthite in the framework of DFT+$U$, based on the new single-crystal structural data from Ref.~\cite{IshikawaYoshida15}. Their new model involved four leading magnetic interactions: antiferromagnetic $J$ and $J_2$ and ferromagnetic $J^\prime$ and $J_1$, as shown in Fig.\,\ref{Fig:Volborthite_NewModel}\,(a,b). This model results in a system of coupled trimers, contrary to the previously proposed frustrated chains, and it can explain the $\frac{1}{3}$-plateau in the magnetization \cite{IshikawaYoshida15} as a result of polarized ``up-up-down'' states formed on magnetic trimers. Each trimer is connected ferromagnetically to its four nearest neighbours and antiferromagnetically to its two second neighbours, as illustrated in Fig.~\ref{Fig:Volborthite_NewModel}\,(d). In weak magnetic fields below the 1/3 plateau, this model supports a bond-nematic phase due to the condensation of two-magnon bound states.

In the most recent theoretical paper, Chern \textit{et al.} \cite{ChernSchaffer17} considered a similar model with four leading exchange interactions and argued that it could host 12 distinct $\mathbb{Z}_2$ spin-liquid phases with fractionalized excitations. The natural candidate for one such phase would be the cooperative paramagnetic state that exists immediately above the two AFM phase transitions in temperature, where finite thermal Hall conductivity has been revealed recently \cite{WatanabeSugii16}. The theory relates it to the existence of a spinon Fermi surface, characteristic of a $U(1)$ spin liquid due to the effective Lorentz force exerted on the deconfined spinons from the $U(1)$ gauge field coupled to the external magnetic field \cite{KatsuraNagaosa10}.

The understanding of high-field magnetic phases in volborthite also received a second impetus from the successful growth of high-quality single crystals \cite{IshikawaYoshida15}. A wide 1/3 magnetization plateau has been observed in fields between 28 and 74~T and associated with the ferrimagnetic spin structure of phase~III. Below 23~T, $^{51}$V NMR measurements evidenced an incommensurate order within phase~II. But between $\sim$\,23~T and $B_{\rm s2}$, a novel intermediate phase (``phase \textit{N}'') with linear field dependence of the magnetization has been found. Judging from the heavily broadened NMR spectrum within this phase, it is characterized by an inhomogeneous distribution of the internal fields, indicative of disorder among static spin moments. This is likely related to the arrangement of the crystal water molecules between the kagome planes \cite{IshikawaYoshida15}. Finally, the nuclear spin-lattice relaxation rate $1/T_1$, measured by $^{51}$V NMR in the ferrimagnetic phase above 28~T, helped to estimate the excitation gap with a large $g$ factor of $\sim\!5.5\pm0.7$, which the authors associate with a magnon bound state \cite{YoshidaNawa17}. A possible spin-nematic phase at high fields, evidenced by a slowing down of spin fluctuations due to the condensation of magnon bound states, has therefore been proposed.

The ordering patterns among $d_{x^2-y^2}$ and $d_{3z^2-r^2}$ copper orbitals that attracted so much attention to volborthite over the last two decades also play a role in a number of other kagome minerals. For instance, an orbital arrangement similar to that of volborthite but with a different orientation of the $d_{x^2-y^2}$ orbitals on the Cu(2) sites has been proposed for KCu$_3$As$_2$O$_7$(OH)$_3$ \cite{OkamotoIshikawa12}, as illustrated in Fig.~\ref{Fig:KagomeOrbitals}\,(e). This synthetic compound has the same monoclinic space group $C2/m$ as volborthite, implying a structural distortion of the kagome planes, and orders antiferromagnetically below $T_{\rm N}\approx7.2$~K. Here the $J_1$ interaction between the Cu(2) sites was proposed to be AFM, in contrast to volborthite, and would be therefore frustrated for any sign of the $J_{\rm ic}$ interaction between Cu(1) and Cu(2). Low-temperature magnetic diffraction patterns revealed that below $T_{\rm N}$ KCu$_3$As$_2$O$_7$(OD)$_3$ develops a complex incommensurate helical structure that leads to multiferroic properties \cite{NilsenOkamoto14}. A follow-up study of the magnetic phase diagram in applied fields up to 20~T found a metamagnetic transition related to the rotation of the helix plane away from the easy plane around $B_{\rm c}\approx3.7$~T \cite{NilsenSimonet17}.

In the structurally related mineral vesignieite, BaCu$_3$(VO$_4$)$_2$(OH)$_2$ \cite{Guillemin55, MaHe90} (space group $B2/m$), the spins are carried exclusively by the $d_{3y^2-r^2}$ orbitals, as shown in Fig.~\ref{Fig:KagomeOrbitals}\,(c)~\cite{OkamotoIshikawa12}. This reduces the spatial anisotropy in the kagome planes, but in contrast to the highly symmetric trigonal structure of herbertsmithite, the monoclinic distortion still remains finite. A higher-symmetry polymorph of the same compound with structurally perfect kagome layers, termed $\beta$-vesignieite, could also be synthesized \cite{YoshidaMichiue12}, but we focus here on the original $\alpha$-vesignieite that can be found in nature. It has been initially reported to order mag\-ne\-tically and form a ``$\mathbf{q}=0$'' type of spin structure with in-plane spin components oriented at nearly 120$^\circ$ with respect to each other \cite{YoshidaOkamoto13}. Another study proposed instead a more exotic state below 9~K with a coexistence of dynamical and frozen moments \cite{ColmanBert11}. It has been also suggested that the intrinsic ground state of vesignieite could resemble a spin liquid, similar to herbertsmithite \cite{OkamotoYoshida09, ZhangOhta10, QuilliamBert11}. An important difference, however, comes from the stronger DMI \cite{ColmanBert11, QuilliamBert11}. In the $S=1/2$ KAFM model with only one NN Heisenberg exchange $J$, a quantum phase transition to an ordered state has been predicted for the critical value of the DMI, $D_{\rm C}=0.1J$ \cite{CepasFong08}. In herbertsmithite, the estimates fall in the range $0.044 < D/J < 0.08$ \cite{ZorkoNellutla08, ElShawishCepas10}, placing it on the spin-liquid side of the QCP. In vesignieite, however, this ratio was estimated in the uncertainty range $0.1 < D/J < 0.19$ \cite{QuilliamBert11, ZorkoBert13}, placing it just on the other side of the QCP. Based on the ESR line width, Zorko \textit{et al.} \cite{ZorkoBert13} found that the DMI is highly anisotropic with the dominant in-plane component $D_\parallel\approx0.19J$ and smaller out-of-plane component $D_\perp\approx0.07J$, comparable to herbertsmithite. The large $D_\parallel$ parameter suppresses quantum spin fluctuations and promotes long-range ordering below $T_{\rm N}\approx9$~K rather than a spin-liquid state.

\begin{figure}[t!]
\begin{center}
\includegraphics[width=\linewidth]{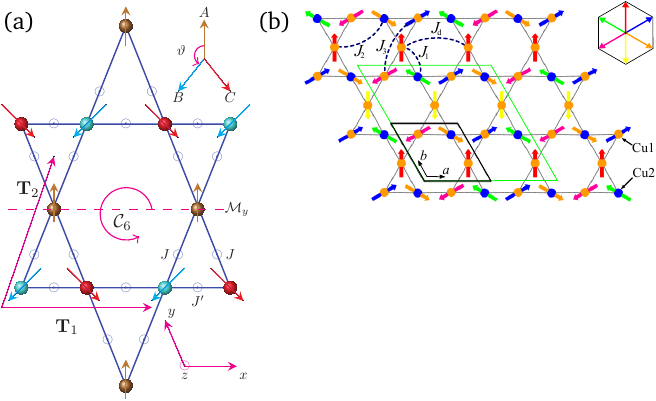}
\end{center}
\caption{(a)~The noncollinear ``$\mathbf{q}=0$'' magnetic order suggested for vesignieite by Owerre~\cite{Owerre17} as the ground state of a generic model with an anisotropic DMI, $D_\parallel$ and $D_\perp$, and two Heisenberg exchange constants, $J$ and $J^\prime$. (b)~An alternative multi-$\mathbf{q}$ structure with the dominant $J_3$ interaction, proposed by Boldrin \textit{et~al.}~\cite{BoldrinFak18}.}\label{Fig:VesignieiteModel}
\end{figure}

A realistic magnetic model that considers the anisotropic DMI in addition to the two inequivalent exchange constants $J$ and $J^\prime$ along the sides of isosceles triangles, as illustrated here in Fig.~\ref{Fig:VesignieiteModel}\,(a), has been studied theoretically by Owerre \cite{Owerre17}. He proposed that the application of magnetic field to such a system in the out-of-plane direction would result in a noncoplanar chiral spin texture with Berry curvature and nontrivial topological magnetic excitations. This can explain why the thermal Hall effect in volborthite was observed only in high magnetic fields \cite{WatanabeSugii16}. The author suggests that similar effects can be expected in volborthite and another distorted kagome mineral edwardsite, Cd$_2$Cu$_3$(SO$_4$)$_2$(OH)$_6$\,$\cdot$\,4H$_2$O (space group $P2_1/c$) \cite{ElliottBrugger10}, which orders with a canted AFM structure below 4.3~K and has the same orbital arrangement as herbertsmithite \cite{IshikawaOkamoto13}. However, high-field magnetization measurements on edwardsite showed a saturated value of only 1/3 of the full expected moment up to 50~T, and the analysis of specific heat suggests that only 1/3 of the spins contribute to the magnetic order, while the other 2/3 form singlets and are therefore magnetically inactive~\cite{FujihalaZheng14}. The DMI in edwardsite may be even larger than in vesignieite, and the presence of 4 inequivalent Cu sites results in a rather complex magnetic model consisting of approximately isosceles $J$-$J^\prime$-$J^\prime$ triangles that are arranged to form linear trimers with $J$, which can explain the weakly ferromagnetic canted $\sim$\,120$^\circ$ spin structure \cite{FujihalaZheng14}.

Contrary to the previously suggested models, the most recent INS measurements, carried out on a synthetic deuterated powder sample of vesignieite, indicate that the magnetic model is actually dominated by the third-nearest-neighbour AFM Heisenberg exchange $J_3$ \cite{BoldrinFak18}. The contributions from NN and NNN interactions turned out to be much smaller, to an extent that the experimental spin-wave spectrum is well described by a $J_3$-only model with a small symmetric exchange anisotropy. According to the proposed classification of ordered states on the kagome lattice \cite{MessioLhuillier11}, this model should lead to another type of a 120$^\circ$ magnetic order, shown in Fig.~\ref{Fig:VesignieiteModel}\,(b), which is characterized by a star of propagation vectors equivalent to $\{\frac{1}{2}00\}$. In this rather unusual triple-$\mathbf{q}$ structure, the Fourier components of a given $\mathbf{q}$ vector are nonzero for only one site, as suggested by the colouring of arrows. This substantially reduces the coupling between the ordered moments on different sites, as the dominant $J_3$ interaction acts within each of the three sublattices individually, which are then coupled by much weaker frustrating interactions $J_1=J$ and $J_2=J^\prime$. In the same work \cite{BoldrinFak18}, a small spin gap of $\sim$\,0.5~meV was observed in the spin-wave spectrum of vesignieite, which was attributed to a symmetric exchange anisotropy on $J_3$, which amounts to $\delta/J_3\approx-0.006$.

Finally, just as it was in the case of volborthite, an improvement in sample quality allowed Boldrin \textit{et al.} \cite{BoldrinKnight16} to observe an additional lowering of the lattice symmetry in recent synchrotron powder diffraction experiments on vesignieite. The x-ray data could be described by the trigonal $P3_{1}21$ space group with additional weak Bragg reflections that are absent in the initially proposed structure. An isostructural compound with Sr instead of Ba, now known as Sr-vesignieite, has also been synthesized \cite{BoldrinWills15}. One exciting implication for both compounds, suggested by these results, is the orbital frustration on the Cu(2) site, where the $d_{x^2-y^2}$ and $d_{3z^2-r^2}$ orbitals are degenerate. Therefore, the Cu(2) spin fluctuates between two orbital arrangements, leaving all oxygens available to mediate superexchange with neighbouring Cu(1) ions. This nontrivial coupling among the spin and orbital degrees of freedom, evidenced by the crystallographic signatures of the dynamic Jahn-Teller effect \cite{BoldrinWills15}, appears to be an essential ingredient of the magnetic Hamiltonian that has never been previously considered. Future studies on single crystals of vesignieite may prove essential for unraveling the details of its low-temperature physics. While they are difficult to prepare via direct chemical reactions, it was recently reported that a topochemical transformation from the volborthite to vesignieite structure can be used to obtain mm-sized single crystals of high quality \cite{IshikawaYajima17}.

Among the less-studied copper minerals with a deformed kagome lattice, one can also mention bayldonite, PbCu$_3$(AsO$_4$)$_2$(OH)$_2$ (space group $C2/c$) \cite{GhoseWan79}, for which the importance of dynamic Jahn-Teller effects was already emphasized \cite{BurnsHawthorne96}. Another recently discovered mineral engelhauptite, KCu$_3$(V$_2$O$_7$)(OH)$_2$Cl, space group $P6_3/mmc$ \cite{PekovSiidra15}, can be viewed as an analogue of volborthite resulting from the replacement of water molecules by the equal amounts of K$^+$ and Cl$^-$ ions. To the best of my knowledge, their magnetic properties have not been systematically investigated so far.

\begin{figure}[t]
\begin{center}
\includegraphics[width=0.6\linewidth]{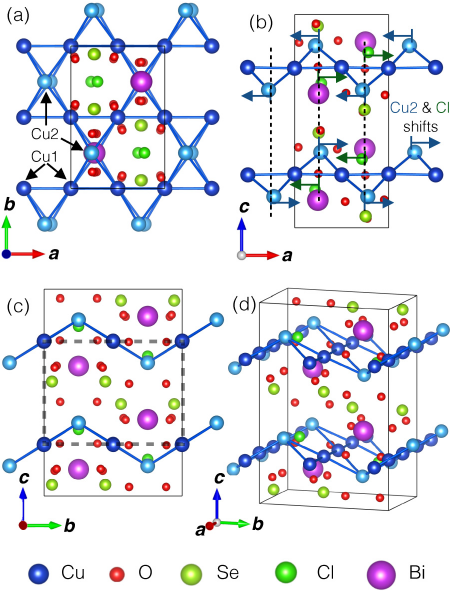}
\end{center}
\caption{Crystal structure of francisite with buckled kagome planes. After Constable \textit{et~al.} \cite{ConstableRaymond17}.}\label{Fig:Francisite}
\end{figure}

\subsection{Francisite: buckled kagome planes with bond frustration}

\begin{figure}[t]
\begin{center}
\includegraphics[width=0.58\linewidth]{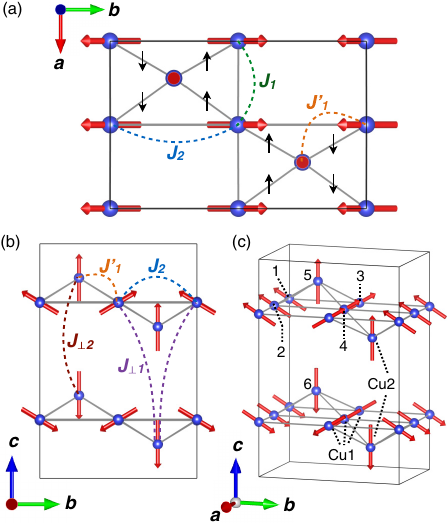}
\end{center}
\caption{The refined magnetic structure and the microscopic magnetic model proposed for francisite. After Constable \textit{et~al.}~\cite{ConstableRaymond17}.}\label{Fig:FrancisiteAFM}
\end{figure}

Another type of distortion that can be found in kagome compounds is the 3D buckling of magnetic kagome planes. It is particularly pronounced in the layered cuprate mineral francisite, Cu$_3$Bi(SeO$_3$)$_2$O$_2$Cl \cite{PringGatehouse90}. It has the orthorhombic space group $Pmmn$ at ambient conditions \cite{NazarchukKrivovichev00, MilletBastide01} but undergoes a structural phase transition at $T^\ast=115$~K to a lower-symmetry $Pcmn$ phase due to a distortion that doubles the unit cell along the $c$ direction \cite{ConstableRaymond17, GnezdilovPashkevich17} and leads to the appearance of multiple additional infrared phonon modes \cite{MillerStephens12}. The physical origin of this transition has been attributed to an antiferroelectric distortion and is reportedly absent in the Br analogue, Cu$_3$Bi(SeO$_3$)$_2$O$_2$Br \cite{PregeljZaharko12, PrishchenkoTsirlin17, GnezdilovPashkevich17}. The spin 1/2 is carried by Cu$^{2+}$ ions on two inequivalent sites: Cu(1) that forms structural chains at the centre of the magnetic layer and Cu(2) that occupies positions alternately above and below the Cu(1) plane, so that a kagome-like quasi-2D structure is formed as a result (see Fig.~\ref{Fig:Francisite}). Weak~inter\-layer interactions stabilize a long-range noncollinear magnetic order below $T_{\rm N}=25$~K \cite{MilletBastide01, ConstableRaymond17}, which is shown in Fig.~\ref{Fig:FrancisiteAFM}.

The microscopic magnetic model for francisite has been addressed in a number of recent theoretical and experimental works \cite{PregeljZaharko12, RousochatzakisRichter15, NikolaevMazurenko16, PrishchenkoTsirlin17}. In an attempt to explain the~canted AFM structure determined from single-crystal neutron diffraction \cite{PregeljZaharko12}, Rousochatzakis \textit{et al.} argued that the minimal model must include anisotropic DMI, in addition to the in- and out-of-plane Heisenberg exchange, which they estimated from DFT calculations \cite{RousochatzakisRichter15}. Later, Nikolaev \textit{et al.} \cite{NikolaevMazurenko16} elaborated on their spin Hamiltonian, replacing it with an electronic one, with the advantage that the individual anisotropic terms no longer had to be separated but were treated implicitly in the electronic Hamiltonian. This alternative \textit{ab initio} approach based on the Hartree-Fock mean-field approximation for the electronic Hamiltonian resulted in a slightly different set of parameters for the effective magnetic model by taking into account hybridization effects and the spin-orbit interaction. In this model, the canted zero-field ground state arises from a competition among ferromagnetic NN and antiferromagnetic NNN interactions within the kagome planes, whereas weaker anisotropic terms fix the spin directions and contribute to the anisotropic behaviour of the magnetization. Therefore, the interaction regime in francisite can be considered similar to that of kapellasite in terms of the bond frustration imposed by the competition among the FM and AFM interactions, while the network of interactions is altered by the stronger buckling of kagome layers.

It has to be noted that both theory papers treated francisite in its high-temperature $Pmmn$ structure, neglecting the lattice distortion present in the $Pcmn$ low-temperature phase that was only understood a year later \cite{ConstableRaymond17, PrishchenkoTsirlin17, GnezdilovPashkevich17}. In particular, the observation of a weak ferroelectric polarization loop below $T^\ast$ led to a suggestion that the low-temperature structure is more likely characterized by the polar $P2_1mn$ than the originally proposed nonpolar $Pcmn$ space group \cite{GnezdilovPashkevich17}. However, direct structural refinement based on synchrotron x-ray diffraction measurements did not confirm this hypothesis \cite{PrishchenkoTsirlin17}. Most likely, only a small fraction (about 10\%) of the sample adopts the polar structure, which is only 3~meV/f.u. higher in energy, depending on the local pattern of Cl and Cu(2) displacements due to residual disorder \cite{GnezdilovPashkevich17}. The Br analogue of francisite, which avoids this structural transition, might be therefore a cleaner example for realization of the proposed theoretical models.

\begin{figure}[t!]
\begin{center}
\includegraphics[width=\linewidth]{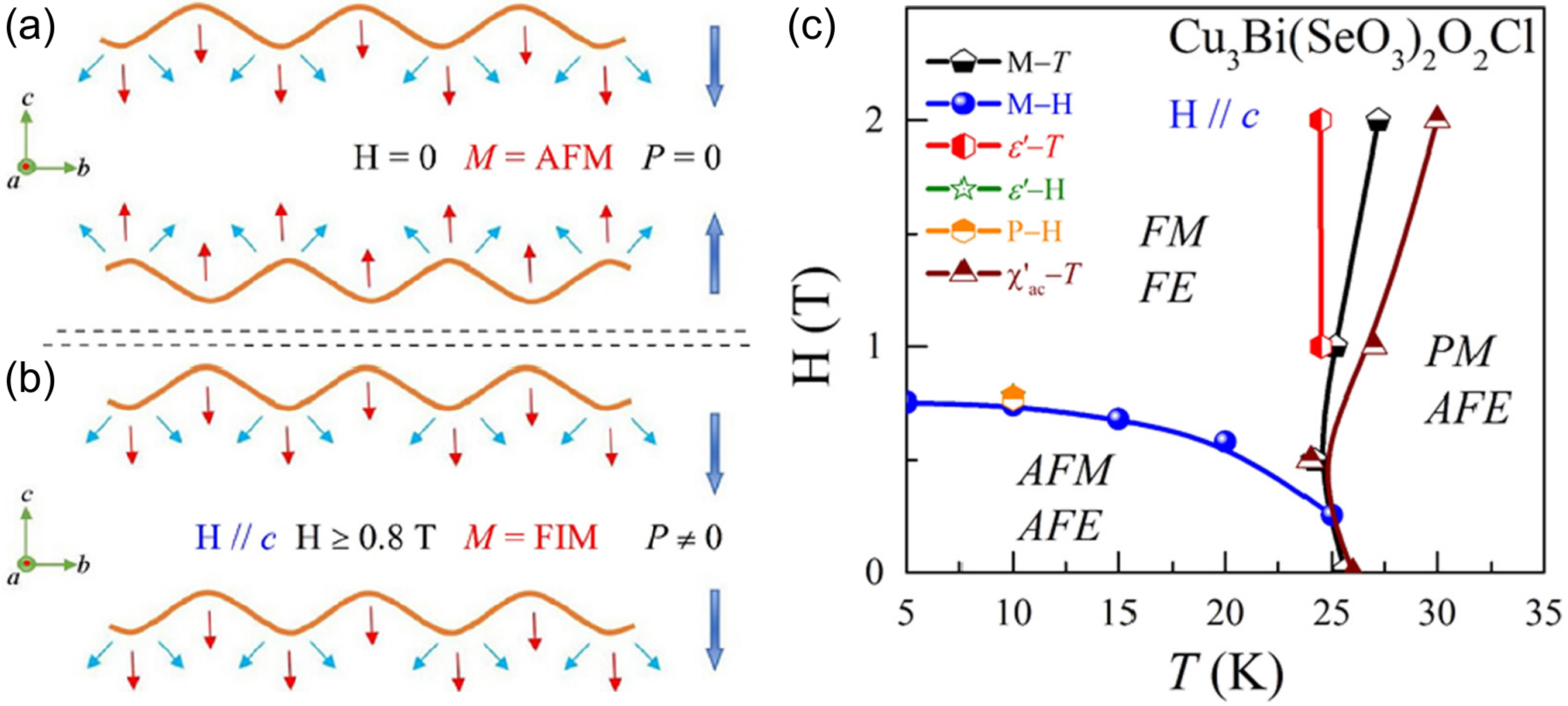}
\end{center}
\caption{The field-induced metamagnetic transition in francisite from an AFM state (a) to a ferri\-magnetic structure (b). The magnetic phase diagram (c) shows both phases in the temperature\,--\,magnetic-field plane. After Wu \textit{et~al.}~\cite{WuChandrasekhar17}.}\label{Fig:FrancisiteFiM}
\end{figure}

The most recent works focused on the multiferroic behaviour of francisite and the field-induced metamagnetic spin-flip transition to a ferrimagnetic structure, which has been observed at the critical field $B_{\rm c}\approx0.8$~T applied along the $c$ axis \cite{WuChandrasekhar17}. The spins within every Cu plane are arranged in a commensurate noncollinear structure with a net FM polarization. In zero magnetic field, these planes are arranged antiferromagnetically along the $c$ direction, so that the FM moments are compensated, as shown in Fig.~\ref{Fig:FrancisiteFiM}\,(a). Above the critical field, all planes align with their moments parallel to the field direction as a result of a spin-flip transition seen as a clear step in the magnetization curve, and the ferrimagnetic structure shown in Fig.~\ref{Fig:FrancisiteFiM}\,(b) is realized above $B_{\rm c}$. The ordering temperature remains nearly unchanged across the transition. The high-field ferrimagnetic phase of francisite showed interesting ferroelectric behaviour in magnetodielectric and pyroelectric current measurements \cite{WuChandrasekhar17}. The phase diagram showing both phases as a function of temperature and magnetic field is reproduced in Fig.~\ref{Fig:FrancisiteFiM}\,(c).

In recent years, a group from Moscow State University systematically investigated the possibility of replacing Ba atoms in the francisite structure with Y or nonmagnetic lanthanides \cite{ZakharovZvereva14, ZakharovZvereva16jac, MarkinaZakharov17} and with magnetic Sm \cite{ZakharovZvereva16}. All investigated Cu$_3R$(SeO$_3$)$_2$O$_2X$ ($R$\,=\,Y,\,La,\,Eu,\,Lu; $X$\,=\,Cl,\,Br) compounds with nonmagnetic $R$ ions showed the same type of magnetic structure as francisite, with only slightly higher N\'eel temperatures between 31 and 38~K. A single metamagnetic transition was found at a higher critical field that varied between 2.4 and 3.0~T depending on the composition. This signifies different strengths of the interlayer exchange interaction, while the essential physics is qualitatively preserved. Unsurprisingly, the magnetic behaviour of the Sm compound with the large Sm$^{3+}$ spin and orbital magnetic moments turned out to be much more complex. This compound develops AFM order below $T_{\rm N}=35$~K and shows a spin-reorientation transition at $T_{\rm C}=8.5$~K. In magnetic field, the system undergoes multiple metamagnetic phase transitions before the magnetization saturates in moderate fields of several teslas \cite{ZakharovZvereva16}.

\begin{figure}[t!]
\begin{center}
\includegraphics[width=0.72\linewidth]{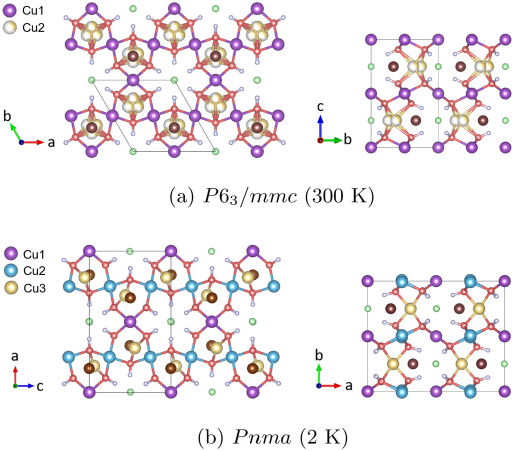}
\end{center}
\caption{The crystal structure of barlowite at room temperature (top) and at low temperature (bottom), viewed perpendicular (left) and parallel (right) to the kagome planes. Fluorine, bromine, oxygen, and hydrogen atoms are shown by green, brown, red, and white spheres, respectively. (a)~At ambient temperature, the Cu(2) ions between the kagome layers (yellow spheres) are disordered over three equivalent sites. (b)~Below 250~K, their tilt directions order, resulting in the broken equivalency of magnetic copper sites in the kagome planes (purple and blue spheres). After Tustain~\textit{et~al.}~\cite{TustainNilsen18}.}\label{Fig:BarlowiteStructure}
\end{figure}

\subsection{Claringbullite and barlowite: three-dimensionally coupled kagome planes}

\begin{figure}[t!]
\begin{center}
\includegraphics[width=0.95\linewidth]{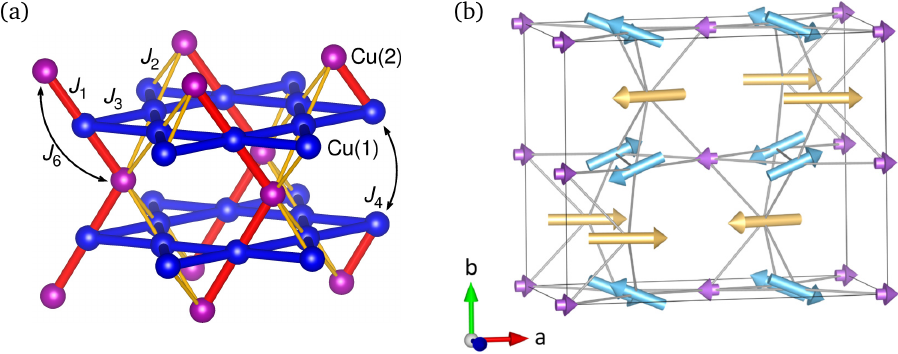}
\end{center}
\caption{(a)~The model of 3D coupled kagome planes that was proposed for barlowite under the assumption of the high-symmetry hexagonal structure realized at ambient temperature. After Jeschke \textit{et~al.}~\cite{JeschkeSalvatPujol15}. (b)~The magnetic structure of barlowite, refined in the orthorhombic $Pnma$ low-temperature phase by Tustain~\textit{et al.}~\cite{TustainNilsen18}.}\label{Fig:BarlowiteModel}
\end{figure}

Mineral claringbullite, Cu$_4$(OH)$_6$FCl \cite{FejerClark77, BurnsCooper95}, and its isostructural Br analogue barlowite, Cu$_4$(OH)$_6$FBr \cite{ElliotCooper10, ElliottCooper14}, crystallize in a hexagonal structure (space group $P6_3/mmc$) in which 3/4 of the Cu atoms on the Cu(1) positions form structurally perfect kagome planes with AA stacking, whereas the remaining 1/4 goes to the Cu(2) intersites \cite{HanSingleton14, JeschkeSalvatPujol15, SmahaHe18}. More recent studies based on x-ray and neutron diffraction indicate that at low temperatures, high-purity synthetic barlowite undergoes an additional lowering of the crystal symmetry to an orthorhombic structure with either $Cmcm$ \cite{PascoTrump18} or $Pnma$ \cite{FengWei18, TustainNilsen18} space group. This structural transition, which happens at $\sim$250~K, is associated with positional ordering in the tilt directions of the interlayer Cu$^{2+}$ ions, as shown in Fig.\,\ref{Fig:BarlowiteStructure}, and leads to the broken equivalency of magnetic copper sites within the kagome planes. Several other phase transitions at intermediate temperatures were reported from specific-heat data \cite{HanSingleton14}. Magnetic susceptibility gives the Curie-Weiss temperature of $\Theta_{\rm CW}=-136$~K. The presence of Cu$^{2+}$~ions at interlayer positions determines the magnetic behaviour of these systems at low temperatures by providing paths for ferromagnetic interlayer couplings \cite{JeschkeSalvatPujol15}. The resulting 3D network of interactions, which is shown (for the high-symmetry phase) in Fig.~\ref{Fig:BarlowiteModel}\,(a), stabilizes a canted AFM ground state with the ordering temperature $T_{\rm N}=15.4$~K and a weak FM moment $\mu<0.1\mu_{\rm B}$/Cu \cite{HanSingleton14, JeschkeSalvatPujol15, HanIsaacs16, FengWei18, PascoTrump18, RanjithKlein18}. The actual magnetic structure, as it was recently refined from neutron diffraction on deuterated barlowite, is illustrated in Fig.\,\ref{Fig:BarlowiteModel}\,(b). It shows that the weak ferromagnetic moment results from the canting of interlayer Cu$^{2+}$ ions in the $Pnma$ low-temperature structure, whereas two nonequivalent moments in the kagome planes are antiferromagnetically compensated \cite{TustainNilsen18}.

The ferromagnetic coupling between the kagome layers was initially assumed to be weak \cite{HanSingleton14}. However, exchange parameters obtained from DFT and GGA+$U$ calculations in the hexagonal crystal structure indicate that the dominant interlayer exchange constant $J_1$ is either slightly larger or comparable to the nearest-neighbour AFM exchange on the kagome triangles, $J_3\approx15.3$~meV \cite{JeschkeSalvatPujol15}. A controversy exists about the magnitude of the DMI in barlowite and about the role played by the Cu(2) magnetic moments. At first, from the observed difference between the experimentally determined spin canting angle of $\sim$\,4.5$^\circ$ and the orientation of the basal plane of the Cu octahedra, a substantial in-plane DMI of the order of $0.1J_3$ has been estimated \cite{JeschkeSalvatPujol15}. The ferromagnetic moment, evidenced by a hysteresis in the magnetization, could be therefore naturally explained by the canted AFM structure in the kagome layers \cite{ElhajalCanals02}. Another group, however, re-estimated the canting angle to a much smaller value $\leq0.2^\circ$, which resulted in a vanishingly small DMI $\lesssim0.006\,J$ \cite{HanIsaacs16}. They concluded that the ordered moment at $T\leq15$~K might actually come from the interlayer spins, not excluding the possibility that the kagome-lattice physics remains practically unaffected. Correspondingly, signatures of persistent spin dynamics within the ordered state were observed in the weak temperature dependence ($\propto \sqrt T$) of the spin-lattice relaxation rate $1/T_1$, measured using $^{79,81}$Br NMR~\cite{RanjithKlein18}. The new $Pnma$ crystal structure reported by K.~Tustain \textit{et al.} \cite{TustainNilsen18} allows for all three components of the DMI for each of the six nearest-neighbour exchanges by symmetry. Therefore, a thorough analysis using a combination of quantum-chemistry calculations and single-crystal neutron spectroscopy measurements would be necessary to verify the resulting Hamiltonian in its full complexity.

While the details of the microscopic magnetic model of barlowite and claringbullite may require additional work to be settled, the main interest to these systems is motivated by the possibility to substitute nonmagnetic ions onto the Cu(2) sites \cite{HanSingleton14, LiuZou15, GuterdingValenti16, FengLi17, FengYi19, FengWei18, PascoTrump18}. This should suppress the interlayer coupling and result in a system similar to herbertsmithite but with much less structural disorder. If the interpretation of Han \textit{et al.} \cite{HanIsaacs16} about the smallness of DMI is correct, one would end up with a KAFM compound that offers better conditions for the realization of a spin-liquid ground state. The feasibility of such a selective substitution of Cu(2) ions with Mg or Zn, as well as with larger Cd and Ca ions, has been systematically verified using first-principles calculations \cite{LiuZou15, GuterdingValenti16}. The Mg- and Zn-substituted compounds were suggested as the best candidates due to the higher site selectivity and smaller lattice distortion. The first successful synthesis of Zn-barlowite, ZnCu$_3$(OH)$_6$FBr, was soon demonstrated by Feng \textit{et al.} \cite{FengLi17}, and the synthesis of Zn-claringbullite followed shortly afterwards \cite{FengYi19, PascoTrump18}. These new compounds show no magnetic order down to 50~mK and exhibit clear signatures of a spin gap in the uniform spin susceptibility obtained from $^{19}$F NMR measurements. The magnetic-field dependence of the gap energy shows signatures of fractionalized spin-$\frac{1}{2}$ excitations, as one would expect for a topological quantum spin-liquid state. Most recently, Feng \textit{et al.} \cite{FengWei18} systematically investigated the evolution of magnetism in Cu$_{4-x}$Zn$_x$(OH)$_6$FBr, bridging the two limiting cases of the 3D antiferromagnetism in barlowite ($x=0$) and the spin-liquid physics in Zn-barlowite ($x=1$). They found that the bulk AFM order is destroyed already around $x\approx0.4$, while remnant magnetic correlations with a reduced transition temperature survive up to $x\approx0.8$. This provides an opportunity to tune the system to a QCP that separates magnetic order from a spin-liquid phase.

\subsection{Brochantite and averievite: new routes to a kagome spin liquid}

\begin{figure}[b]
\begin{center}
\includegraphics[width=0.66\linewidth]{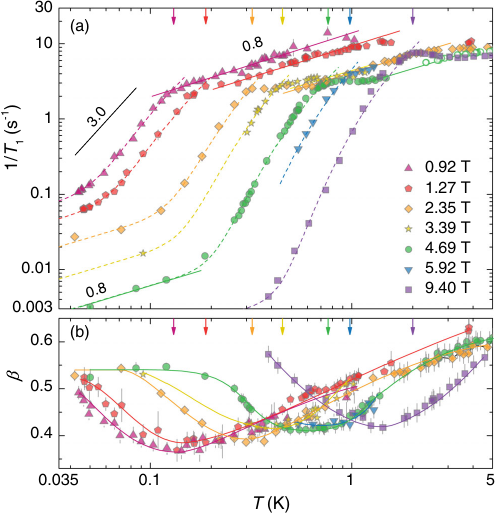}
\end{center}
\caption{(a)~Temperature dependence of the $^2$D NMR spin-lattice relaxation rate $1/T_1$ in Zn-brochantite for various applied magnetic fields. The data show a gradual opening of a gap in the spinon excitation spectrum. (b)~Stretching exponents $\beta$ from the magnetization recovery curves, characterizing the distribution of relaxation times in
the NMR experiment. After Gomil\v{s}ek \textit{et~al.}~\cite{GomilsekKlanjsek17}.}\label{Fig:Zn-brochantite}
\end{figure}

One more kagome compound that deserves to be mentioned here derives from the copper-sulfate mineral brochantite, Cu$_4$SO$_4$(OH)$_6$, which crystallizes in the monoclinic $P2_1/a$ space group \cite{VilminotRichardPlouet06}. While the parent brochantite orders magnetically below 7.5~K with a weakly canted AFM structure, most interest in the field of quantum magnetism is attracted to its synthetic modification known as Zn-brochantite, in which 1/4 of the Cu atoms have been replaced with nonmagnetic Zn. It was immediately recognized that this compound is a promising candidate for the realization of a quantum spin liquid, because it displays no magnetic ordering down to 50~mK despite a much higher Weiss temperature of $-79$\,K \cite{LiPan14}. From the $T$ dependence of the susceptibility, a gapless RVB spin-liquid ground state with a spinon Fermi surface has been suggested \cite{LiPan14, GomilsekKlanjsek16a, GomilsekKlanjsek16b, GomilsekKlanjsek17}. However, Rietveld refinement of synchrotron powder x-ray diffraction data revealed that the kagome layers in this system are both highly distorted and strongly buckled \cite{LiPan14}.

The assumption about the spin-liquid state of Zn-brochantite has been recently corroborated in a series of experiments by M.~Gomil\v{s}ek \textit{et al.} that combined local-probe techniques with neutron scattering \cite{GomilsekKlanjsek16a, GomilsekKlanjsek16b, GomilsekKlanjsek17}. Their $\mu$SR measurements, performed down to 21~mK, revealed a considerable increase in the $1/T^{\mu}_1$ relaxation rate and the $\mu^+$ Knight shift below $\sim$\,5~K with a saturation below 0.6~K. In the $^2$D NMR data, a nonmonotonic behaviour of $1/T^{\rm NMR}_1$ with anomalies around 15, 5, and 0.76~K and a power-law dependence at higher temperatures was observed. This led to a conclusion that two spin-liquid regimes, SL1 and SL2, are realized at different temperatures \cite{GomilsekKlanjsek16a}. Neutron spectroscopy revealed an $\omega/T$ scaling behaviour of the dynamic susceptibility at high temperatures \cite{GomilsekKlanjsek16a}, which can be explained by the proximity to a QCP driven by the magnetic anisotropy, as in the case of herbertsmithite. Additional NMR measurements, presented in Fig.~\ref{Fig:Zn-brochantite}, suggested that the gapless spin liquid is intrinsically unstable against spinon pairing (similar to Cooper pairing in superconductors) under the application of arbitrary small magnetic field, which opens a full or partial spin gap seen in the temperature dependence of the spin-lattice relaxation rate~\cite{GomilsekKlanjsek17}.

A similar route to the realization of a kagome spin liquid has been suggested by Botana \textit{et~al.} \cite{BotanaZheng18} for the mineral averievite, Cu$_5$V$_2$O$_{10}$(CsCl). This copper-oxide mineral, discovered in the 90's on Kamchatka \cite{VergasovaStarova98}, is composed of Cu$^{2+}$ kagome layers, separated by Cu$^{2+}$-V$^{2+}$ honeycomb layers. It is a member of a more general family of compounds, Cu$_5$V$_2$O$_{10}$($MX$), that were later synthesized by Queen \cite{QueenPhD09} with $MX$\,=\,RbCl, CsCl, and CsBr. The parent averievite compound has a Curie-Weiss temperature of $\Theta_{\rm CW}\approx185$~K and orders antiferromagnetically below $T_{\rm N}\approx24$~K due to the significant interlayer coupling via the honeycomb copper ions \cite{BotanaZheng18}. However, selectively substituting Zn$^{2+}$ on these copper sites suppresses the AFM order, suggesting that the kagome planes become decoupled. This opens several new avenues for the synthesis of promising spin-liquid candidates based on the averievite structure. In a recent theory work, Volkova and Marinin \cite{VolkovaMarinin18} compared the sign and strength of magnetic interactions in averievite, calculated using the so-called ``crystal chemistry method'', with those of structurally related minerals NaCu$_5$O$_2$(SeO$_3$)$_2$Cl$_3$ (ilinskite) and K$_2$Cu$_5$Cl$_8$(OH)$_4\cdot2$H$_2$O (avdononite) to analyze their potential for the realization of a quantum spin liquid on the kagome lattice.\vspace{-2pt}

\section{Other layered quasi-2D lattices}

\subsection{The variety of two-dimensional crystal structures}

\begin{figure}[t]
\begin{center}
\includegraphics[width=0.55\linewidth]{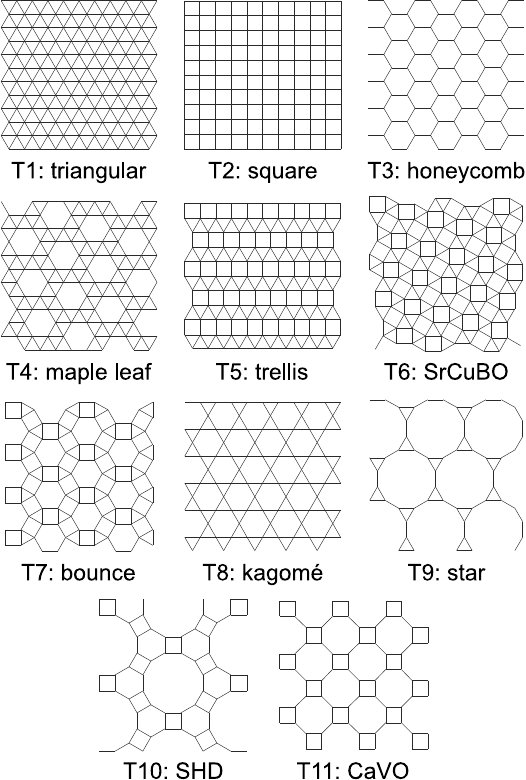}
\end{center}
\caption{The 11 Archimedean tilings, after Richter \textit{et~al.} \cite{RichterSchulenburg04} and Farnell \textit{et~al.}~\cite{FarnellGoetze14, FarnellGoetze18}.}\label{Fig:Archimedean}
\end{figure}

The vast majority of quasi-2D crystals form regular periodic structures known as Archimedean lattices \cite{RichterSchulenburg04, FarnellGoetze14, Yu15, FarnellGoetze18}. The simplest and most symmetric of them, built from identical regular polygons, derive from the well known triangular (T1), square (T2), and honeycomb (T3) tilings. Together~with the kagome (T8) lattice discussed in the previous chapter, they constitute the only four Archimedean tilings in which all NN bonds are equivalent. Triangular lattice, found in the minerals delafossite and mcconnellite, is an emblem of geometric frustration, where it occurs already in the simplest model with just a single AFM interaction on all NN bonds. In contrast, square and honeycomb lattices with only NN interactions are not geometrically frustrated, yet bond frustration may arise upon inclusion of further-neighbour exchange couplings \cite{MisguichLhuillier13, SchmidtThalmeier17}. This statement can be generalized to all 11 Archimedean lat\-tices, shown in Fig.~\ref{Fig:Archimedean}, of which only four are bipartite (i.e., consist of only even-sided regular polygons): square, honeycomb, square-hexa\-gonal-dodecagonal (T10), and CaVO~(T11). A magnetic square lattice is found, for example, in the mineral diaboleite,~Pb$_2$Cu(OH)$_4$Cl$_2$ \cite{TsirlinJanson13}. The T10 lattice consists of squares, hexagons and dodecagons (hence the name, SHD) and has only been investigated theoretically \cite{TomczakSchulenburg01}, because no real material with this structure is known \cite{RichterSchulenburg04}. The name CaVO derives from the compound CaV$_4$O$_9$, in which the T11 lattice, consisting of squares and octagons, is realized \cite{TaniguchiNishikawa95}. Some non-Archimedean periodic bipartite lattices, in particular the so-called dice lattice \cite{JagannathanMoessner06, ValdesLebrecht07}, which is dual to the kagome \cite{NikolicSenthil03}, have also been theoretically considered.

The remaining 7 lattices contain triangles and are therefore geometrically frustrated. The trellis (T5) lattice with its pronounced unidirectional structure can be considered as an arrangement of coupled two-leg spin ladders, which tend to dimerize across the rungs by forming singlets, if the exchange interactions on the rung bonds are sufficiently strong. This model is realized in the $A$V$_2$O$_5$ vanadates and in hollandite-type compounds \cite{IsobeKoishi06} that are isostructural to the mixed-valent manganese mineral hollandite, BaMn$_8$O$_{16}$ \cite{Miura86}. The T6 lattice got its name from the layered material SrCu$_2$(BO$_3$)$_2$ \cite{KageyamaOnizuka98, KageyamaYoshimura99} and is essentially equivalent to the Shastry-Sutherland lattice discussed in Chapter~\ref{Chap:Dimers}. As it was already mentioned in section \ref{Sec:KagomeIntro}, the kagome (T8) and star (T9) lattices with only Heisenberg-type AFM interactions on NN bonds have no semiclassical ordered ground states because of a combination of strong frustration with low coordination number. In contrast, the triangular AFM lattice and all other Archimedean lattices order in the classical limit \cite{RichterSchulenburg04, FarnellGoetze14, FarnellGoetze18}. Among them, the maple-leaf (T4) lattice, which results from a 1/7 site depletion of the triangular lattice, is intermediate from a frustration viewpoint between the triangular and kagome lattices and is thought to be very close to the classically disordered ground state~\cite{SchmalfusTomczak02}. In spite of its complexity, this lattice is realized in synthetic iron phosphonates \cite{CaveCoomer06} and in the naturally occurring copper minerals spangolite, Cu$_6$Al(SO$_4$)(OH)$_{12}$Cl\,$\cdot$\,3H$_2$O \cite{MerlinoPasero92}; sabelliite, Cu$_2$ZnAsO$_4$(OH)$_3$ \cite{OlmiSabelli95}; fuettererite, Pb$_3$Cu$_6$TeO$_6$(OH)$_7$Cl$_5$ \cite{KampfMills13}; and the most recently discovered bluebellite, Cu$_6$IO$_3$(OH)$_{10}$Cl, and mojaveite, Cu$_6$TeO$_4$(OH)$_9$Cl \cite{MillsKampf14} (for a review, see Ref.~\cite{Norman18}).

Finally, the bounce (T7) lattice can be seen as a bond-depleted modification of the maple-leaf lattice. Indeed, the two lattices can be considered as limiting cases of a more general spin model with unequal $J_1$ and $J_2$ interactions on bonds that separate neighbouring triangles \cite{RichterSchulenburg04, FarnellDarradi11}. The maple-leaf limit is realized for $J_2/J_1=1$, while the bounce lattice corresponds to $J_2/J_1=0$. Although these ``pure'' limits both have magnetically ordered ground states, the generalized model exhibits a quantum-critical transition at $\alpha_{\rm c}=J_2/J_1\approx1.45$ to a quantum orthogonal-dimer singlet state without magnetic long-range order \cite{FarnellDarradi11}. In this regime, application of magnetic field would result in magnetization plateaus, similar to those found in the Shastry-Sutherland or triangular-lattice models \cite{RichterSchulenburg04}.

\subsection{Delafossite and mcconnellite: classical spins on the triangular lattice}\label{Sec:Delafossite}

\begin{figure}[b!]
\begin{center}
\includegraphics[width=\linewidth]{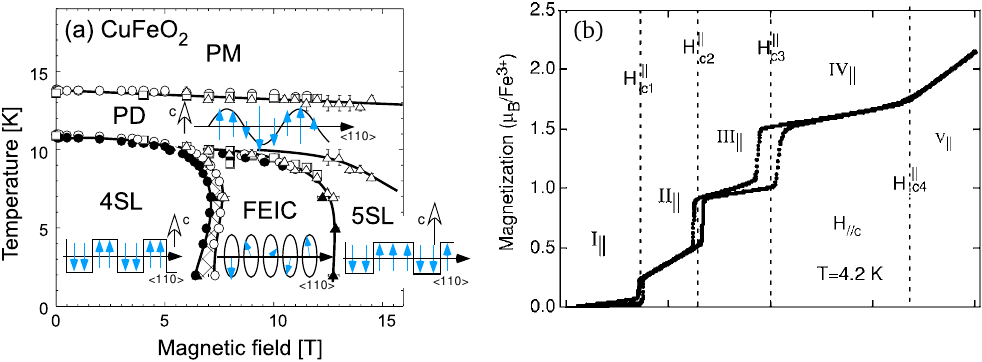}\vspace{-3pt}
\end{center}
\caption{(a)~The magnetic phase diagram of delafossite, CuFeO$_2$, with a schematic illustration of ordered phases \cite{MitsudaMase00}. (b)~Magnetic field dependence of the magnetization with a sequence of metamagnetic phase transitions. After Terada \textit{et al.} \cite{Terada14, TeradaNarumi07}.}\label{Fig:Delafossite_PhaseDiagram}
\end{figure}

\begin{figure}[t]
\begin{center}
\includegraphics[width=\linewidth]{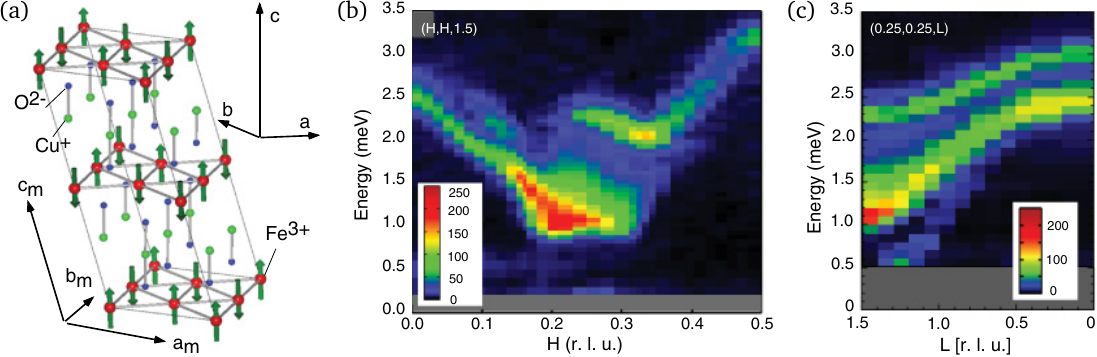}\vspace{-1pt}
\end{center}
\caption{(a)~The four-sublattice magnetic structure of delafossite, CuFeO$_2$ \cite{MitsudaYoshizawa91, MekataYaguchi92}. (b,\,c)~INS intensity maps along the $\left(H\,H\,\frac{3}{2}\right)$ and $\left(\frac{1}{4}\,\frac{1}{4}\,L\right)$ directions in momentum space \cite{NakajimaMitsuda11}. After Nakajima \textit{et~al.}~\cite{NakajimaSuno11}.\vspace{-1pt}}\label{Fig:Delafossite}
\end{figure}

The delafossite group of minerals serves as a perfect realization of the triangular magnetic lattice. The hexagonal delafossite structure (space group $R\overline{3}m$) derives from the cubic rocksalt (NaCl) structure, if two different cations statistically distributed on the cation site undergo an order-disorder transition, segregating into alternate (111) planes to form a layered structure \cite{OhtaniHonjo87, KippVanderah90}. In the minerals delafossite, CuFeO$_2$ \cite{Pabst38}, and mcconnellite, CuCrO$_2$ \cite{DannhauserVaughan55}, that share this crystal structure, the magnetic Fe$^{3+}$ ($S=5/2$) and Cr$^{3+}$ ($S=3/2$) ions form a perfect triangular lattice separated by nonmagnetic monovalent Cu$^+$ layers. The magnetic order in CuFeO$_2$ and CuCrO$_2$ has been known for several decades \cite{MuirWiedersich67, Apostolov69, DoumercWichainchai86, MitsudaYoshizawa91, MekataYaguchi92, MekataYaguchi93}. In CuFeO$_2$, two magnetic transitions at $T_{\rm N1}\approx16$~K and $T_{\rm N2}\approx11$~K were observed in neutron diffraction and thermodynamic measurements, see Fig.\,\ref{Fig:Delafossite_PhaseDiagram}\,(a). A collinear up-up-down-down AFM ground state with spins parallel to the $c$ axis and a propagation vector $\mathbf{Q}_{\uparrow\uparrow\downarrow\downarrow}=(\frac{1}{4}\,\frac{1}{4}\,\frac{3}{2})$, as shown in Fig.~\ref{Fig:Delafossite}\,(a), was proposed \cite{MitsudaYoshizawa91, MekataYaguchi92, MekataYaguchi93}. The magnetic structure of the intermediate-temperature phase between $T_{\rm N1}$ and $T_{\rm N2}$ is a sinusoidally amplitude-modulated structure with a $(q\,q\,\frac{3}{2})$ propagation vector that changes with temperature \cite{MitsudaKasahara98}. Two magnetic phase transitions at $T_{\rm N1}\approx24.2$~K and $T_{\rm N2}\approx23.6$~K are also observed in CuCrO$_2$. However, this compound exhibits a noncollinear incommensurate spin helix below $T_{\rm N}\approx24$~K, propagating in the $[HH0]$ direction, with moments lying in the $[HHL]$ plane \cite{FrontzekEhlers12, EhlersPodlesnyak13}. The formation of this spin-spiral structure is associated with magnetoelectric effects \cite{KimuraNakamura08, SekiOnose08, SodaKimura10, MunFrontzek14, TokuraSeki14, Terada14}. In CuFeO$_2$, similar noncollinear order and ferroelectric behaviour can be induced by partial Al substitution on the Fe site \cite{SekiYamasaki07, KanetsukiMitsuda07}.

In high magnetic fields applied parallel to the $c$ axis, delafossite undergoes a sequence of first-order metamagnetic phase transitions at $B^\parallel_{\rm c1}\approx7$~T, $B^\parallel_{\rm c2}\approx13$~T, $B^\parallel_{\rm c3}\approx20$~T, and $B^\parallel_{\rm c4}\approx34$~T, seen as sharp magnetization steps with a hysteretic behaviour \cite{FukudaNojiri98, MitsudaMase00, TeradaNarumi07, LummenStrohm09, LummenStrohm10}. The nearly flat plateau regions in between correspond to different magnetic phases, labeled I$_\parallel$, II$_\parallel$, III$_\parallel$, IV$_\parallel$, and V$_\parallel$ in Fig.~\ref{Fig:Delafossite_PhaseDiagram}\,(b). In particular, the 1/5 magnetization plateau, corresponding to the phase III$_\parallel$, is consistent with the $\uparrow\uparrow\uparrow\downarrow\downarrow$ type of magnetic order \cite{LummenStrohm10, NakajimaTerada13}, while the 1/3 magnetization plateau in phase IV$_\parallel$ can be attributed to an $\uparrow\uparrow\downarrow$ (up-up-down) spin arrangement \cite{TeradaNarumi07, FishmanYe08, LummenStrohm09, LummenStrohm10}. The intermediate phase II$_\parallel$ is a noncollinear incommensurate spin spiral \cite{KimuraLashley06} that exhibits ferroelectric properties \cite{MitamuraMitsuda07, Terada14}. As noted by Haraldsen \textit{et al.} \cite{HaraldsenSwanson09}, these metamagnetic transitions, driven either by magnetic field or chemical substitution, can arise from spin-wave instabilities that are preceded by the softening of spin-wave modes at the would-be ordering wave vectors. Indeed, neutron-spectroscopy measurements in the collinear AFM state of CuFeO$_2$ revealed two local minima in the spin-wave dispersion that coincide with the incommensurate ordering wave vectors of its CuFe$_{1-x}$Al$_x$O$_2$ daughter compound \cite{TeradaMitsuda04, TeradaMitsuda07, YeFernandezBaca07}. These soft modes can be also recognized in Fig.~\ref{Fig:Delafossite}\,(b).

The magnetic order in CuFeO$_2$ is also sensitive to pressure. Under hydrostatic pressure, the collinear low-temperature phase turns first into a proper-screw magnetic ordering between 3 and 4~GPa, similar to the field-induced ferroelectric phase. Above 4~GPa, a second pressure-induced incommensurate phase appears, representing a combination of proper-screw and cycloidal spin spiral configurations \cite{TeradaKhalyavin14}. The intermediate-temperature phase is stabilized by pressure in such a way that $T_{\rm N1}$ increases to $\sim25$\,K (nearly twofold) at 8~GPa \cite{TeradaKhalyavin14}. Application of uniaxial pressure in the $ab$ plane also stabilizes the intermediate-temperature phase due to a partially relieved geometric frustration, as can be seen in $T_{\rm N1}$ increasing by as much as 5~K at 600~MPa \cite{TamatsukuriAoki16}. These results suggest that the spin-lattice coupling should be taken into account in describing the ordered phases in delafossite. The importance of spin-lattice effects also follows from the observation of spin-driven bond order in the 1/5-magnetization plateau phase by means of INS measurements under applied magnetic fields \cite{NakajimaTerada13}.

The spin model that was initially proposed to describe the magnetic interactions in CuFeO$_2$ included four Heisenberg-type exchange interactions and a single-ion anisotropy term \cite{Fishman08}. From a global fit of this model to the INS data, Ye \textit{et al.} \cite{YeFernandezBaca07} obtained the experimental estimates for the in-plane AFM interactions $J_1\approx1.14$~meV, $J_2\approx0.50$~meV, $J_3\approx0.65$~meV, the out-of-plane coupling $J_z\approx0.33$~meV, and the single-ion anisotropy $D\approx0.17$~meV. In the theoretical phase diagram of an Ising-spin triangular-lattice antiferromagnet \cite{TakagiMekata95}, these parameters fall within the region of the 4-sublattice $\uparrow\uparrow\downarrow\downarrow$ phase, which is consistent with the ground state of CuFeO$_2$. However, follow-up single-crystal neutron scattering data measured in a single-domain AFM state, obtained by ``detwinning'' a single crystal under a small uniaxial pressure \cite{NakajimaMitsuda11, NakajimaSuno11}, revealed considerable deviations from this model. A selection of these INS data is reproduced in Fig.~\ref{Fig:Delafossite}. This discrepancy has been attributed to the effect of spin-lattice coupling, manifested in the weak trigonal-to-monoclinic lattice distortion that takes place below $T_{\rm N}$ \cite{YeRen06, TeradaMitsuda06}. In particular, the low-energy spin-wave mode at the AFM zone centre is split into two branches that were observed previously in ESR studies in zero magnetic field \cite{FukudaNojiri98, KimuraNishihagi10}. Moreover, the higher-energy branch cannot be reproduced by spin-wave calculations and was attributed to an electromagnon excitation~\cite{SekiKida10}. To describe the dispersion of this electric-field-active magnon mode, coupling to lattice or charge degrees of freedom must be considered. This led T.~Nakajima \textit{et al.} \cite{NakajimaSuno11} to propose a more elaborate spin model that takes into account the spin-driven lattice distortion and can successfully describe the available INS measurements. In this model, every exchange interaction $J_i^{\phantom{\prime}}$ is split into $J_i^{\phantom{\prime}}$ and $J_i^\prime$, and the NN interaction $J_1^{\phantom{\prime}}$ into $J_1^{\phantom{\prime}}$, $J_1^\prime$, and $J_1^{\prime\prime}$, to account for a distortion of the triangular lattice that arises as a result of the structural transition. In addition to the single-ion anisotropy $D$, a small in-plane anisotropy $E\approx0.035$~meV has been also included. From a fit to the experimental data, the largest effect that the monoclinic distortion has on the splitting of exchange constants was observed for the NN interaction, with $J_1^{\phantom{\prime}}\approx0.455$~meV, $J_1^\prime\approx0.422$~meV, and $J_1^{\prime\prime}\approx0.150$~meV. This example illustrates that even a slight structural distortion of less than 0.4\% \cite{YeRen06, TeradaMitsuda06} can cause a threefold difference in the initially equivalent exchange parameters, effectively lifting the vast degeneracy of the strongly frustrated triangular lattice.

\begin{figure}[t]
\begin{center}
\includegraphics[width=0.58\linewidth]{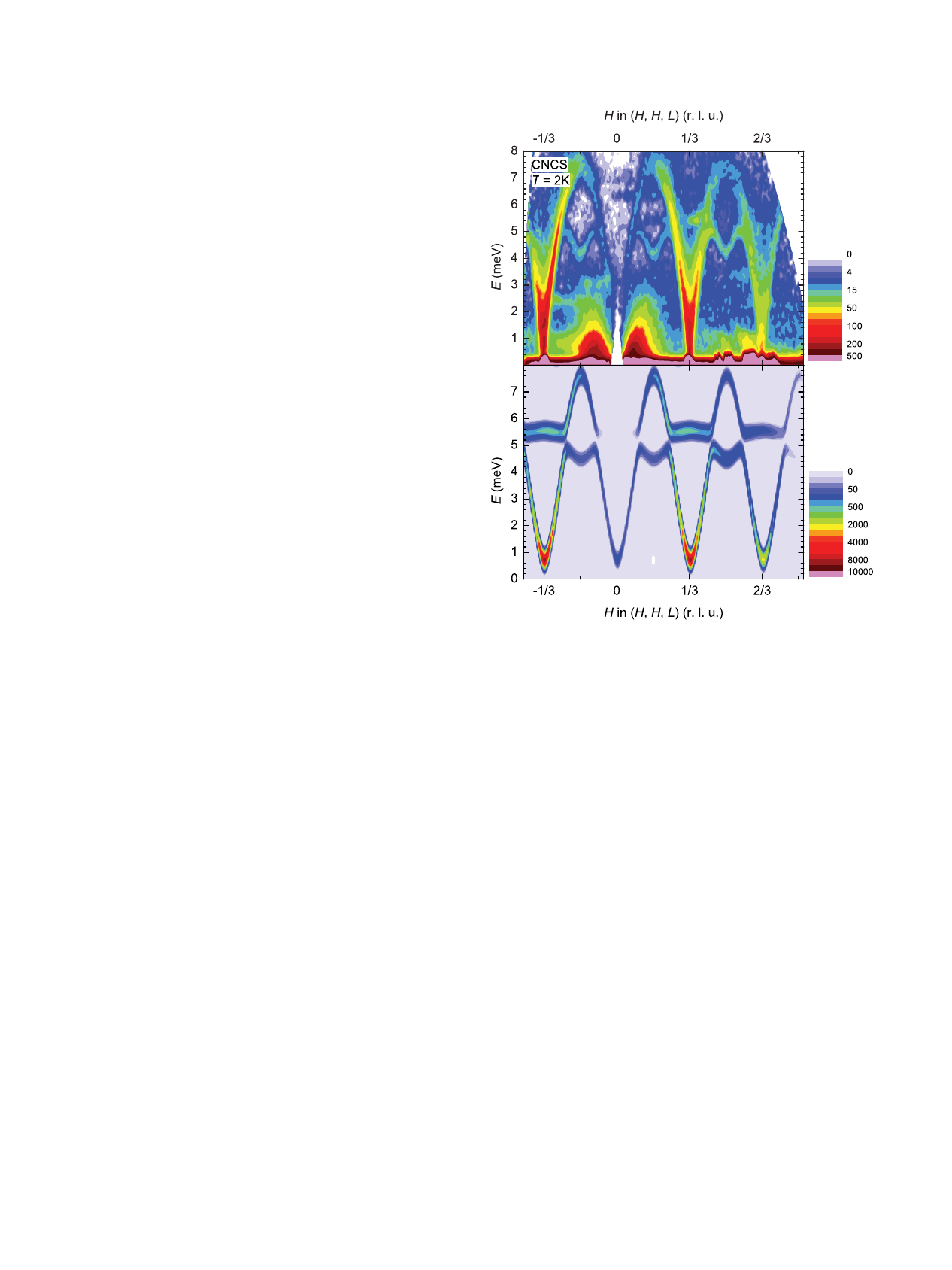}
\end{center}
\caption{Spin waves in mcconnellite, measured on a synthetic single crystal, in comparison to the results of a spin-wave calculation. After Frontzek \textit{et~al.}~\cite{FrontzekHaraldsen11}.}\label{Fig:Mcconnellite}
\end{figure}

In the isostructural CuCrO$_2$, the spin-wave spectrum has been also mapped out on synthetic single crystals by Poienar \textit{et al.} \cite{PoienarDamay10} and Frontzek \textit{et al.} \cite{FrontzekHaraldsen11}. In contrast to CuFeO$_2$, this ferroelectric compound undergoes no structural phase transition across $T_{\rm N}$, hence the $J_1$-$J_2$-$J_3$-$J_z$ model, supplied with $D_x$ and $D_z$ anisotropy terms, can successfully describe the whole excitation spectrum, as shown in Fig.~\ref{Fig:Mcconnellite}. The much larger NN interaction $J_1\approx2.8$~meV, as compared to CuFeO$_2$, reduces the $D/J_1$ ratio and the influence of further-neighbour interactions, stabilizing a proper-screw magnetic structure with a nearly commensurate $\left(\frac{1}{3}\!-\!\delta~\frac{1}{3}\!-\!\delta~0\right)$ propagation vector \cite{FrontzekEhlers12, EhlersPodlesnyak13} that became known as the incommensurate ``Y~state'' \cite{LinBarros14}. Application of high magnetic fields leads to the appearance of multiple field-induced magnetic phases in CuCrO$_2$ that have been investigated both experimentally and theoretically \cite{MunFrontzek14, LinBarros14}. For an in-plane magnetic field, $\mathbf{B}\perp\mathbf{c}$, a transition from a proper-screw to a cycloidal-spiral phase occurs at $B_{\rm f}\approx5.3$~T. For fields applied parallel to the $c$ axis, a sequence of metamagnetic transitions has been observed at much higher fields exceeding 40~T. These high-field phases include a noncoplanar umbellate spin structure, a commensurate Y~state, and a collinear up-up-down phase \cite{LinBarros14}.

In a more recent neutron-spectroscopy work, Park \textit{et al.} \cite{ParkOh16} investigated the high-energy part of the spin-wave spectrum of CuCrO$_2$, above the energy range covered in previous studies \cite{PoienarDamay10, FrontzekHaraldsen11}. They found an additional collective mode around 12.5~meV that has a mixed magnon-phonon character. As in the case of CuFeO$_2$, the spin-lattice coupling has a substantial influence on the magnetic excitations, and the inclusion of exchange-striction effects in the theoretical model can accurately capture the new features of the INS data \cite{ParkOh16}. Kajimoto \textit{et al.} \cite{KajimotoTomiyasu15} also pointed out the presence of a diffusive quasielastic component in the neutron-scattering spectrum with a characteristic momentum dependence. This component appears only at elevated temperatures and presumably originates from scattering on uncorrelated triangular clusters that persist both above and below the magnetic ordering temperature. This feature could be generic for triangular-lattice antiferromagnets.

Finally, it is worth noting that the mineral crednerite, CuMnO$_2$ \cite{ToepferTrari95}, is the $S=2$ analog of delafossite. However, it crystallizes in a lower-symmetry monoclinic ($C2/m$) structure that undergoes a spin-driven transition to a triclinic $C\overline{1}$ phase as magnetic order sets in at $T_{\rm N}\approx65$~K. The monoclinic symmetry of the lattice is due to a Jahn-Teller distortion of the Mn$^{3+}$ ($t_{\rm 2g}^3{\kern.5pt}e_{\rm g}^1$) cation that lifts the $e_{\rm g}$ orbital degeneracy \cite{DamayPoienar09, UshakovStreltsov14}, deforming the lattice so that it forms isosceles-triangle layers with reduced frustration. Upon the formation of the collinear AFM order with propagation vector $\left(-\frac{1}{2}\,\frac{1}{2}\,\frac{1}{2}\right)$, strong magnetoelastic coupling \cite{VecchiniPoienar10, ZorkoKokalj15} relieves the remaining frustration through the monoclinic-to-triclinic transition. Damay \textit{et~al.} \cite{DamayPoienar09} therefore argued that the magnetism in CuMnO$_2$, with largely different exchange interactions on the sides of isosceles triangles, is better considered in the framework of a frustrated square-lattice model with a nearest-neighbour coupling $J_1$ and a diagonal interaction $J_2$. For $J_2/J_1>1/2$, this model is known to stabilize the collinear AFM phase, assisted by any arbitrarily weak coupling to the lattice \cite{BeccaMila02}. Powder neutron-spectroscopy measurements revealed a large low-temperature spin-wave gap of about 6~meV, comparable with the ordering temperature, and persistent 2D short-range magnetic correlations \cite{TeradaTsuchiya11}.

For the sake of completeness, it should be noted that a triangular lattice of $\frac{5}{2}$-spins is also realized in the mineral yavapaiite, KFe(SO$_4$)$_2$ \cite{BramwellCarling96, SerranoGonzalez98}. However, in this particular mineral the exchange coupling along one side of the triangles ($J^\prime$) is larger than along the two others ($J$) due to a monoclinic distortion, so that it actually represents a realization of the so-called row model \cite{Zhitomirsky96}. Nevertheless, instead of the theoretically expected helical spin structure, predicted for the row model with $J^\prime/J>1/2$ in zero magnetic field, yavapaiite realizes a more complex sine-wave modulated phase with the same wave vector as a result of substantial magnetic anisotropy. In contrast, its synthetic analogs RbFe(SO$_4$)$_2$ and CsFe(SO$_4$)$_2$ that crystallize in the trigonal structure (space group $P\overline{3}$) have equal magnetic bonds and were studied as realizations of classical-spin triangular lattices with the expected 120$^\circ$ helical spin structure \cite{InamiAjiro96, SerranoGonzalez98, SerranoGonzalez99, Inami07}.

\subsection{Devilline: distorted spin-$\frac{1}{2}$ triangular lattice}

\begin{figure}[b!]
\begin{center}
\includegraphics[width=\linewidth]{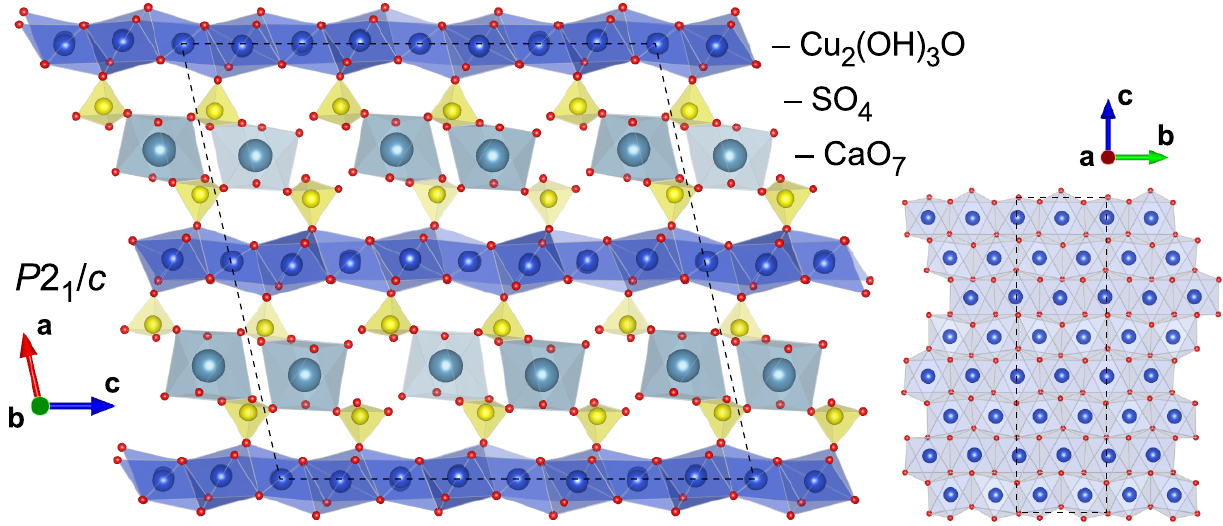}
\end{center}
\caption{The crystal structure of devilline, visualized according to the results of structural refinement by Sabelli \textit{et~al.}~\cite{SabelliZanazzi72} using \textsc{Vesta} software \cite{MommaIzumi11}. A fragment of the $^{~\,\,2}_\infty$[Cu$_2$(OH)$_3$O]$^-$ sheet, viewed along the $a$ axis, is shown at the right. The dashed lines outline the unit cell.}\label{Fig:Devilline}
\end{figure}

Among the quasi-2D layered compounds, the monoclinic copper sulphate-hydroxide mineral devilline, CaCu$_4$(SO$_4$)$_2$(OH)$_6$\,$\cdot$\,3H$_2$O \cite{SabelliZanazzi72}, represents a remarkable realization of somewhat distorted and corrugated triangular-lattice layers, well separated by nonmagnetic spacers. Each copper ion, in a 4+2 coordination, is linked by six edges to its neighbours, forming  $^{~2}_\infty$[Cu$_2$(OH)$_3$O]$^-$ sheets parallel to (100), as shown in Fig.~\ref{Fig:Devilline}. Adjacent sheets are connected by calcium ions in sevenfold coordination, by SO$_3^{2+}$ tetrahedra, and by a system of hydrogen bonds. The large unit cell of devilline contains eight inequivalent Cu sites, and the distortions within the sheets lead to a variation in the Cu--Cu distances ranging from 3.04 to 3.27~\AA. This mineral has a Pb analogue lautenthalite, PbCu$_4$(SO$_4$)$_2$(OH)$_6$\,$\cdot$\,3H$_2$O \cite{MedenbachGebert93}, along with several other close structural relatives: Cu$_5$(SO$_4$)$_2$(OH)$_6$\,$\cdot$\,4H$_2$O (kobyashevite) \cite{PekovZubkova13}, Cu$_4$Mn$^{2+}$(SO$_4$)$_2$(OH)$_6$\,$\cdot$\,4H$_2$O (campigliaite) \cite{Sabelli82}, Cu$_4$Cd(SO$_4$)$_2$(OH)$_6$\,$\cdot$\,4H$_2$O (niedermayrite) \cite{GiesterRieck98}, Cu$_4$Zn(SO$_4$)$_2$(OH)$_6$\,$\cdot$\,6H$_2$O (ktenasite) \cite{MelliniMerlino78}, and Cu$_4$Ca(SO$_4$)$_2$(OH)$_6$·3H$_2$O (serpierite) \cite{SabelliZanazzi68}. For a comparative analysis of the structure and crystal chemistry of sulfate minerals, an interested reader can refer to recent reviews \cite{HawthorneKrivovichev00, SchindlerHuminicki06}.

So far, experimental realizations of the quantum spin-$\frac{1}{2}$ triangular lattice have been proposed only in Ba$_3$CoSb$_2$O$_9$ \cite{ShirataTanaka12, ZhouXu12, SusukiKurita13, MaKamiya16} or in rare-earth compounds, such as YbMgGaO$_4$ \cite{LiLiao15, LiChen15, ShenLi16, XuZhang16, LiAdroja17prl, LiAdroja17, PaddisonDaum17, ZhuMaksimov17, ZhangMahmood18} or Yb-delafossites \cite{BaenitzSchlender18}, where due to a large crystal-field splitting, the ground-state Kramers doublet is well separated from higher-energy multiplets and can be considered as an effective $S=1/2$ state at sufficiently low temperatures. The high current interest in spin-$\frac{1}{2}$ triangular-lattice magnets is dictated by the reports proposing that it may support a quantum spin-liquid ground state \cite{ZhuWhite15, IaconisLiu18}. It is therefore highly desirable to find an implementation of the real spin-$\frac{1}{2}$ triangular lattice in cuprates, and devilline-group minerals can be a perfect starting point in this search. Some of them, e.g. niedermayrite or ktenasite, have more symmetric crystal structures, and the unavoidable distortions in the copper planes vary depending on the composition. It would be worthwhile to estimate numerically, to what extent the interaction network in these compounds is affected by the distortions of the triangular-lattice copper layers, and whether these distortions are sufficiently weak to preserve the underlying physics. Low-temperature magnetic characterization measurements on these minerals would also be highly desirable.

\begin{figure}[b]
\begin{center}
\includegraphics[width=0.5\linewidth]{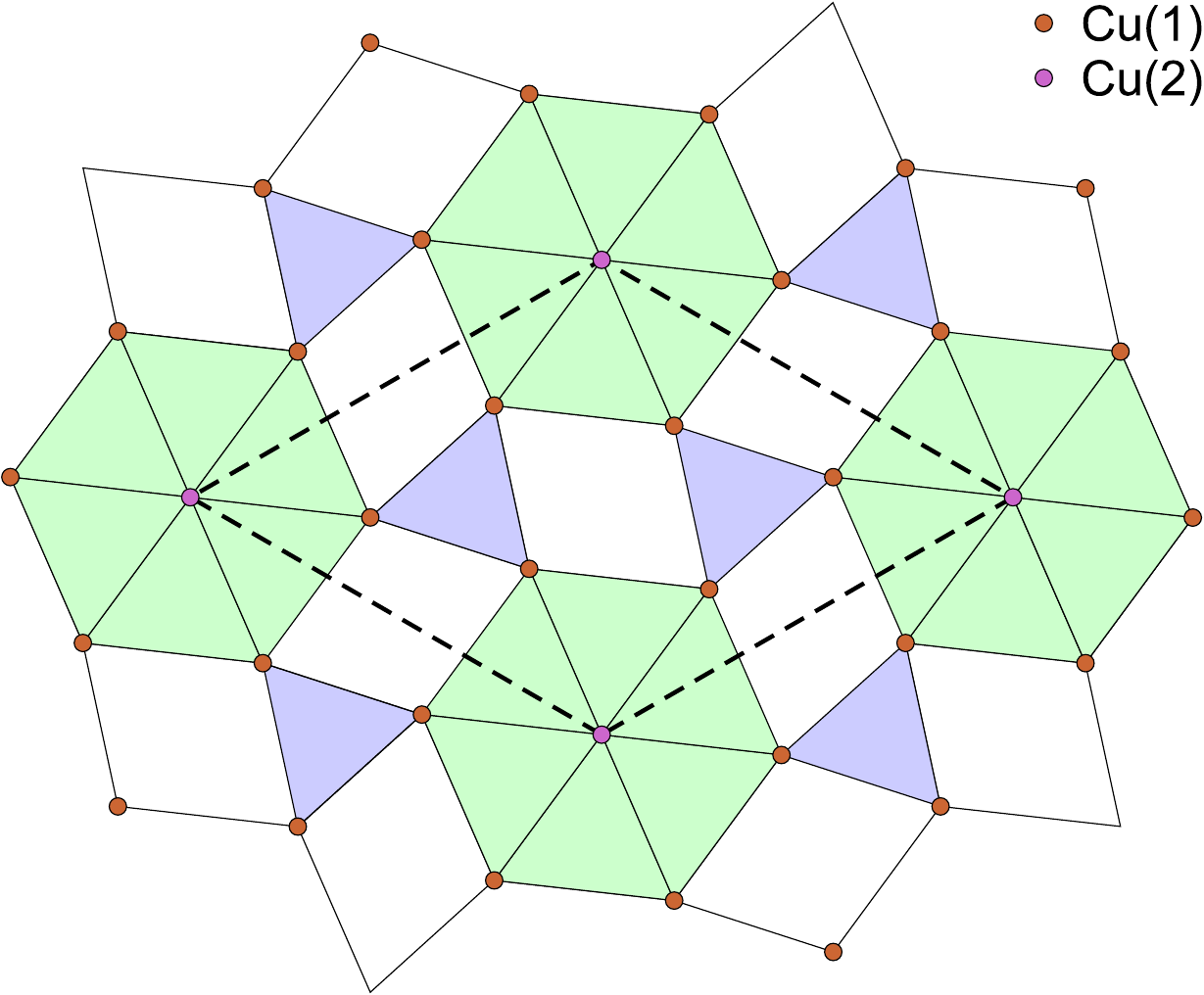}
\end{center}
\caption{The structure of magnetic copper layers in schulenbergite, viewed along the trigonal axis, according to the structural refinement of Mumme \textit{et~al.}~\cite{MummeSarp94}. Circles represent Cu atoms, two types of equilateral triangles are highlighted with colour. The dashed lines outline the unit cell.}\label{Fig:Schulenbergite}
\end{figure}

A very similar triangular-lattice layered structure has been reported in two polymorphs of the copper hydroxyl nitrate Cu$_3$(NO$_3$)(OH)$_3$, gerhardtite and rouaite. These compounds have been successfully synthesized \cite{YoderBushong10}, which should hopefully enable their low-temperature investigations in the nearest future. The hydrated version of the same compound, known as the mineral likasite, Cu$_3$(NO$_3$)(OH)$_5$\,$\cdot$\,2H$_2$O, can also represent interest as a realization of frustrated one-dimensional chains consisting of corner-sharing tetrahedra of Cu$^{2+}$ ions \cite{Effenberger86}.

Another highly symmetric layered structure with a triangular motif in the copper planes is realized in the trigonal sulphate mineral schulenbergite, Cu$_7$(SO$_4$)$_2$(OH)$_{10}$\,$\cdot$\,3(H$_2$O) \cite{MummeSarp94}. Here the layers represent strongly distorted triangular-lattice planes, in which every unit cell contains equilateral triangles of different sizes, which are expected to be the source of geometric frustration. Cu atoms occupy two inequivalent Wyckoff sites: Cu(1) is shared by two larger triangles and one smaller triangle, whereas Cu(2) sits in the centre of the hexagon formed by six larger triangles, as shown in Fig.~\ref{Fig:Schulenbergite}.

\subsection{Spangolite: a maple-leaf lattice antiferromagnet}\label{Sec:MapleLeaf}

\begin{figure}[b]
\begin{center}
\includegraphics[width=0.51\linewidth]{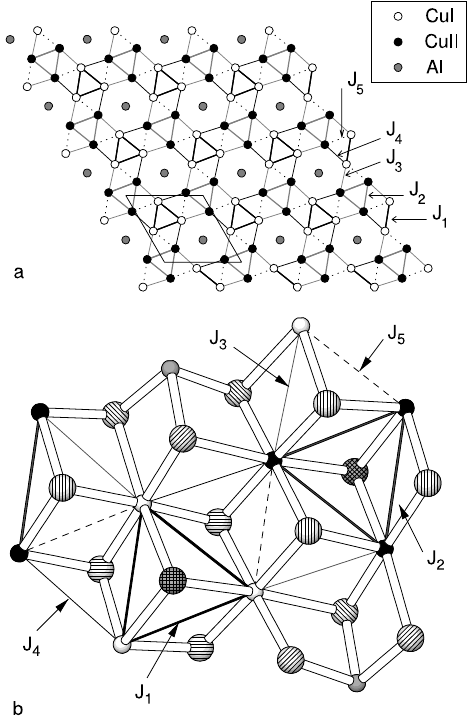}
\end{center}
\caption{(a)~The structure of distorted maple-leaf layers in spangolite. (b)~The five magnetic interactions acting within ($J_1$, $J_2$) and between ($J_3$, $J_4$, $J_5$) the trimers. After Fennell \textit{et~al.}~\cite{FennellPiatek11}.}\label{Fig:Spangolite}
\end{figure}

As already mentioned previously, there is a handful of naturally occurring copper minerals, some of them discovered only recently, that realize the spin-$\frac{1}{2}$ magnetic maple-leaf (T4) lattice \cite{Norman18}. Surprisingly, this complex lattice appears to be more common in the mineral world than its prototype triangular (T1) lattice. However, none of these minerals has been so far grown artificially in the form of single crystals, and the number of published low-temperature studies of their magnetic properties remains very limited. In 2011, Fennell \textit{et al.} \cite{FennellPiatek11} presented the results of a structural and basic magnetic characterization of the layered hydrated copper sulfate mineral spangolite, Cu$_6$Al(SO$_4$)(OH)$_{12}$Cl\,$\cdot$\,3H$_2$O, which is found in the form of finely grained thin turquoise plates. For the purpose of physical characterization, they chose a natural specimen from the Blanchard mine, Socorro, Colorado. The refinement of single-crystal x-ray diffraction data resulted in a structure with the trigonal space group $P3_1c$, consisting of well separated slightly distorted triangular-lattice layers in which 1/7 of the Cu$^{2+}$ ions are replaced with nonmagnetic Al$^{3+}$, in agreement with the earlier result of Hawthorne \textit{et al.} \cite{HawthorneKimata93}. The resulting magnetic superlattice is very close to a maple-leaf lattice with a small distortion that separates the magnetic ion positions into two structurally inequivalent sites, arranged in two sets of pure trimers as shown in Fig.~\ref{Fig:Spangolite}.

The application of GKA rules to the spangolite structure suggests that the superexchange interactions $J_1$ and $J_2$ within both trimers must be antiferromagnetic, resulting in a strong frustration \cite{FennellPiatek11}. The trimers are coupled with AFM interactions $J_3$, $J_4$, and a FM interaction $J_5$ along different exchange paths shown in Fig.~\ref{Fig:Spangolite}\,(b).

Measurements of the magnetic susceptibility show a broad maximum around 50~K that is typical of a spin-dimer system. No signs of magnetic ordering were seen down to 0.1~K, despite the high Curie-Weiss temperature of $\Theta_{\rm CW}=-38$~K. The effective magnetic moment obtained from the susceptibility was 4.55~$\mu_{\rm B}/{\rm f.u.}$, that is significantly reduced from the anticipated value of 10.39~$\mu_{\rm B}/{\rm f.u.}$ for one formula unit with six Cu$^{2+}$ spins. From the low-temperature Curie tail, a $\sim$\,7.5\% population of defective spins in the natural spangolite specimen could be estimated. The existing experimental evidence strongly suggests that spangolite has a singlet ground state, yet the microscopic details of its formation still await to be clarified.

The presence of two inequivalent sublattices in spangolite certainly represents an unnecessary complication that may prevent this mineral from playing the desired role of the model system for the realization of theoretically proposed models on the perfect maple-leaf lattice \cite{SchmalfusTomczak02, FarnellDarradi11}. It has been noted \cite{FennellPiatek11} that the emerald-green mineral sabelliite, Cu$_2$ZnAsO$_4$(OH)$_3$, with a single-sublattice maple-leaf lattice of Cu$^{2+}$ ions \cite{OlmiSabelli95} could be a better candidate for this role. Other minerals with the same structural motif are fuettererite \cite{KampfMills13}, bluebellite, and mojaveite \cite{MillsKampf14}, whose detailed magnetic characterization would be definitely desirable.\vspace{-3pt}

\subsection{Diaboleite: a spin-$\frac{1}{2}$ square lattice without frustration}

The tetragonal mineral diaboleite, Pb$_2$Cu(OH)$_4$Cl$_2$ \cite{CooperHawthorne95}, forms a defect perovskite-related structure of quasi-2D Cu$^{2+}$ square-lattice planes with the dominant AFM coupling, $J_1\approx 3.3$~meV, on the NN bonds and a much weaker NNN coupling, $J_2\approx0.04$~meV \cite{TsirlinJanson13}. The layers are weakly interacting via two interlayer exchange constants, $J_\perp\approx0.086$~meV and $J_\perp^\prime\approx0.034$~meV, which stabilize an AFM-ordered state below $T_{\rm N}\approx11$~K. The material is therefore analogous to the layered perovskite PbVO$_3$, which is also characterized by AFM interactions, yet with a much larger $J_2/J_1$ ratio of $\sim$0.35, close to the critical region of the $J_1$-$J_2$ frustrated square lattice \cite{TsirlinBelik08, NathTsirlin08}. In contrast, the negligibly small $J_2$ in diaboleite essentially deprives it of magnetic frustration.

\begin{figure}[t]
\begin{center}
\includegraphics[width=0.7\linewidth]{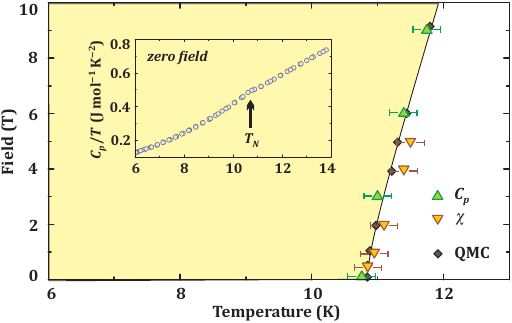}
\end{center}
\caption{Magnetic-field dependence of~the N\'eel temperature in diaboleite from QMC simulations, specific heat, and magnetic susceptibility measurements. After Tsirlin \textit{et~al.}~\cite{TsirlinJanson13}.\vspace{-2pt}}\label{Fig:Diaboleite}
\end{figure}

Application of high magnetic fields leads to a monotonic increase in the ordering temperature, as shown in Fig.~\ref{Fig:Diaboleite}, which is typical among square-lattice antiferromagnets \cite{NathTsirlin08, SenguptaBatista09, TsirlinNath11}. A recent theoretical study \cite{LiWu17} also estimated the $g$ factors of Pb$_2$Cu(OH)$_4$Cl$_2$ for fields parallel and perpendicular to the planes, which showed a sizeable anisotropy in the $g$ factor of about 14\%.

In natural diaboleite crystals, powder x-ray diffraction and chemical analysis revealed the presence of stacking faults and about 5\,--\,10\% of copper vacancies~\cite{TsirlinJanson13}. They lead to local structural distortions but do not disrupt the couplings between the remaining Cu sites. While the magnetic transition remains sharp, the vacancies influence the low-temperature behaviour of the magnetic susceptibility below the broad hump seen around 30~K.

As one can see, in spite of the model character of diaboleite that serves as an example of the textbook square-lattice model with just a single dominant AFM interaction $J_1$, the low-temperature physical measurements on this mineral remain limited. One can therefore expect that it will draw more interest in future as a subject for more detailed studies using neutron spectroscopy and local magnetic probes.

\subsection{Shattuckite: a combination of CuO$_2$ planes and ribbons}

Another layered mineral that fell under recent scrutiny is shattuckite, Cu$_5$(SiO$_3$)$_4$(OH)$_2$~\cite{Kawahara76}. It forms blue translucent crystals with an orthorhombic structure (space group \textit{Pcab}) that contains corrugated CuO$_2$ planes separated by twisted CuO$_2$ ribbons. This combination of 2D and 1D structural motifs results in a complex weakly frustrated network of magnetic interactions. Below the magnetic ordering temperature, $T_{\rm N}=7$~K, an AFM order with a weak FM canting has been reported. The spin canting results in hysteresis loops in the magnetization with a small spontaneous magnetization of 0.075 $\mu_{\rm B}/{\rm f.u.}$ \cite{KoshelevZvereva16}. At higher magnetic fields, no evidence for a spin-flop transition was evidenced up to 9~T. The AFM transition can be also seen in specific heat as a rounded maximum, which broadens and shifts to slightly higher temperatures in magnetic field.

\begin{figure}[b]
\begin{center}
\includegraphics[width=\linewidth]{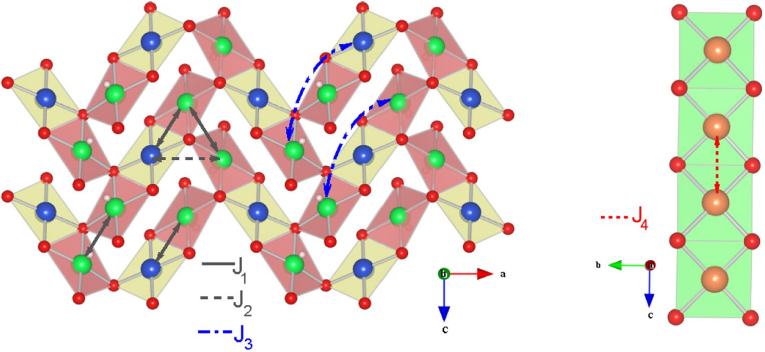}
\end{center}
\caption{Exchange paths for magnetic interactions within the CuO$_2$ square planes (left) and twisted CuO$_2$ ribbons (right) in shattuckite. After Koshelev \textit{et~al.}~\cite{KoshelevZvereva16}.}\label{Fig:Shattuckite}
\end{figure}

From the temperature dependence of the magnetization in the paramagnetic state, the Curie-Weiss temperature of $-13.5$~K and the Curie constant of 2.23~emu\,K/mol have been estimated \cite{KoshelevZvereva16}. An electron paramagnetic resonance characteristic of Cu$^{2+}$ ions with a weakly anisotropic $g$-factor, $(g_\parallel-g_\perp)/g_\perp\approx0.06$, was observed on a powder sample obtained by grinding a natural shattuckite mineral from Kaokoveld Plateau, Namibia \cite{KoshelevZvereva16}. A higher value of the $g$-factor anisotropy, $(g_\parallel-g_\perp)/g_\perp\approx0.18$, was reported in an earlier study on a sample from Ajo, Arizona, USA \cite{SarmaReddy82}. These results suggest that the ground state of the Cu$^{2+}$ ion corresponds to the unpaired $d_{x^2-y^2}$ orbital in a tetragonal crystal field imposed by the distorted square planar environment of oxygen ligands. The experimental crystal-field parameters calculated from the positions of optical absorption bands can be found in Ref.~\cite{SarmaReddy82}, where they are compared to those of azurite, malachite, chalcanthite, and plancheite. The optical and ESR spectra of shattuckite have been later calculated theoretically \cite{Feng12}, yielding a satisfactory match to the experimentally measured $g$-factors and optical absorption transitions.

To estimate the relevant magnetic interactions in shattuckite, A.\,V.~Koshelev \textit{et al.} performed electronic-structure calculations in the framework of a self-consistent spin-polarized DFT \cite{KoshelevZvereva16}. Their spin model is illustrated in Fig.~\ref{Fig:Shattuckite}. There are three structurally inequivalent magnetic Cu sites in the crystal structure, of which Cu(1) and Cu(2) form a corrugated layer of edge-sharing and corner-sharing CuO$_4$ plaquettes, while Cu(3) ions form edge-sharing ribbons running along the $\mathbf{c}$ direction that are connected by nonmagnetic SiO$_4$ tetrahedra along the $b$ axis. All dominant exchange interactions within the planes are reportedly antiferromagnetic: $J_1\approx14$~meV, $J_2\approx6$~meV, and $J_3\approx3$~meV. On the other hand, the NN interaction $J_4\approx-1$~meV within the ribbons is ferromagnetic and weak, comparable with the $J_5\approx-1$~meV coupling between the chains and the layers. However, the mutual cancelation of $J_5$ interactions due to symmetry considerations virtually decouples the chains from the layers. Similarly, the interchain Cu(3)\,--\,Cu(3) interaction was estimated to be very small (beyond the accuracy limit of DFT) but could tentatively stabilize a long-range AFM order on the Cu(3) sublattice. The presented hierarchy of exchange parameters allows us to classify the shattuckite mineral as a quasi-2D magnetic system with weak frustration. The competing $J_2$ interaction between Cu(1) and Cu(2) ions in the neighbouring corner-sharing plaquettes and the further-neighbour interaction $J_3$ are too weak to introduce significant frustration to the system. As a result, the AFM structure resulting from these interactions alone would remain collinear. The most likely origin of the weak ferromagnetism at $T<T_{\rm N}$ is due to the DMI along the Cu(1)\,--\,O\,--\,Cu(1) and Cu(1)\,--\,O\,--\,Cu(2) pathways within the planes that are allowed by symmetry and may lead to a small spin canting, according to Ref.~\cite{KoshelevZvereva16}.

\subsection{Langite and wroewolfeite: coexistence of magnetic order and quantum disorder}

There are two naturally occurring polymorphs of the layered hydrated copper hydroxyl sulphate mineral Cu$_4$(OH)$_6$SO$_4\cdot2$H$_2$O: langite (space group $Pc$) \cite{GalyJaud84, GentschWeber84} and wroewolfeite (space group $Pm$) \cite{DunnRouse75, HawthorneGroat85}. Both consist of weakly coupled two-dimensional magnetic layers composed of alternating edge- and corner-sharing copper chains, separated by sulphate groups and water molecules. The structure is therefore similar to that of brochantite but with a larger interlayer separation that determines the quasi-2D magnetic behaviour. The same structural layer units are also found in the mineral posnjakite, Cu$_4$(OH)$_6$SO$_4$\,$\cdot$\,H$_2$O \cite{MelliniMerlino79}, to which wroewolfeite is converted upon loss of water. Wroewolfeite has been synthesized artificially \cite{DabinettHumberstone08}, but its magnetic properties were not yet systematically investigated. This may be partly due to the metastability of both posnjakite and wroewolfeite phases, which readily lose water and are converted to brochantite under unfavourable conditions \cite{DabinettHumberstone08, ZittlauShi13}.

\begin{figure}[b]
\begin{center}
\includegraphics[width=\linewidth]{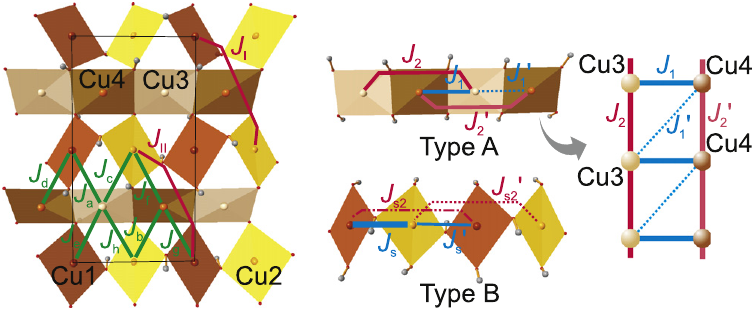}
\end{center}
\caption{The 2D network of magnetic interactions in langite. After Lebernegg \textit{et~al.}~\cite{LeberneggTsirlin16}.}\label{Fig:Langite}
\end{figure}

Meanwhile, the langite structure has been recently addressed by Lebernegg \textit{et al.} \cite{LeberneggTsirlin16}. Its magnetic layers contain four structurally inequivalent Cu sites, of which Cu(1) and Cu(2) form slightly buckled corner-sharing CuO$_4$ chains, whereas Cu(3) and Cu(4) form linear edge-sharing CuO$_4$ chains. Langite orders antiferromagnetically below a N\'eel temperature of $T_{\rm N}\approx5.7$~K as a result of a complex frustrated network of exchange interactions shown in Fig.~\ref{Fig:Langite}. On the one hand, the edge-sharing (type~A) chains are formed by four AFM interactions: the NN couplings $J_1\approx3.3$~meV and $J_1^\prime\approx0.8$~meV between Cu(3) and Cu(4) ions and two nearly equal NNN interactions $J_2\approx J_2^\prime\approx3.1$~meV on the Cu(3)--Cu(3) and Cu(4)--Cu(4) bonds. This makes this chain equivalent to an AFM ladder model with legs formed by Cu(3) and Cu(4) ions, connected by $J_1$ rungs and a much weaker diagonal exchange interaction $J_1^\prime$, as shown schematically in Fig.~\ref{Fig:Langite} (right panel). On the other hand, the corner-sharing (type~B) chains are characterized by FM interactions $J_{\rm s}\approx-6.4$~meV and $J_{\rm s}^\prime\approx-2.0$~meV. In addition, there are multiple interchain interactions $J_{\rm a}\ldots\,J_{\rm h}$ with different signs, so that the chains are not independent but form part of a complex 2D magnetic network. Specific-heat and magnetization data indicate that the spin lattice of langite splits into two sublattices with predominantly FM and AFM couplings at low temperatures \cite{LeberneggTsirlin16}. The former develops long-range magnetic order with a saturation field of $\sim$\,12~T, whereas the latter evades long-range magnetic order due to frustration. Lebernegg \textit{et al.} argue that this separation into two magnetic sublattices must be generic for magnetic minerals that combine edge- and corner-sharing CuO$_4$ chains in their crystal structure.

\section{Three-dimensional frustrated lattices}

\subsection{Magnetic frustration on 3D crystal lattices}

While the majority of 3D spin systems tend to develop magnetic order at sufficiently low temperatures, signatures of magnetic frustration can be seen in the unusual behaviour of their spin fluctuations, which are expected to be stronger than in the case of conventional ferro- or antiferromagnets. A separate matter of \mbox{present-day} interest in quantum magnetism is the search for magnetic systems that would display magnon fractionalization, resembling the emergence of spinon excitations in spin chains but in a 3D network of magnetic interactions \cite{ZhouKanoda17, CastelnovoMoessner12, KimchiAnalytis14,  HermannsOBrien15, OBrienHermanns16, WanKim16, ChillalIqbal18}. One example of an actively studied 3D spin-liquid candidate with fractionalized excitations, PbCuTe$_2$O$_6$ \cite{KoteswararaoKumar14, KhuntiaBert16, ChillalIqbal18}, mimics the hyperkagome structure of the natural mineral choloalite \cite{PowellThomas94}.

\begin{figure}[b]
\begin{center}
\includegraphics[width=0.52\linewidth]{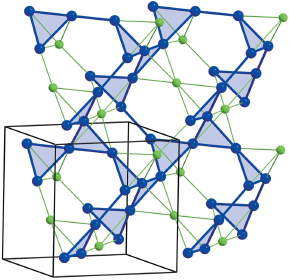}
\end{center}
\caption{The hyperkagome lattice (thick blue lines) and its relationship to the pyrochlore lattice (thin green lines). The cubic unit cell is shown in black. After Talanov~\textit{et~al.}~\cite{TalanovShirokov15}.}\label{Fig:Hyperkagome}
\end{figure}

In three dimensions, the pyrochlore lattice plays a similar role to the triangular lattice in 2D, serving as the emblem of geometric frustration for AFM interactions between the nearest neighbours \cite{GaulinGardner11, GaulinGardner13}. It can be viewed as an arrangement of corner-sharing tetrahedra or as an alternating sequence of stacked kagome and triangular layers. It is realized, in particular, in materials with the cubic pyrochlore or spinel ($A{\kern1.5pt}B_2{\kern.5pt}X_4$) crystal structure, if the magnetic ion occupies the $B$ site \cite{TakagiNiitaka11}. In naturally occurring minerals, distorted pyrochlore lattices are realized, for instance, in atacamite and melanothallite, which will be discussed in the following. For isotropic Heisenberg interactions between the spins, magnetic frustration can arise either due to the AFM nearest-neighbour exchange coupling or because of the competition of nearest- and further-nearest-neighbour interactions. Inclusion of multiple exchange terms in the Hamiltonian may stabilize various ordered ground states, such as ferro- or antiferromagnetism, single- or multi-$\mathbf{q}$ spin spirals, nematic order, or some more exotic phases \cite{ReimersBerlinsky91, OkuboNguyen11, TymoshenkoOnykiienko17}. In the presence of considerable single-ion anisotropy, noncollinear 2-in-2-out, 3-in-1-out or all-in-all-out spin configurations can be realized in every tetrahedron, resulting in AFM or spin-ice ground states \cite{BramwellGingras01, BramwellHarris01, BramwellGiblin09, Gingras11, CastelnovoMoessner12}, among other possibilities. This field of research is definitely too broad and goes beyond the scope of the present review.

A regular 1/4 site depletion of the pyrochlore lattice can lead to the formation of the 3D hyperkagome, apart from 2D kagome or kagome-staircase lattices \cite{Henley09, MendelsBert16}. The hyperkagome lattice and its relationship to the parent pyrochlore lattice are illustrated in Fig.~\ref{Fig:Hyperkagome}. The spin-$\frac{1}{2}$ AFM nearest-neighbour Heisenberg model on the hyperkagome lattice has been studied by various authors \cite{HopkinsonIsakov07, LawlerParamekanti08, ZhouLee08, LawlerKee08, BergholtzLaeuchli10, SinghOitmaa12}, who showed that it hosts a valence-bond crystal state due to geometric frustration. Inclusion of various anisotropic interactions or weak further-neighbour coupling terms lifts the massive degeneracy of the ground state and stabilizes various noncollinear and noncoplanar magnetic phases that compete in the phase diagram \cite{ChenBalents08, KimchiVishwanath14, Shindou16, MizoguchiHwang16, BuessenTrebst16, HuangKim17}. The hyperkagome lattice of classical spins is realized in Fe and Mn garnets, some of which are known as minerals, e.g. Mn$_3$Al$_2$(SiO$_4$)$_3$ (spessartine) \cite{Prandl73, GeigerArmbruster97, HarenWoensdregt01, LauKlimczuk09} and Fe$_3$Al$_2$(SiO$_4$)$_3$ (almandine) \cite{AnovitzEssene93, DachsGeiger12}. Both compounds develop long-range AFM order at low temperatures, but their magnetic structures and spin-dynamical properties have not been investigated in detail to the best of my knowledge.

In many other 3D crystal structures, even in the absence of structural motifs supporting geometric frustration, bond frustration can still be present as a result of competing interactions. Moreover, frustration may arise due to anisotropic compass-type interactions \cite{NussinovBrink15}, prominently manifested in the Kitaev-Heisenberg models \cite{Kitaev06, ChaloupkaJackeli10, ReutherThomale11} that were recently generalized to various tricoordinated 3D lattices \cite{LeeSchaffer14, HermannsTrebst14, LeeKim15, HalaszPerreault17}.

\subsection{Atacamite and other polymorphs of Cu$_2$(OH)$_3$Cl: distorted pyrochlore structure}\label{Sec:Atacamite}

\begin{figure}[t]
\begin{center}
\includegraphics[width=\linewidth]{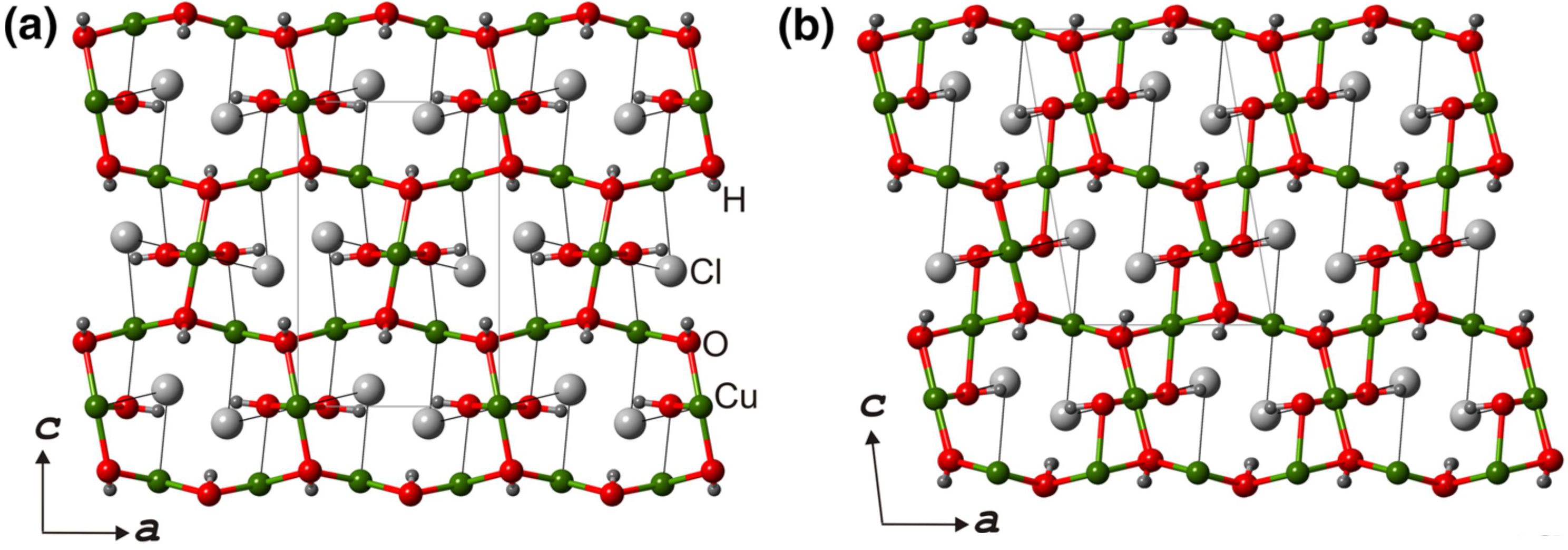}
\end{center}
\caption{Comparison of the crystal structures of atacamite (a) and clinoatacamite (b), viewed along the $b$ axis. After Krivovichev \textit{et~al.}~\cite{KrivovichevHawthorne17}.\vspace{-1pt}}\label{Fig:Atacamite}
\end{figure}

Copper hydroxychloride, Cu$_2$(OH)$_3$Cl, has multiple polymorphs and plays a central role in the field of quantum magnetism as the parent compound of the famous kagome antiferromagnet herbertsmithite, ZnCu$_3$(OH)$_6$Cl$_2$, and its relatives \cite{PuphalZoch18}. These compounds, in which Zn replaces one quarter of the Cu atoms to stabilize the rhombohedral structure featuring kagome planes \cite{Norman16}, were discussed in Chapter~\ref{Chap:Kagome}. Multiple polymorphic crystal forms of Cu$_2$(OH)$_3$Cl are known, including atacamite (orthorhombic, space group $Pnma$), paratacamite (rhombohedral, space group $R\overline{3}$), clinoatacamite, and botallackite (both monoclinic, space group $P2_1/n$). The structure of paratacamite was shown to be Zn-stabilized \cite{BraithwaiteMereiter04}, so in fact it does not belong to the sequence of pure copper-hydroxychloride phases according to the present consensus. The other three polymorphs form the so-called Ostwald cascade of metastable phases: botallackite--atacamite--clinoatacamite \cite{KrivovichevHawthorne17}, whose relative stability is now well understood \cite{PollardThomas89, MalcherekMihailova17}. Botallackite, or $\alpha$-Cu$_2$(OH)$_3$Cl, represents an ``ephemeral'' phase that crystallizes first under most conditions but then quickly recrystallizes to the more stable polymorphs unless the solutions responsible for its crystallization are removed or dried out. It therefore represents the rarest of the naturally occurring Cu$_2$(OH)$_3$Cl minerals. Atacamite, or $\beta$-Cu$_2$(OH)$_3$Cl, is a much more common metastable phase, which has been known as a mineral since the 18$^{\rm th}$ century. Finally, the $\gamma$-polymorph clinoatacamite, discovered only in 1996 by Jambor \textit{et al.} \cite{JamborDutrizac96}, is the thermodynamically stable phase at ambient temperatures. The crystal structures of these two phases are similar, as illustrated in Fig.~\ref{Fig:Atacamite}. In some of the earlier works, including Ref.~\cite{PollardThomas89}, before clinoatacamite became known as a separate mineral species, it was sometimes confused with paratacamite \cite{KrivovichevHawthorne17}.

As the most common and well known mineral, atacamite attracted considerable attention of the solid-state physics community, yet its magnetic properties are still a subject of debate. Its crystal structure contains two crystallographically independent Cu atoms in the Jahn-Teller distorted octahedral coordination. The first one is bonded to two Cl atoms, forming a Cu(OH)$_4$Cl$_2$ octahedron, while the other one is bonded to a single Cl atom to form a Cu(OH)$_5$Cl octahedron. This difference leads to a highly distorted pyrochlore structure that can be alternatively viewed as a set of Cu--O--Cu chains connected by CuO ladders \cite{ZhengMori05}. It has been argued based on first-principles DFT calculations that the magnetism in both atacamite and botallackite is quite similar and could be described by an effective quasi-1D uniform AFM chain model with spin frustration arising from the NNN interactions \cite{KangLee09}. The similarity of these two polymorphs is also indicated by the similar N\'eel temperatures of 9.0 and 7.2~K, respectively \cite{ZhengMori05}. On the other hand, the experimental situation with respect to the low-temperature magnetic properties of these minerals remains controversial. Synthetic atacamite was previously reported to exhibit spin-glass behaviour \cite{MoriYamaguchi95}, whereas AFM behaviour was observed in both natural and synthetic samples in more recent studies \cite{ZhengOtabe04, ZhengMori05, ZenmyoKubo13}. In particular, a proton NMR study found no evidence of the disordering or spin-glass moments but rather suggested an ``all-in all-out'' type of long-range order~\cite{ZenmyoKubo13}. Contrasting results were obtained from $\mu$SR measurements, which pointed towards a disordered ground state in atacamite and a long-range ordered state in botallackite \cite{ZhengMori05}. On the other hand, a very recent neutron-diffraction study on natural atacamite single crystals originating from the Poona Mine in Australia revealed long-range magnetic order described by the propagation vector $\mathbf{q}=\left(\frac{1}{2}\,0\,\frac{1}{2}\right)$ \cite{HeinzeBeltranRodriguez18}. However, the magnetic structure could not be solved unambiguously, and it is still unclear if the magnetic structure proposed by Zenmyo \textit{et~al.} \cite{ZenmyoKubo13} is consistent with the observed ordering vector. The final conclusion about the magnetic ground state of atacamite will therefore remain open to debate until a conclusive magnetic structure determination by means of neutron diffraction on pure synthetic crystals becomes available.

Interestingly, a similar controversy also exists about the spin-1 analogue of atacamite, the synthetic frustrated antiferromagnet Ni$_2$(OH)$_3$Cl with a distorted-pyrochlore structure \cite{ZhengHagihala09, MaegawaOyamada10}. It exhibits a clear AFM transition at $T_{\rm N}=4$~K seen both in magnetic susceptibility and specific heat, and a long-range magnetic order was evidenced by neutron diffraction. Nevertheless, no signature of an order was detected by $\mu$SR \cite{ZhengHagihala09}. Proton NMR revealed a significant broadening of the spectra below $T_{\rm N}$, signalling the development of static internal fields at the $^1$H sites, yet a fraction of frozen spins with slow dynamics was found to coexist with the paramagnetic state up to 20~K \cite{MaegawaOyamada10}. Other hydroxyhalides of transition metals, in particular $\alpha$-Cu$_2$(OH)$_3$Br and $\alpha$-Cu$_2$(OH)$_3$I, have also been synthesized \cite{ZhengYamashita09}. Similar to botallackite, which orders magnetically below $T_{\rm N}\approx7.2$~K, its Br- and I-analogs exhibit long-range AFM order with $T_{\rm N}\approx10$ and 14~K, respectively.

Finally, magnetic characterization of clinoatacamite \cite{ZhengKawae05, ZhengKubozono05} and some other compounds of the $M_2$(OH)$_3$Cl ($M$\,=\,Mn,\,Fe,\,Co,\,Ni) transition-metal series \cite{ZhengKawae06, ZhengHagihala07, HagihalaZheng07} revealed a co-existence of long-range magnetic order with disordered or fluctuating moments. Results of the magnetization, specific-heat, and $\mu$SR experiments have shown that the magnetic order, which occurs in clinoatacamite below $T_{\rm c1}\approx18.1$~K, is responsible for a surprisingly small entropy release of only $0.05R\ln2$/Cu \cite{ZhengKawae05, ZhengKubozono05}. Then, another abrupt transition into a metastable spin-glass-like state occurs at $T_{\rm c2}\approx6.5$~K, accompanied by a large specific-heat anomaly. After that, partial long-range order coexists with fluctuating or disordered moments due to strong frustration, causing a partial depolarization of muon spins down to 20~mK. This kind of coexistence is seen as a common feature of the $M_2$(OH)$_3$Cl compounds, as it also occurs likewise in the case of antiferromagnetic Fe$_2$(OH)$_3$Cl and ferromagnetic Co$_2$(OH)$_3$Cl \cite{ZhengKawae06, ZhengHagihala07}. On the other hand, such coexistence is absent in the isostructural hydroxybromides~\cite{HagihalaZheng07}.

\begin{figure}[b!]
\begin{center}\vspace{-0.5pt}
\includegraphics[width=0.5\linewidth]{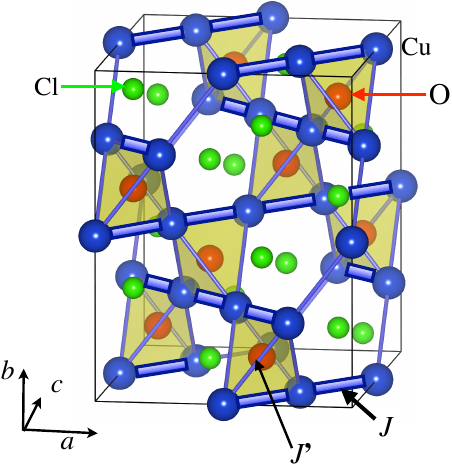}\vspace{-3pt}
\end{center}
\caption{The crystal structure of melanothallite with two dominant magnetic interactions. After Nishiyama \textit{et~al.}~\cite{NishiyamaOyamada11}.}\label{Fig:Melanothallite}
\end{figure}

More recently, the nature of the successive magnetic transitions in $\gamma$-Cu$_2$(OH)$_3$Cl was readdressed by Raman spectroscopy \cite{LiuZheng13}. Both transitions at $T_{\rm c1}$ and $T_{\rm c2}$ have prominent signatures in the Raman response as a result of the spin-phonon coupling. Moreover, the new data point towards the existence of spin fluctuations in the intermediate phase on a picosecond time scale, which falls out of the $\mu$SR time window. These fluctuations give rise to a broad continuum in the Raman spectra, providing a reasonable explanation for the previously reported small entropy release across the upper transition. The authors conclude by stating that future NMR and single-crystal neutron studies should further clarify the magnetic structure and the nature of the partial order in clinoatacamite \cite{LiuZheng13}.

\subsection{Melanothallite: a pyrochlore-lattice antiferromagnet with multiferroic properties}

Another cuprate mineral with a pyrochlore-like structure of the magnetic sublattice is the oxychloride Cu$_2$OCl$_2$, known as melanothallite \cite{KrivovichevFilatov02}. It crystallizes in an orthorhombic structure with the space group \textit{Fddd}, in which Cu$^{2+}$ occupies a single Wyckoff position, octahedrally coordinated by oxygen and chlorine ions. The copper ions form a 3D network of corner-sharing distorted tetrahedra that can be seen as a deformed pyrochlore lattice, shown in Fig.~\ref{Fig:Melanothallite}. This mineral first came to attention because of the anomalous negative thermal expansion along the $b$ axis above room temperature \cite{KrivovichevFilatov02}. The first low-temperature study revealed an AFM transition in specific heat and magnetic susceptibility at $T_{\rm N}\approx70$~K and some signatures of a low-dimensional behaviour at higher temperatures \cite{OkabeSuzuki06, KawashimaOkabe07}. Immediately afterwards, a $\mu$SR study was reported \cite{KawashimaOkabe07}, in which clear oscillations of the muon polarization were observed below $T_{\rm N}$, evidencing a long-range magnetically ordered state. However, the $\mu$SR spectra showed an excessive damping that had been initially attributed to the effects of frustration.

The first attempt to determine the spin structure in the ordered state of Cu$_2$OCl$_2$ was based on Cu-NMR measurements. The study of Nishiyama \textit{et al.} \cite{NishiyamaOyamada11} suggested that melanothallite realizes a commensurate all-in-all-out type of magnetic ordering. This conclusion follows from rather indirect evidence, namely from the comparison of the electric field gradient estimated from the measured quadrupolar-resonance frequency with that calculated within a point-charge model, combined with the knowledge about the AFM nature of the NN interaction, resulting from a susceptibility measurement. Moreover, the spin-lattice relaxation rates, measured with $^{35}$Cl- and $^{63}$Cu-NMR below $T_{\rm N}$, are proportional to the first power of temperature, $T_1^{-1}\propto T$, in contrast to the power-law behaviour $T^\alpha$ with $\alpha\geq2$ that is expected for a conventional ordered magnet with spin-wave excitations. This indicates that large spin fluctuations due to geometric frustration must be present in the system \cite{NishiyamaOyamada11}.

More recent direct measurements of the spin structure in melanothallite by neutron diffraction \cite{ZhaoFernandezDiaz16} have discarded the assumption about the commensurate all-in-all-out ground state, proclaiming instead an incommensurate spiral-like phase with a propagation vector $(0.827~0~0)$. As in many other helimagnetic materials \cite{TokuraSeki10, TokuraSeki14}, this type of incommensurate noncollinear order leads to multiferroic properties \cite{ZhaoFernandezDiaz16}. Thus, synthetic melanothallite exhibits ferroelectricity in measurements of the dielectric constant and pyroelectric polarization as functions of temperature. It therefore represents the first known transition-metal oxyhalide with a multiferroic behaviour that persists up to relatively high temperatures.

A similar lattice with a structure approximating the pyrochlore lattice of $\frac{1}{2}$-spins is also found in the mineral paramelaconite, Cu$_4$O$_3$. Alternatively, its structure can be seen as alternating chains of edge-shared CuO$_4$ plaquettes. In contrast to Cu$_2$OCl$_2$, this compound has a more pronounced difference between the interchain and intrachain interactions, which reduces the frustration and places it closer to a low-dimensional spin-chain antiferromagnet \cite{PinsardGaudart04, OkabeSuzuki06}. Paramelaconite develops long-range AFM order around 40~K, which has been described as a commensurate noncollinear spin structure with a reduced ordered moment of $\sim$\,0.46\,$\mu_{\rm B}$ and the propagation vector $\mathbf{q}=\left(\frac{1}{2}\,\frac{1}{2}\,\frac{1}{2}\right)$, which is rather unusual for a pyrochlore lattice \cite{PinsardGaudart04}.\vspace{-2pt}

\subsection{Choloalite: a 3D quantum spin liquid on a hyper-hyperkagome lattice}\label{Sec:Choloalite}

\begin{figure}[b]
\begin{center}
\includegraphics[width=0.64\linewidth]{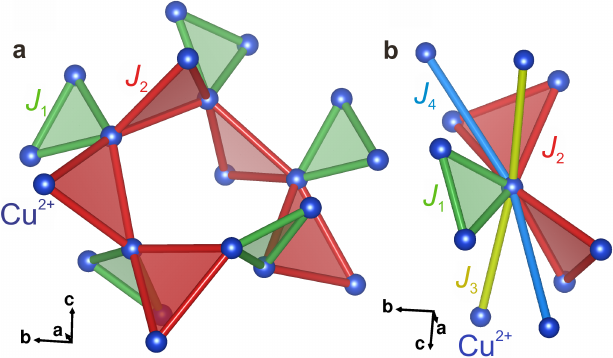}
\end{center}
\caption{Magnetic sublattice and the model of frustrated magnetic interactions in PbCuTe$_2$O$_6$. After Chillal \textit{et~al.} \cite{ChillalIqbal18}.}\label{Fig:Choloalite}
\end{figure}

A rich variety of minerals is represented by tellurium oxycompounds, some of them realizing unique crystal structures, as summarized in a recent review article by Christy, Mills, and Kampf \cite{ChristyMills16}. Among them, the mineral choloalite, PbCuTe$_2$O$_6$ \cite{PowellThomas94, LamGroat99}, has a cubic structure with the chiral space group $P4_132$. Its magnetic sublattice of Cu$^{2+}$ ions represents a complex 3D network consisting of two kinds of corner-sharing triangles with different Cu--Cu bond lengths of 4.37 and 5.60~\AA. The larger triangles form a hyperkagome lattice, to which individual smaller triangles are attached, as shown in Fig.~\ref{Fig:Choloalite}. Several years ago, is was realized that this mineral could represent a new candidate compound for hosting a 3D quantum spin-liquid state, as it does not order magnetically down to subkelvin temperatures \cite{KoteswararaoKumar14, KhuntiaBert16}, unlike its sibling compound SrCuTe$_2$O$_6$ that shows two magnetic transitions at $T_{\rm N1}\approx5.5$~K and $T_{\rm N2}\approx4.5$~K \cite{AhmedTsirlin15, KoteswararaoPanda15}. The magnetic specific heat of high-purity synthetic choloalite shows a broad maximum at 1.1~K, followed by a weak kink at 0.87~K of unknown origin \cite{KoteswararaoKumar14}.

The dominant AFM interactions among the Cu$^{2+}$ ions generate a strongly frustrated hyper-hyperkagome network of $S=1/2$ spins, but the magnetic connectivity is additionally enhanced by weaker interactions that partially relieve this frustration. Through a combination of magnetization measurements with NMR and $\mu$SR, it was shown that spin fluctuations persist down to 20~mK without spin freezing \cite{KhuntiaBert16}. Further, the sublinear power-law dependence of the spin-lattice relaxation rate, $T_1^{-1}(T)$, evidences a nonsinglet ground state with a gapless excitation spectrum. From the NMR perspective, anomalies that were observed around 1~K can be related to the dramatic slowing down of the spin dynamics, possibly arising from some kind of instability of the spinon Fermi surface \cite{KhuntiaBert16}.

Inelastic neutron scattering measurements, performed on both powder and single-crystal samples in the group of B. Lake \cite{ChillalIqbal18}, also present strong evidence for the manifestation of a quantum spin liquid in PbCuTe$_2$O$_6$. This is seen, first of all, in the presence of a diffuse continuum suggestive of fractionalized spinon excitations. In zero magnetic field, the scattering is concentrated on a sphere with a radius $|\mathbf{Q}|\approx0.8$\,\AA$^{-1}$ in momentum space, dispersing towards lower $|\mathbf{Q}|$ at higher energies. In the same work, DFT calculations have been used to re-estimate exchange interactions, and the results suggest that the two leading frustrated interactions, $J_1$ and $J_2$, are of almost equal strength. This is at variance with the original results of Koteswararao \textit{et al.} \cite{KoteswararaoKumar14}, where the hyperkagome interaction $J_2$ was found to be much stronger than all other interactions. However, the new set of parameters can be well reconciled with the experimental observations. By applying the pseudo-fermion functional renormalization group (PFFRG) approach to the spin model with four exchange parameters, $J_1\approx J_2>J_3>J_4$, the absence of long-range magnetic order could be confirmed for the $J_1/J_2$ ratios between 0.975 and 1.08, and the calculated value of $J_1/J_2\approx1.056$ for PbCuTe$_2$O$_6$ falls within this range \cite{ChillalIqbal18}. Although infinite classical ground-state degeneracy is strictly present only for vanishing $J_3$ and $J_4$ couplings, it has been suggested that these interactions are sufficiently weak in PbCuTe$_2$O$_6$, so that even in the full model considering further-neighbour interactions, long-range magnetic order is fully suppressed by quantum fluctuations.\vspace{-2pt}

\subsection{Magnetic minerals of the perovskite group}

It would be an inexcusable omission not to mention the perovskite-group minerals with their rich and complex structural hierarchy that was recently summarized by Mitchell, Welch, and Chakhmouradian \cite{MitchellWelch17}. While the studies of synthetic perovskite-structured compounds have been definitely too extensive to be covered in the scope of the present review, I would like to concentrate here only on some naturally occurring compounds, having no pretensions of comprehensiveness. The classical perovskite has a formula $ABX_3$, where $A$ and $B$ are cations, arranged in two interpenetrating simple-cubic sublattices, and $X$ is the anion arranged on an fcc sublattice. In the inverse-perovskite (or antiperovskite) compounds, the cations are replaced by anions, and vice versa, resulting in an $A_3BX$ structure \cite{Krivovichev08}. In spite of their diversity, the majority of the naturally occurring perovskite-group minerals are either nonmagnetic or include magnetic cations only as impurities. Nonetheless, magnetism can occur whenever a magnetic transition-metal or rare-earth ion occupies a cation site. This leads to a simple-cubic lattice of magnetic ions in the case of perovskites, that lacks geometric frustration, or to a geometrically frustrated fcc lattice in the case of antiperovskites. Additional interesting physics can then result from the lowering of crystal symmetry due to the cation or vacancy ordering.

\begin{figure}[t]
\begin{center}
\includegraphics[width=0.765\linewidth]{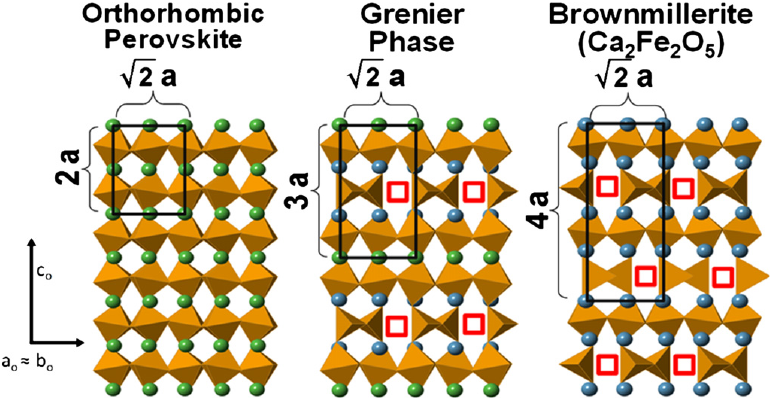}
\end{center}
\caption{Relationship between the orthorhombic perovskite, Grenier, and brownmillerite crystal structures. The red squares represent ordered oxygen vacancies. After Price \textit{et~al.}~\cite{PriceBrowning15}.}\label{Fig:Brownmillerite}
\end{figure}

For example, magnetic properties of the anion-deficient compounds that correspond to the minerals of the brownmillerite subgroup, such as Ca$_2$Fe$_2$O$_5$, attracted recent attention \cite{KagomiyaHirota17, PhanTho18}. The mineral Ca$_2$Fe$_2$O$_5$ has an orthorhombic vacancy-ordered structure (see Fig.~\ref{Fig:Brownmillerite}) and orders magnetically far above room temperature, below $T_{\rm N}\approx730$~K. The magnetic structure is slightly canted, resulting in weak ferromagnetism \cite{KagomiyaHirota17}. This canting is evidenced by the M\"ossbauer spectroscopy measurements and can be explained by the presence of DMI that is allowed due to the lowered crystal symmetry. The magnetic properties can be tuned by substituting La for Ca in Ca$_{2-x}$La$_x$Fe$_2$O$_5$, which tends to destabilize the brownmillerite structure in favour of the Grenier phase, LaCa$_2$Fe$_3$O$_8$ \cite{PhanTho18}. The ferromagnetic hysteresis loops tend to get larger upon La doping, showing a monotonic increase in the coercivity field between $x=0$ and $x=1$.

Magnetic ordering is also expected in $A$-site vacant double hydroxyperovskites that form either cubic or tetragonal structures, such as CuSn(OH)$_6$ (mushistonite), FeGe(OH)$_6$ (stottite), FeSn(OH)$_6$ (natanite), FeSn(OH)$_5$O (jeanbandyite), and MnSn(OH)$_6$ (wickmanite and tetrawickmanite) \cite{MitchellWelch17}. In both structures, magnetic ions are arranged into a face-centred cubic or tetragonal sublattice, which has a potential for geometric frustration. However, to the best of my knowledge, the magnetic properties of these hydroxide perovskites have not been investigated until now, neither theoretically nor experimentally. The same likely holds for the naturally occurring stoichiometric $B$-site ordered oxide double-perovskite Ca$_2$NbFe$^{3+}$O$_6$ (latrappite).

\section{Molecular magnets}

\subsection{Magnetism in spin clusters}

After having discussed one-, two-, and three-dimensional magnetic interaction networks that are found in natural minerals, we now turn to zero-dimensional structures represented by nearly noninteracting spin clusters. The simplest possible example of such a cluster is a spin dimer, which we already discussed in Chapter~\ref{Chap:Dimers}. However, larger magnetic clusters that consist of more than two spins can also be found in minerals. They often form regular geometric arrangements of ions that can be viewed as magnetic ``molecules'', arranged into a crystal by nonmagnetic spacer atoms. Such clusters would not be stable as standalone molecules outside of the crystal, which offers a natural connection between mineralogy and the highly active research field of molecular magnetism.

\begin{figure}[b]
\begin{center}
\includegraphics[width=0.53\linewidth]{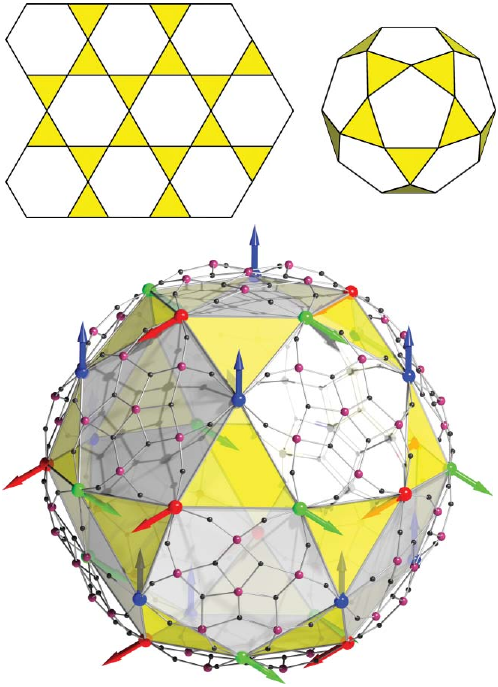}
\end{center}
\caption{An icosidodecahedron as a molecular analog of the kagome lattice (top). Its realization in the form of a Mo$_{72}$Fe$_{30}$ cluster is shown below. After K\"ogerler \textit{et~al.}~\cite{KoegerlerTsukerblat10}.}\label{Fig:Icosidodecahedron}
\end{figure}

Molecular nanomagnets and single-molecule magnets have been in the focus of condensed-matter research for several decades \cite{Schnack04, GatteschiSessoli06, SchnalleSchnack10, FurrerWaldmann13, BartolomeLuis14}. The finite number of spins in the cluster leads to a discrete spectrum of magnetic energy levels. Unlike conventional magnets, magnetic clusters cannot develop a long-range magnetic order. Instead, the ground state is protected by an energy barrier for spin reversal, so that magnetic relaxation slows down below some characteristic blocking temperature $T_{\rm B}$ that depends on the energy gap separating the ground state from the next excited energy level. This guarantees that the magnetic moments at sufficiently low temperatures acquire a certain static ground-state configuration, by analogy with an ordered state of an infinite system. The magnetic excitation spectrum of such a molecule corresponds to a discrete set of transitions to excited states with different total spin quantum number $S$, of which only a subset of transitions with $|{\Delta}S|\leq1$ can be probed by inelastic neutron scattering. This principle has been most vividly illustrated by Caciuffo \textit{et al.} \cite{CaciuffoAmoretti98} in an INS experiment performed on a butterfly-shaped iron molecular cluster with eight Fe$^{3+}$ ions.

Of particular interest are molecular nanomagnets that show geometric frustration \cite{SchmidtRichter05, SchroederNojiri05, KoegerlerTsukerblat10}. This sit\-u\-a\-tion occurs whenever the cluster contains antiferromagnetically interacting classical or quantum spins that are arranged in equilateral triangles or, more generally, form closed loops with an odd number of vertices. For instance, a ring of $N$ spins with only nearest-neighbour AFM interactions would be frustrated for every odd $N$ \cite{Schnack04}. In 3D, geometrically frustrated clusters can be exemplified by regular or quasiregular polyhedra that contain odd-sided facets. For instance, the Fe$_{30}$ spin icosidodecahedron depicted in Fig.~\ref{Fig:Icosidodecahedron} has been experimentally realized in the giant magnetic keplerate molecule Mo$_{72}$Fe$_{30}$ \cite{MuellerLuban01, GarleaNagler06}, one of the largest single-molecule magnets synthesized to date. It can be thought of as an analog of the kagome lattice on a sphere, consisting of corner-sharing equilateral triangles that are separated by pentagons (rather than hexagons in the planar kagome case)~\cite{RousochatzakisLaeuchli08}. A~cuboctahedron, realized in the natural mineral tsch\"ortnerite (see section \ref{Sec:Tschortnerite}), is the only other example of such an arrangement, where the corner-sharing triangles are separated by squares \cite{SchnackSchnalle09, SchnackWendland10}.

The presence of frustration may lead to ground-state degeneracy or to the presence of multiple magnetic configurations of the cluster that are very close in energy. In experiments, this can be seen as the appearance of discrete jumps in the field dependence of the magnetization \cite{Schnack04}, because even relatively weak magnetic fields (as compared to the saturation field) are sufficient to tip the energy balance towards another competing ground-state configuration~\cite{SchroederNojiri05}. From the theory point of view, the benefit of magnetic clusters is that their energy levels can be computed very accurately using exact diagonalization (at least for not too big clusters that preclude this approach due to computational constraints) \cite{SchmidtRichter05}. This offers not only a perfect opportunity for direct comparisons between theory and experiment but also allows theorists to predict magnetic properties of various spin clusters that have not been experimentally realized until now. For example, magnetization and specific heat have been calculated for clusters with icosahedral and dodecahedral symmetries under assumption of only nearest-neighbour Heisenberg AFM interactions \cite{Konstantinidis05, Konstantinidis15}.

\subsection{Olivenite and libethenite: a weakly interacting network of AFM tetramer units}

\begin{figure}[b!]
\begin{center}
\includegraphics[width=0.55\linewidth]{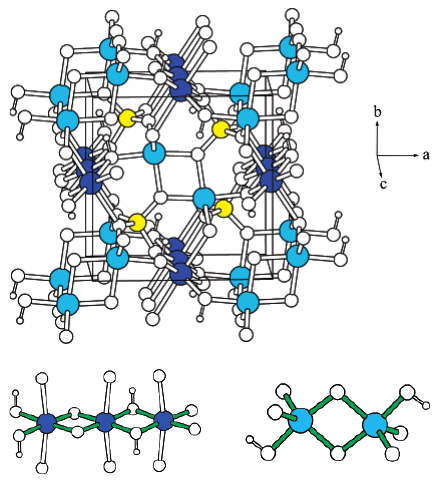}
\end{center}
\caption{The crystal structure of adamite-type minerals with its two building blocks: an edge-sharing [Cu(1)$_2$O$_6$(OH)$_2$]$_\infty$ chain and an edge-sharing Cu(2)$_2$O$_6$(OH)$_2$ dimer. After Belik \textit{et~al.}~\cite{BelikKoo07}.}\label{Fig:Libethenite}
\end{figure}

The minerals olivenite, Cu$_2$(AsO$_4$)(OH) \cite{Heritsch38}, and libethenite, Cu$_2$(PO$_4$)(OH) \cite{Cordsen78}, share the orthorhombic adamite-type structure (space group $Pnnm$) that is built up from linear chains formed by edge-sharing Cu(1)O$_4$(OH)$_2$ octahedra and dimers of edge-sharing Cu(2)O$_4$(OH) trigonal bipyramids \cite{RojoMesa96, BelikKoo07}, as shown in Fig.~\ref{Fig:Libethenite}. At the same time, specific heat, magnetization, and magnetic susceptibility measurements on synthetic libethenite samples indicated the absence of any long-range magnetic order down to 1.8~K and a spin gap of about 12~meV \cite{BelikKoo07}, suggesting the presence of magnetic clusters with a spin-singlet ground state. From the spin-dimer analysis based on extended H\"uckel tight-binding calculations, Belik \textit{et al.} \cite{BelikKoo07} concluded that the magnetic couplings both along the chains and within the dimers are negligibly weak, in accordance with the GKA rules for the Cu(1)--O--Cu(1) bond angles of 96.7$^\circ$ and 100.3$^\circ$, respectively. The strongest AFM superexchange interaction $J_1\approx12$~meV actually occurs between the Cu(1) and Cu(2) atoms mediated by the OH group with a large Cu(1)--O--Cu(2) bridging angle of 122.9$^\circ$, and the next-strongest interaction $J_2$ is given by the Cu(2)--O$\,\cdots\,$O--Cu(2) super-superexchange between the structural dimers. Therefore, the magnetic model of libethenite involves weakly interacting rhombic tetramer clusters coupled by $J_1$. This gives the natural explanation of the spin gap as the separation between the spin-singlet ground state and the excited triplet state of the cluster.

This scenario has been verified more recently by $^{31}$P NMR \cite{KuoLue08} and powder neutron spectroscopy measurements \cite{MatsudaDissanayake15}. The temperature-dependent shift of the NMR line points towards low-dimensional magnetism and can be perfectly described by the isolated spin-tetramer model. The spin-lattice relaxation rate measurements confirmed the presence of a 12~meV spin gap, which coincides with the leading exchange parameter $J_1$ \cite{KuoLue08} and perfectly agrees with the earlier results of Belik \textit{et al.} \cite{BelikKoo07}. This energy also coincides with the lowest dispersionless magnetic excitation ($E_1$) measured on a powder sample using neutron spectroscopy, which corresponds to the transition from the singlet ground state to the first excited triplet state. However, the second higher-energy excitation ($E_2$), which is expected at twice this energy ($\sim$\,24~meV) according to the simple tetramer model, was observed at a lower energy of $E_2=20$~meV. It corresponds to the transition from the ground state to the second excited triplet state. The observed ratio $E_2/E_1\approx5/3$ could be explained by including diagonal couplings of the order of 4~meV within the tetramer \cite{MatsudaDissanayake15}.

Apart from the copper-based spin-$\frac{1}{2}$ systems represented by olivenite and libethenite, the same adamite-type crystal structure is also realized in several magnetic systems with higher half-integer values of the spin. These include the synthetic compounds Co$_2$(PO$_4$)(OH) \cite{RojoMesa96, RojoMesa02} and Co$_2$(AsO$_4$)(OH) \cite{PedroRojo10} with $S=3/2$ and the mineral eveite, Mn$_2$(AsO$_4$)(OH), with $S=5/2$. However, their magnetic properties are drastically different. The Co compounds are known as 3D antiferromagnets \cite{RojoMesa96, RojoMesa02, PedroRojo10}, whereas partial substitution of Cu for Co in the (Co$_{1-x}$Cu$_x$)$_2$(OH)PO$_4$ series of solid solutions suppresses the ordering temperature and leads to the onset of low-dimensional magnetic properties \cite{PedroRojo07}. This crossover has been explained theoretically by a combination of structural distortions and changes in the localized band structure \cite{KarmakarYakhmi12}. Compounds with a partial substitution of Co$^{2+}$ ($S=3/2$) with Ni$^{2+}$ ($S=1$) \cite{PedroRojo06} and with the substitution of (PO$_4^{3-}$) by (AsO$_4^{3-}$) \cite{PedroRojo12} have been also studied. The corresponding Mn compounds, however, received much less attention with respect to their magnetic properties and would deserve a dedicated investigation.

\subsection{Boleite: Cu$_{24}$ clusters in the form of truncated cubes}

\begin{figure}[b]
\begin{center}
\includegraphics[width=0.89\linewidth]{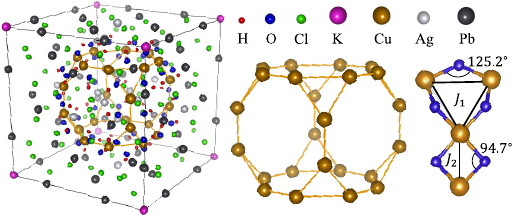}\vspace{-2pt}
\end{center}
\caption{The crystal structure of boleite: The unit cell (left), the Cu$_{24}$ cluster (middle), and its fragment showing relevant exchange interactions and bond angles (right). After Dreier \textit{et~al.}~\cite{DreierHolm18}.}\label{Fig:Boleite}
\end{figure}

The complex halide mineral boleite with the formula KPb$_{26}$Ag$_9$Cu$_{24}$(OH)$_{48}$Cl$_{62}$ \cite{Rouse73, CooperHawthorne00} forms large deep-blue cubic crystals. Its crystal structure, with the space group $Pm\overline{3}m$, has a large cubic unit cell with the lattice parameter $a=15.29$~\AA. All atoms in the unit cell, except for the Cu$^{2+}$ ions, are nonmagnetic. There are 24 structurally equivalent magnetic Cu$^{2+}$ ions per unit cell that form a cluster shaped as a truncated cube (see Fig.~\ref{Fig:Boleite}). The large distance between such clusters ensures that they are magnetically decoupled and can be seen as isolated molecular magnets. One expects two dominant superexchange interactions that act along the sides of the triangles, $J_1$, with the Cu--O--Cu bridging angle of 125.2$^\circ$, and along the rungs connecting them, $J_2$, with the bridging angle of 94.7$^\circ$ \cite{DreierHolm18}. According to the GKA rules, $J_1$ must be antiferromagnetic and therefore strongly frustrated, whereas $J_2$ is expected to be smaller due to the proximity of the Cu--O--Cu bridging angle to the point where the FM and AFM exchange contributions are compensated \cite{LeberneggTsirlin14}.

Magnetic properties of boleite were recently studied by Dreier \textit{et al.} \cite{DreierHolm18} using several natural mm-sized single crystals. They compared magnetic susceptibility measurements with the results of exact diagonalization to reveal the ground state of the Cu$_{24}$ cluster, quantify the exchange interactions, and understand the low-temperature spin dynamics. According to their results, the spin system in boleite is characterized by two distinct temperature scales. At temperatures below $\sim$\,100~K, the spins on individual triangles (trimers), bound by strong intratrimer interactions $J_1\approx17.9$~meV, freeze into a state that represents a linear combination of basis states of the form $\left|\uparrow\downarrow\downarrow\right\rangle$ and equivalent cyclic permutations thereof, resulting in an effective \mbox{$S=1/2$} degree of freedom on every triangle. At much lower temperatures, $T\lesssim5$~K, these effective spins that sit at the vertices of the cube undergo dimerization due to the weaker interactions $J_2\approx3.3$~meV. This results in a singlet ground state of the whole cluster, separated from the nearest magnetic excited state by a small spin gap of the order of 1~meV. The authors used a combined projectional and trimer (CPT) model with these parameters and could successfully describe the measurements in the whole measured range of temperatures \cite{DreierHolm18}.

Unfortunately, the need to rely on natural single crystals, that may vary in their chemical composition and structural quality, and the inability to synthesize the deuterated version of boleite artificially for neutron-scattering experiments have so far precluded more thorough investigations of their spin dynamics. In particular, the value of the spin gap has not been measured directly. Also, no magnetization plateaus or similar anomalies could be revealed, most probably due to the insufficient chemical and structural homogeneity of the natural samples. Neutron-scattering experiments, described in the appendix B of Ref.~\cite{DreierHolm18}, were essentially unsuccessful as they were unable to detect any magnetic signal, most likely because of the large incoherent-scattering background from hydrogen. Reportedly, even the use of polarized neutrons could not overcome this problem. On the other hand, the search for higher-energy excitations at the scale of $J_1$ proved to be very difficult due to the large number of optical phonons in this energy region. Notwithstanding, boleite certainly represents a unique natural realization of a molecular-magnet system, realizing a quantum $S=1/2$ magnetic cluster that has not been reported so far in any other compound. Therefore, the synthesis of chemically pure artificial boleite would certainly open up a new research direction for a number of experimental techniques, including high-field magnetization, local probes, and neutron spectroscopy.

Two minerals that are closely related to boleite are pseudoboleite, Pb$_{31}$Cu$_{24}$Cl$_{62}$(OH)$_{48}$ \cite{GiuseppettiMazzi92}, and cumengeite, Pb$_{21}$Cu$_{20}$Cl$_{42}$(OH)$_{40}$\,$\cdot$\,6H$_2$O \cite{HawthorneGroat86}. They have less symmetric tetragonal unit cells in which Cu$^{2+}$ ions occupy two inequivalent general Wyckoff positions. In pseudoboleite, the Cu$_{24}$ clusters have approximately the same shape and size as in boleite but are arranged in a body-centred tetragonal lattice, therefore one expects that the essential physical properties of individual clusters should remain the same. In cumengeite, the Cu$_{20}$ clusters are flattened, as would result from collapsing the truncated cube by eliminating the rungs parallel to the $c$ direction.

\subsection{Tsch\"ortnerite: cuboctahedral Cu$_{12}$ clusters}\label{Sec:Tschortnerite}

\begin{figure}[b]
\begin{center}\vspace{-1pt}
\includegraphics[width=0.495\linewidth]{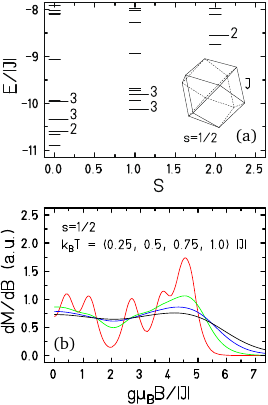}\vspace{-2pt}
\end{center}
\caption{(a)~Calculated low-lying energy levels for a regular cuboctahedral cluster of antiferromagnetically interacting spin-$\frac{1}{2}$ ions with an exchange constant $J$ \mbox{(see inset)}. Numbers near selected levels denote their multiplicities. (b)~Differential susceptibility of the same cluster as a function of applied field for several temperatures. After Schnack \textit{et~al.}~\cite{SchnackSchnalle09}.\vspace{-3pt}}\label{Fig:Cuboctahedra}
\end{figure}

In 1993, a new mineral of the zeolite group \cite{ArmbrusterGunter01} with a conspicuously large cubic unit cell (space group $Fm\overline{3}m$, lattice parameter $a=31.62$\,\AA) was discovered by the mineral collector Jochen Tsch\"ortner at the Bellberg volcano near Mayen, Eifel, Germany. It was named tsch\"ortnerite after a detailed structural characterization published 5~years later \cite{EffenbergerGiester98}. The mineral is very rare and is found only in small cubic light-blue transparent crystals with a maximum size up to 0.15\,mm, which can be described by the chemical formula Ca$_4$(K,Ca,Sr,Ba)$_3$Cu$_3$Al$_{12}$Si$_{12}$O$_{48}$(OH)$_8\cdot20$H$_2$O. Despite the very complex structure and chemistry of tsch\"ortnerite, its putative spin-$\frac{1}{2}$ magnetic sublattice is surprisingly simple and symmetric. The copper ions occupy a single Wyckoff position and form well separated Cu$_{12}$ clusters in the form of regular cuboctahedra, surrounded by nonmagnetic atoms. This mineral can therefore serve as a paradigmatic example of a crystal consisting of nearly noninteracting frustrated magnetic molecules with several nearest-neighbour superexchange paths, similar to the above-discussed boleite.

The relatively small size of the clusters allows for an exact diagonalization of its complete energy spectrum \cite{SchnackSchnalle09, SchnackWendland10}, which is reproduced here in Fig.~\ref{Fig:Cuboctahedra}\,(a). As it is common for many geometrically frustrated low-dimensional quantum magnets \cite{SchmidtSchnack05}, the ground state of the molecule is a spin singlet, and there are several low-lying excited singlet states below the first triplet. The rather high symmetry of the cluster leads to a high degeneracy among its energy levels. The lowest of the $S=0$ excitations defines an energy gap of $\sim{\kern.7pt}0.23{\kern.7pt}|J|$, whereas the first $S=1$ excitation is found at $\sim{\kern.7pt}0.77{\kern.7pt}|J|$ and is ninefold degenerate. The corresponding differential susceptibility as a function of an applied magnetic field, $g\mu_{\rm B}B/|J|$, was calculated for different values of the normalized temperature, $k_{\rm B}T/|J|$, as plotted in Fig.~\ref{Fig:Cuboctahedra}\,(b).\vspace{-2pt}

\section{Itinerant magnetism}

\subsection{Mackinawite: the natural prototype of an iron-based superconductor}

Practically all minerals discussed to this point were insulators. Indeed, there are very few examples of naturally occurring stoichiometric metallic minerals that show magnetic properties. I deliberately leave aside the discussion of nickel-iron alloys, for example, that occur in meteorites \cite{HerndonRowe74, StaceyBanerjee74}, as they are non-stoichiometric, well understood and described, and represent little interest from the point of view of fundamental solid-state physics. However, it is well known that some of the binary iron-chalcogenide compounds, for instance the metastable tetragonal iron-sulfide phase (space group $P4/nmm$) that is isostructural to the mineral mackinawite, $t$-(Fe,Ni)$_{1+x}$S$_{1-y}$, are metallic \cite{DeveyGrauCrespo08, SubediZhang08, TrescaGiovannetti17, KuhnKidder17}. Tetragonal iron sulfide recently became famous as an unconventional superconductor with a $T_{\rm c}$ of $\sim$\,4.5~K \cite{LaiZhang15, ZhangLiu17}, showing structural similarity to the superconducting iron arsenides \cite{Johnston10, Stewart11}. However, one should not forget that the interest in this compound existed much earlier in geochemistry and mineralogy \cite{LennieRedfern95, LennieVaughan96, LennieRedfern97, MulletBoursiquot02}. Unfortunately, naturally occurring mackinawite usually contains significant admixtures of other transition metals (Co, Ni, Cu) that can intercalate between the iron sheets, occupying interstitial sites \cite{KwonRefson15, WilkinBeak17}. Because it is known that as much as 10\% of Co or Ni impurities can suppress superconductivity in iron chalcogenides \cite{MizuguchiTomioka09}, it is doubtful that superconductivity can be found in natural mackinawite samples.

\begin{figure}[t]
\includegraphics[width=\linewidth]{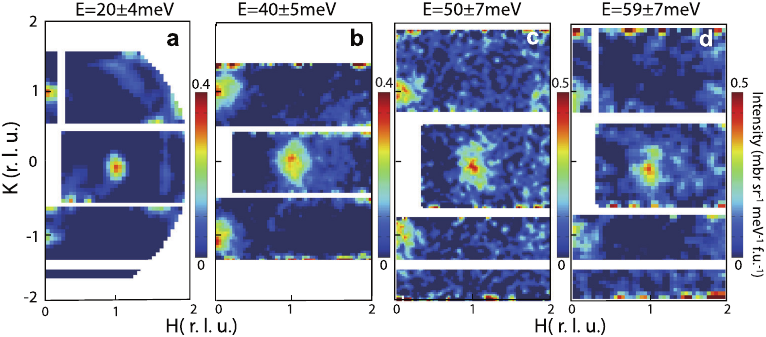}
\caption{Spin excitations in tetragonal FeS, measured by time-of-flight neutron spectroscopy at different energy transfers, as indicated above each panel. After Man \textit{et~al.}~\cite{ManGuo17}.}\label{Fig:FeS}
\end{figure}

Already in 1971, M\"ossbauer measurements on natural mackinawite were performed down to 4.2~K \cite{VaughanRidout71}. However, these early data showed no signatures of magnetic order, and a low-spin state of iron was concluded. Three decades later, low-temperature M\"ossbauer measurements were repeated on synthetic $t$-Fe$_{1+x}$S$_{1-y}$, revealing clear signatures of magnetic order \cite{MulletBoursiquot02}. From more recent studies \cite{KwonRefson11, KuhnKidder17}, it became known that this order represents a commensurate AFM structure with the propagation vector $\left(\frac{1}{4}\,\frac{1}{4}\,0\right)$ and a N\'eel temperature $T_{\rm N}\approx116$~K. It has been even suggested that this magnetic state may stabilize a topologically nontrivial band structure with symmetry-protected surface states \cite{HaoZheng17}. Remarkably, around the same time, other authors \cite{KirschnerLang16, ManGuo17} suggested that the magnetic order in mackinawite originates from impurities and is not an intrinsic property of pure $t$-FeS. In high-quality single crystals, neutron scattering found no signatures of static magnetic order down to 3~K \cite{ManGuo17}. Moreover, inelastic neutron scattering data, reproduced here in Fig.~\ref{Fig:FeS}, revealed dynamic spin-stripe fluctuations residing at the usual $(\pi,0)$ wave vector that corresponds to the nesting of the hole and electron Fermi surfaces, as in most other iron-based superconductors \cite{Dai15, Inosov16}. These fluctuations do not couple to superconductivity, suggesting that FeS is a weakly correlated analog of its sibling compound FeSe. This conclusion follows from a much smaller enhancement of the effective electron mass and from a much larger energy scale for spin fluctuations \cite{ManGuo17}.\vspace{-2pt}

\section{Summary and outlook}

The numerous case studies presented in this review illustrate that the world of minerals offers an inexhaustible source of materials and crystal structures as a playground for the realization of various models in quantum magnetism and for trying computational approaches of modern quantum chemistry. A considerable number of works summarized here were undertaken during the last two decades, marking an appearance of a new interdisciplinary field of research that bridges mineralogy with low-temperature condensed-matter physics. For most of the studied compounds, it was impossible to comprehend the underlying physics that determines their magnetic properties from just a superficial examination of their structure. On many occasions, the application of modern experimental and computational methods offered new surprises, and the ultimate understanding of the magnetic properties of most minerals was gained from a long sequence of trial and error that extended over several decades. At the same time, one has to admit that apart from a mere handful of examples presented here, the majority of known minerals have not been so far subjected even to the simplest low-temperature physical characterization, not even to magnetic susceptibility or magnetization measurements. In this respect, physicists have barely scratched the surface of the vast variety of magnetic minerals that could serve as subjects for future studies. There are ongoing systematic low-temperature investigations of natural samples from the world's largest mineral collections \cite{Feder14}. It is therefore interesting to speculate here about the potential interest that other naturally occurring crystal structures with magnetic ions could represent for the condensed-matter physics community.

There are abundant examples of copper oxysalt minerals realizing various spin-$\frac{1}{2}$ magnetic sublattices at low temperatures that have not, as of yet, received much attention from the point of view of their low-temperature physical properties. A general structural classification of cuprates illustrated with multiple examples has been discussed, for instance, by Leonyuk \textit{et al.} \cite{LeonyukBabonas98, LeonyukMaltsev01}, and the structural details resulting from Jahn-Teller effects in Cu$^{2+}$ oxysalt minerals have been reviewed by Burns and Hawthorne \cite{BurnsHawthorne96}. Some of these minerals realize very complex crystal structures, as it is the case for chloromenite, Cu$_9$O$_2$(SeO$_3$)$_4$Cl$_6$, that was recently identified as a promising multiferroic material \cite{HsiehWu18}. Its lattice contains five structurally inequivalent copper sites that order antiferromagnetically below the N\'eel temperature of $T_{\rm N}\approx16$~K, that can be further increased with magnetic fields. At a higher temperature of $T_{\rm E}\approx267$~K, an antiferroelectric transition takes place.

\begin{figure}[t!]
\begin{center}
\includegraphics[width=0.45\linewidth]{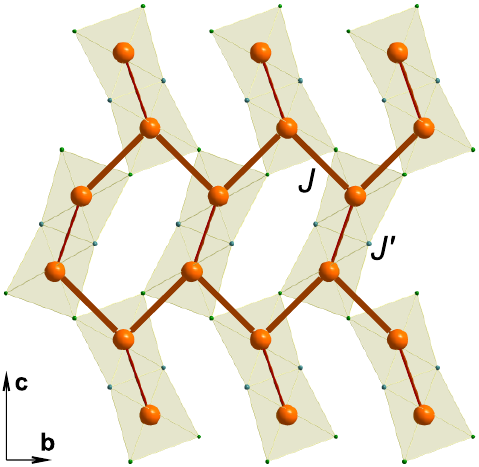}
\end{center}
\caption{The 2D network of structural dimers in nissonite, forming a coupled lattice of zigzag magnetic chains. Larger spheres denote Cu$^{2+}$ ions, small dots are oxygen atoms.\vspace{-2pt}}\label{Fig:Nissonite}
\end{figure}

A low-dimensional structure consisting of Cu$_3$(OH)$_2$ ferrimagnetic ribbons of edge-sharing copper octahedra is realized in the copper-molybdate mineral lindgrenite, Cu$_3$(MoO$_4$)$_2$(OH)$_2$ \cite{BaoKong06}. Neutron-diffraction measurements revealed that below the ordering temperature of $\sim$\,13~K, the magnetic moments on the two inequivalent copper sites both point along the $a$ axis and are antiparallel to each other, showing no signs of geometric frustration \cite{VilminotAndre06}. On the other hand, the dehydrated version of the same mineral, Cu$_3$Mo$_2$O$_9$ \cite{VilminotAndre09}, forms distorted tetrahedral spin chains running parallel to the $b$ axis that support a weakly ferromagnetic noncollinear magnetic order as a result of frustration \cite{HaseKuroe15}.

A much simpler two-dimensional arrangement of structural dimers, shown in Fig.~\ref{Fig:Nissonite}, occurs in the copper phosphate mineral nissonite, Cu$_2$Mg$_2$(PO$_4$)$_2$(OH)$_2$\,$\cdot$\,5H$_2$O, crystallizing in the space group $C2/c$ \cite{GroatHawthorne90}. According to GKA rules, it should be described by two dominant AFM interactions, resulting in coupled AFM zigzag chains. It would be therefore interesting to investigate its magnetic ground state and study its proximity to a magnetic QCP. Another phosphate mineral cornetite, Cu$_3$(PO$_4$)(OH)$_3$, features buckled layers consisting of edge-sharing zigzag chains of Jahn-Teller distorted octahedra, cross-linked by edge-sharing octahedral dimers \cite{EbyHawthorne89}. Edge-sharing sheets consisting of distorted copper-ion polyhedra are also found in pseudomalachite, Cu$_5$(PO$_4$)$_2$(OH)$_4$ \cite{ShoemakerAnderson77}, its isostructural arsenate analog cornwallite, Cu$_5$(AsO$_4$)$_2$(OH)$_4$, and its polymorph cornubite \cite{ArltArmbruster99}. Another arsenate mineral bradaczekite, NaCu$_4$(AsO$_4$)$_3$, features chains formed by edge-sharing CuO$_6$ octahedra and AsO$_4$ groups, linked into 2D sheets \cite{KrivovichevFilatov01}.

\begin{figure}[b]
\begin{center}
\includegraphics[width=0.47\linewidth]{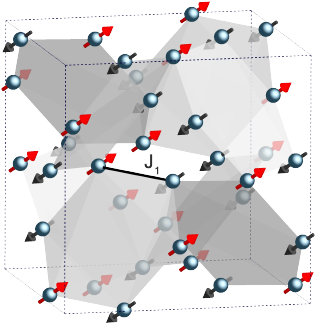}\vspace{-2pt}
\end{center}
\caption{The magnetic structure of Cu$_3$TeO$_6$ after Bao~\textit{et~al.}~\cite{BaoWang18}.}\label{Fig:Cu3TeO6}
\end{figure}

A very rich variety of low-dimensional structures can be also found in copper sulfate \mbox{minerals \cite{HawthorneKrivovichev00, SchindlerHuminicki06}}. To name just a few examples, the mineral leightonite, K$_2$Ca$_2$Cu(SO$_4$)$_4$\,$\cdot$\,2H$_2$O \cite{MenchettiBindi02}, has two sets of linear copper chains that run at 65.5$^\circ$ to each other. In cyanochroite, K$_2$Cu(SO$_4$)$_2$\,$\cdot$\,6H$_2$O \cite{BosiBelardi09}, and in kr\"ohnkite, Na$_2$Cu(SO$_4$)$_2$\,$\cdot$\,2H$_2$O \cite{HawthorneFerguson75}, the Cu$^{2+}$ ions are well sep\-a\-rat\-ed by SO$_4$ groups, water molecules, and alkali-metal ions, resulting in large Cu--Cu distances with no immediate superexchange paths. This could result in exceedingly small magnetic interactions and spin dynamics, \mbox{restricted to very low energies}.

Recent years also saw heightened interest to the copper-tellurium oxide minerals \cite{ChristyMills16, Norman18}. Some of them have been already discussed in sections \ref{Sec:MapleLeaf} and \ref{Sec:Choloalite} as prospective maple-leaf lattice antiferromagnets or promising quantum spin-liquid candidates. Very recently, another copper-tellurium compound Cu$_3$TeO$_6$, naturally occurring as the green mineral mcalpineite \cite{CarboneBasso13}, was identified as a host of topological magnon excitations \cite{YaoLi18, BaoWang18}. The interesting geometry of its magnetic lattice, composed of almost planar regular vertex-shared hexagons of Cu$^{2+}$ $S = 1/2$ spins, has been noted already in 2005, as it was first characterized by magnetic susceptibility, torque magnetometry, and neutron powder diffraction measurements \cite{HerakBerger05}. This structure became known as a ``spin web'' \cite{ManssonPrsa12}. Large synthetic single crystals of Cu$_3$TeO$_6$ have been produced by the PbCl$_2$ flux method \cite{HeItoh14}. The compound orders magnetically below $T_{\rm N}\approx61.7$~K, developing a collinear AFM order as shown in Fig.~\ref{Fig:Cu3TeO6}. Such a magnetic state has a $P{\kern-.5pt}T$ symmetry ($P$ and $T$ being space-inversion and time-reversal operations, respectively), which implies that magnons are expected to exhibit nontrivial topological properties, as it was recently confirmed in INS measurements \cite{YaoLi18, BaoWang18}. In a recent review, Norman \cite{Norman18} also emphasized that the mineral leisingite, MgCu$_2$TeO$_6$(H$_2$O)$_6$ \cite{MargisonGrice97}, is a realization of a spin-$\frac{1}{2}$ honeycomb lattice with an expected compensation of superexchange interactions along the Cu--O--Cu paths, whereas jensenite, Cu$_3$TeO$_6$(H$_2$O)$_2$ \cite{GriceGroat96}, contains distorted brucite-like honeycomb layers in combination with isolated dimers~\cite{ChristyMills16}.

\begin{figure}[t!]
\begin{center}
\includegraphics[width=0.52\linewidth]{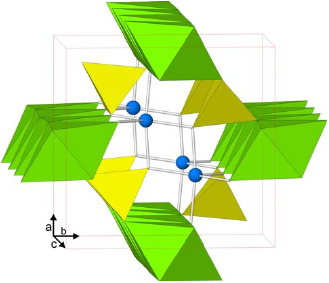}
\end{center}
\caption{The crystal structure of auriacusite after Mills \textit{et~al.}~\cite{MillsKampf10}. The spheres represent Fe$^{3+}$ ions, Cu octahedra are shown in green, and As tetrahedra in yellow.}\label{Fig:Auriacusite}
\end{figure}

\begin{figure}[t]
\begin{center}
\includegraphics[width=0.82\linewidth]{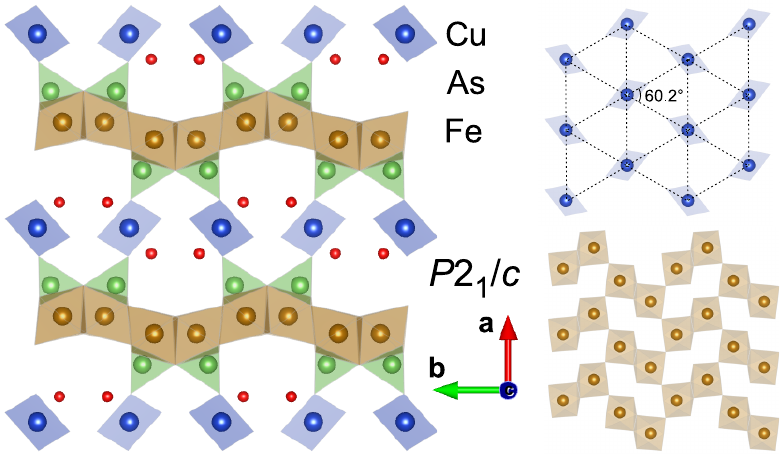}
\end{center}
\caption{The crystal structure of arthurite \cite{KellerHess78}. The right panels show copper (top) and iron (bottom) planes viewed along the $a$ axis.}\label{Fig:Arthurite}
\end{figure}

Finally, crystal structures that combine Cu$^{2+}$ spin-$\frac{1}{2}$ moments with other magnetic ions are also common. For instance, the new arsenate mineral auriacusite of the olivenite group, Fe$^{3+}$Cu$^{2+}$AsO$_4$O, with the structure shown in Fig.~\ref{Fig:Auriacusite}, has a network of copper octahedra, connected via magnetic Fe$^{3+}$ sites \cite{MillsKampf10}. Another arsenate mineral arthurite, CuFe$^{3+}_2$(AsO$_4$)$_2$(OH)$_2$\,$\cdot$\,4H$_2$O, has a monoclinic structure (space group $P2_1/c$), in which copper and iron planes alternate along the $a$ axis \cite{KellerHess78}. The copper planes have a nearly perfect triangular-lattice structure, consisting of isosceles triangles with the base of 5.60\,\AA\ and the sides of 5.58\,\AA, whereas the iron planes consist of structural dimers formed by edge-sharing octahedra, which are in turn arranged by corner sharing into buckled layers as shown in Fig.~\ref{Fig:Arthurite}. Then, a newly discovered mineral iyoite, MnCuCl(OH)$_3$ \cite{NishioHamane17}, is a Mn-Cu ordered analogue of botallackite. Its structure is based on brucite-like sheets orthogonal to the $a$ axis, built from distorted edge-sharing Mn(OH)$_5$Cl and Cu(OH)$_4$Cl octahedral chains that alternate along the $c$ direction. It is therefore expected to show magnetic properties arising from an interaction of quantum and classical spins on two interpenetrating quasi-low-dimensional sublattices.

The examples given here are intended to show that the investigations of quantum magnetism in crystal structures motivated by naturally occurring minerals are far from being complete. Many new studies appear every year, which are concerned with the physical problems that are absolutely central for modern solid-state physics. Future progress will certainly depend on the ability to reproduce complex mineral structures in the form of chemically pure synthetic samples, and the incessant discoveries of new minerals should guide chemists in this endeavor. In this emerging field of knowledge, new results are expected from fruitful collaborations among mineralogists, crystallographers, chemists and crystal growers, experimental solid-state physicists, and condensed-matter theorists.

\section*{Acknowledgments}

I would like to thank H.~Rosner (MPI CPfS, Dresden) and K.~Lefmann (Niels Bohr Institutet, Copenhagen) for stimulating discussions, as well as E.~Kroke, J.~Hunger, M.~Schwarz, U.~Bl\"a\ss, and Th. Schlothauer for the inspiration gained from visiting the mineral collection at the TU Bergakademie Freiberg. I also thank A.~S. Cameron for a thorough proofreading of the manuscript and A.\,A.~Tsirlin for very helpful critical suggestions on its content. This work has been funded by the German Research Foundation (DFG) within the Collaborative Research Center SFB~1143 in Dresden (project C03) and the W\"urzburg-Dresden Cluster of Excellence on Complexity and Topology in Quantum Matter~--~\textit{ct.qmat} (EXC 2147, project-id 39085490).\vspace{-2pt}

\bibliographystyle{tADP}
\bibliography{Minerals}

\end{document}